\documentclass[10pt]{iopart}

\usepackage{iopams}  
\usepackage{bbm}

\usepackage[german,american,english]{babel}
\usepackage{graphicx}
\usepackage{graphics}
\usepackage{bm}
\usepackage{epstopdf}

\newcommand {\bkt} [1] {\langle #1 \rangle}
\newcommand {\dbkt} [2] {\langle #1 | #2 \rangle}

\newcommand {\pd} [2] {\frac{\partial #1}{\partial #2}}
\newcommand {\td} [2] {\frac{d #1}{d #2}}

\newcommand {\beq}{\begin{equation}}
\newcommand {\eeq}{\end{equation}}
\newcommand {\beqn}{\begin{eqnarray}}
\newcommand {\eeqn}{\end{eqnarray}}
\newcommand {\bit}{\begin{itemize}}
\newcommand {\eit}{\end{itemize}}

\newcommand{\ba}{\begin{array}{rl}}
\newcommand{\ea}{\end{array}}
\newcommand{\bc}{\begin{cases}}
\newcommand{\ec}{\end{cases}}

\newcommand{\om}{\iffalse}

\begin{document}

\topical[Transport in two-dimensional topological materials]
{Transport in two-dimensional topological materials: recent developments in experiment and theory}

\author{Dimitrie Culcer$^1$, Aydin Cem Keser$^1$, Yongqing Li$^2$, Grigory Tkachov$^3$}

\address{$^1$ School of Physics and Australian Research Council Centre of Excellence in Low-Energy Electronics Technologies, UNSW Node, 
The University of New South Wales, Sydney 2052, Australia}
\ead{d.culcer@unsw.edu.au}
\ead{a.keser@unsw.edu.au}
\vspace{10pt}

\address{$^2$ Beijing National Laboratory for Condensed Matter Physics, Institute of Physics, Chinese Academy of Sciences, Beijing 100190, China \\
School of Physical Sciences, University of Chinese Academy of Sciences, Beijing 100049, China}
\ead{yqli@iphy.ac.cn}\vspace{10pt}

\address{$^3$ Institute of Physics, Augsburg University, 86135 Augsburg, Germany}
\ead{gregor.tkachov@physik.uni-augsburg.de}

\vspace{10pt}
\begin{indented}
\item[] July 2019
\end{indented}

\begin{abstract}
We review theoretical and experimental highlights in transport in two-dimensional materials focussing on key developments over the last five years. Topological insulators are finding applications in magnetic devices, while Hall transport in doped samples and the general issue of topological protection remain controversial. In transition metal dichalcogenides valley-dependent electrical and optical phenomena continue to stimulate state-of-the-art experiments. In Weyl semimetals the properties of Fermi arcs are being actively investigated. A new field, expected to grow in the near future, focuses on the non-linear electrical and optical responses of topological materials, where fundamental questions are once more being asked about the intertwining roles of the Berry curvature and disorder scattering. In topological superconductors the quest for chiral superconductivity, Majorana fermions and topological quantum computing is continuing apace.  
\end{abstract}

%
%
%
\maketitle
%
\ioptwocol
\tableofcontents

\section{Introduction}

The new millennium has witnessed the seemingly inexorable rise of topological phenomena, culminating with the 2016 Nobel Prize in Physics. The term \textit{topological materials} encompasses a broad range of structures that exhibit topological phases \cite{Niu_Review_2016, CulcerGeresdi}, which can be characterised by a topological invariant that remains unchanged by deformations in the system Hamiltonian, e.g. the $Z_2$ invariant, the Chern number, or more generically the Berry curvature, whose integral over the Brillouin zone yields the Chern number. Topological materials include topological insulators (TI) \cite{Shen_Book, Ando-TI}, Weyl and Dirac semimetals (WSM, DSM)  \cite{Armitage_Weyl_RMP_2018}, transition metal dichalcogenides (TMD) \cite{Dixiao_MoS2_2012}, which are often termed \textit{2D materials} by themselves, graphene subject to a proximity effect \cite{TPE_Nanoscale14, TPE_CPB14}, and topological superconductors  \cite{Shen_Book}. The distinction is not always clear cut, since some categories overlap: for example, certain dichalcogenides can become topological insulators under appropriate circumstances \cite{Okugawa_PRB2014, Collins_Nature2018}, while others, such as WTe$_2$, can be Weyl semimetals.

2D topological materials offer opportunities that did not exist previously. The electron gas is on the surface and is directly accessible, unlike semiconductor heterostructures, where it is buried. This facilitates excellent electrostatic control over the conduction in the 2D channel, which enables transistor applications, as well as accessibility to light for optical applications. The major advantages of TMDs is that they can be made atomically thin and have a direct band gap, making them ideal for optical emitters and detectors. They exhibit strong excitonic effects and piezoelectricity \cite{Rostami_NPJ2DM2018}. WSMs and TMDs can have very high mobilities. Many topological materials exhibit very strong spin-orbit coupling, which, under appropriate conditions, yields dissipationless edge state conduction without large magnetic fields, leading to potential transistor applications through the quantum spin and anomalous Hall effects. An example is provided by WSe$_2$, which has an extraordinarily large spin-orbit coupling, a wide direct band gap, and especially a strong anisotropic lifting of the valley degeneracy in a magnetic field, which makes it ideal for accessing the valley degree of freedom. Likewise, in TI spin-orbit torques have taken off spectacularly, especially since the low mobilities of TIs are not a concern as long as large currents can be achieved by increasing the electron density, as was done in Bi$_2$Te$_3$ \cite{Wang_NC2017}. Topological superconductivity has seen spectacular growth, with exciting developments in achieving chiral superconductivity, Majorana edge modes \cite{Avila_Majorana_2018}, the fractional Josephson effect and unconventional Cooper pairing.

In direct analogy with the rise of graphene, topological materials have matured into a \textit{topological zoo} with broad applications across different fields. In this review we emphasise this breadth of interest while bringing out conceptual unifying features such as spin- and pseudospin-charge coupling, the Berry curvature and inter-band effects and their interplay with disorder, and Cooper pairing in topological materials. We concentrate on the most actively researched two-dimensional transport phenomena in TI, TMD, WSM Fermi arcs and TSC, and include highlights from optical studies, since transport and optical responses are intertwined, and are broadly described by linear response theory. They must frequently be considered on the same footing, for example in determining the non-linear optical response, in which a DC \textit{shift} or \textit{rectification} current is also generated. 

The bulk of the review focuses on surface states. We first present a conceptual overview of the subject, followed by a review of transport in non-superconducting topological materials. The theoretical component of this section centres on insights obtained from linear-response theory and extensions thereof, while the experimental component reviews progress in the laboratory. After introducing the model Hamiltonians and outlining the concepts behind linear response and inter-band coherence, we discuss transport in topological insulators, with particular emphasis on weak localisation and anti-localisation, the quantum Hall effect, spin-orbit torques and their relation to the current-induced spin polarisation and spin-Hall transport, magnetoresistance and the anomalous Hall effect. We pay special attention to the continuing controversy surrounding topological protection and transport by the edge states of topological insulators. The anomalous Hall effect has an extension in transition metal dichalcogenides, which have multiple valleys, and exhibit a valley-Hall effect, which is reviewed next. In Weyl semimetals we focus on Fermi arcs and their manifestation in transport. We discuss the non-linear electrical response, a burgeoning field with important unanswered theoretical questions and potential applications in photovoltaics and solar energy. The latter half of the review is devoted to the latest theoretical and experimental developments in chiral superconductivity, Majorana edge modes and related phenomena in topological superconductivity. 

\section{Background}

In this section we attempt a conceptual summary of topological materials and Weyl-Dirac physics in a condensed matter context, and transport phenomena in these materials. Our aim is to provide basic explanations for commonly encountered terms in the literature.

\textit{Degeneracy points.}
In resonators and crystals, the spectrum can be drawn as a function of wave/crystal momenta. If the energy levels are degenerate at a fixed momentum, a perturbation would lift the degeneracy, a phenomenon called avoided crossings. However in a two band system, the perturbation moves the  degeneracy point in momentum space rather than rendering  the spectrum gapped. This is because, a $2\times 2$ Hermitian matrix can be  decomposed in terms of the identity and the Pauli matrices,
\begin{equation}
\label{hermitian}
H_{2\times 2} = a_0(\mathbf{k}) \sigma_0 + a_i(\mathbf{k}) \sigma_i 
\end{equation}
and for the eigenvalues to coincide viz. $E_1 = E_2 = a_0$, three functions $a_i$ must vanish. If this accidentally happens at the point $\mathbf{k}^*$, then a small perturbation would simply move this point in the Brillouin zone (BZ). The situation was well understood since the early days of quantum mechanics~\cite{degeneracy}. It was later understood for example that the dispersion looks conic around the accidental degeneracy, dubbed diabolical point~\cite{diabolical}. This fact can easily be seen by linearizing Eq.~\ref{hermitian} around $\mathbf{k}^*$. For simplicity, if we assume that this cone is isotropic, the Hamiltonian around the degeneracy point behaves either like $H = v_F \pmb{\sigma}\cdot \mathbf{k}$, or $-v_F \pmb{\sigma}\cdot \mathbf{k}$ up to an additive constant. Save for the value of the Fermi velocity $v_F$, these are the chiral and anti-chiral Weyl partners that make up a Dirac fermion, such as a relativistic electron, albeit in the limit of zero rest mass. 

\textit{Dirac/Weyl equation, chirality.}
Let us briefly explore this analogy with relativistic motion of high energy particles. Dirac equation describes positive/negative energy states describing particles/anti-particles with spin. A particle can possess a spin vector that is parallel/anti-parallel to its orbital motion. This is captured in the helicity operator $\pmb{\sigma} \cdot \mathbf{k}$. Since, massless particles travel at the speed of light, this property is Lorentz invariant and equivalent to `chirality'. Our linearized Hamiltonian \textit{is} proportional to the chirality operator, hence the sign of Fermi velocity gives the chirality. If around the degeneracy point, the cone is not isotropic, we can simply contract or stretch this cone to make it so, and use this straight forward definition. In other words, if we expand $a_i (\mathbf{k}) \sigma_i\approx a_i^j (k_j-k^*_j)\sigma_i$, the sign of the determinant of $a_i^j$ doesn't change if we stretch or rotate the cone, hence $\textrm{sgn}(det(a^i_j))$ defines chirality. 

\textit{Location of Weyl points in BZ.}
We should not take this analogy too literal, because crystals can have a lot of different properties while the space-time vacuum is constrained by many symmetries.   
For example, the excitations in  a crystal do not have to obey a Lorentz symmetry,  hence the cone around the degeneracy point can be anisotropic and arbitrarily tilted.   
Moreover, we can have a degeneracy at $k^*_+$, with positive chirality and a negative chirality counter-part at $k^*_- \neq k^*_+$. So unlike the ordinary (relativistic) massless Dirac fermion where the chiral partners exist at the same point, we can have degeneracy points with either chirality at various locations in the Brillouin zone.

\textit{Total chirality, Nielsen-Ninomiya Theorem.}
Indeed, every positive chirality degeneracy point must come with its negative chirality counter-part as long as the bands are defined on a periodic structure, that is the BZ. This statement is called the Nielsen-Ninomiya theorem. Instead of a rigorous proof, we will give an intuitive explanation. If we cut a $1$D dispersion relation curve $f(k_z)$, with a constant energy line, we intersect the curve  various times. We can convince ourselves that if the band is a smooth curve, which it is, we intersect it at even number of points, moreover, at half of these points, the curve has positive slope and  at the other half, negative. In three dimensions, suppose that the degeneracy points lie on the $k_z$ axis. Linearizing the Hamiltonian in the $k_x-k_y$ direction looks for example like, $H = f(k_z) \sigma_z + k_x \sigma_x + k_y\sigma_y$. Since $f(k_z)$ is periodic in BZ, the linearized Hamiltonian around the degeneracy points are $k_x \sigma_x + k_y \sigma_y \pm k_z \sigma_z$, hence have opposite chirality. 

\textit{Weyl points under discrete symmetries.}
As long as we have center of inversion in the crystal, that is inversion symmetry $\mathcal{I}$, we know where to find the chiral partners. If we have a Weyl point at $\mathbf{k^*}$, say with cone $+\pmb{\sigma}\cdot \mathbf{(k-k^*)}$, we must have $-\pmb{\sigma}\cdot \mathbf{(k+k^*)}$ at $\mathbf{-k^*}$. These are nothing but the $+/-$ chiral Weyl points.  Similar analysis applies if the crystal has mirror planes. Another important discrete symmetry is  time reversal $\mathcal{T}$. Intuitively, when the direction of time is reversed, so does that of momentum and spin. Since chirality is spin component in the direction of momentum, it is invariant under $\mathcal{T}$. In the condensed matter case, instead of spin we are talking about the band index or eigenvector of $\sigma$, usually called pseudo-spin. In this case, $\mathcal{T}$ switches the sign of $\mathbf{k}$ and $\sigma_y$ which leaves chirality invariant.  Therefore if there is $\mathcal{T}$-symmetry, the number of  Weyl points must be a multiple of four: for every pair that are $\mathcal{T}$-partners with positive chirality  another pair with negative chirality must exist so that the total chirality is zero. We must be more careful, when both inversion and time-reversal symmetries present in the system like, as we discuss below.  

\textit{Dirac points and mass gap instability.}

In an actual crystal we have many bands. We can argue that the other bands are further away in energy so we can apply the analysis to those two bands that cross each other and treat the rest as conduction and valence bands. But what if, the bands come in completely degenerate pairs? 
Indeed this is the case, if we have both  $\mathcal{T}$ (there is no magnetization in any form) and  $\mathcal{I}$ (there is a center of inversion). For example, if we have  an electronic spin up state at $\mathbf{k}$, then the time reversed partner state sits at $-\mathbf{k}$ with spin down. 
Moreover, by inversion symmetry, another state with spin down must sit right on top of the one with spin up at $\mathbf{k}$. There is no surprise here, electrons come with spin partners and if there is no magnetic couplings the spin is treated as dummy, hence there are at least two electrons at any momentum.
A perturbation can not open a gap as long as it does not contain spin dependent forces, hence all the analysis so far goes as is, save for a doubling of every Weyl point. However, spin-orbit coupling exists  and sometimes very strong in these materials. Therefore spin degeneracy is already lifted.  Nevertheless, the existence of $\mathcal{IT}$ symmetry still guarantees two-fold degeneracy. This fact is known as Kramer's theorem, that we can informally illustrate as follows. We have already mentioned that, the existence of $\mathcal{T}$, requires that a Weyl point at $k^*$ in BZ is accompanied by another with the same chirality at $-k^*$. On the other hand $I$ requires that we have opposite chirality Weyl point at $k^*$ again! So, after all, in case there is $\mathcal{IT}$, we must have two opposite chirality Weyl fermion sitting on top of each other. This means isolating  the $2\times2$ subspace like we did in Eq.~\ref{hermitian} is not possible, because a generic perturbation applies on a $4\times 4$ subspace that has $\mathcal{IT}$. It turns out that, we can decompose such a matrix in terms  of $5$
generators instead of the $3$ Pauli matrices. Therefore, we need $5$ functions, each taking $3$ momenta as parameters, to vanish. This is not possible unless enforced by additional symmetry or fine tuning. Once it is achieved, the resulting system is called a Dirac semi-metal, owing to the fact that two chiral Weyl partners at the same momentum point make up a Dirac particle with mass zero. Even when a gap opens, the resulting massive Dirac fermion might not be a  trivial insulator. 

\textit{Topological insulator as a Dirac system, gapless boundary modes.}
We can see what happens to the linearized band structure when Weyl fermions of opposite chirality coincide at the same momentum point. A perturbation can open a mass gap $m$ of either sign that renders the linearized Hamiltonian  
\begin{equation}
	H = \left(
	             \begin{array}{cc}
		-v_F\pmb{\sigma} \cdot \mathbf{(k- k^*)} &  m\\
		m & v_F\pmb{\sigma} \cdot \mathbf{(k- k^*)} 
	 \end{array}
	 \right)
\end{equation}
that is the four component Dirac equation with mass $m$.
The real surprise comes, when we impose a boundary to this material where $m$, switches sign. If the boundary is defined by $z = 0$, with $m = |m| \textrm{sgn}(z)$,  it turns out that there must be a localized state 
$$|\psi\rangle = e^{- \int_0^z m(z') dz'}  (|\psi_1\rangle, |\psi_2\rangle)^T$$
where $\psi_1$ and $\psi_2$ are 2-spinors that satisfy
$$|\psi_1\rangle = -i\sigma_z |\psi_2\rangle$$
To solve the Dirac equation,  the two-spinor $\psi_2$ must be an eigenstate of 
\begin{equation}
	H_{boundary} = v_F\sigma_x (k_x-k_x^*) + v_F\sigma_y (k_y-k_y^*)
\end{equation}
This means we have a gapless mode localized at the interface. 
In general, there is a gapless Hamiltonian, proportional to  $\pmb{\sigma}_\parallel \cdot \mathbf{k}_{\parallel}$ parallel to the surface. Ordinary insulators, including the vacuum obey the Dirac equation, with a positive mass. Therefore, if a gapless surface mode does not exist on the boundary of an insulator, we call it ordinary, if it does, the crystal is called a topological insulator (TI). This fact has a microscopic interpretation. The bands are nothing but coalesced atomic orbitals. If strong spin orbit coupling reverses the energy order of atomic orbitals, the band gap is inverted, so does the sign of the Dirac mass $m$. At a boundary with an ordinary insulator, the atomic orbitals should go back to their natural order at which point they have to meet at the same energy, so the gap must close at the boundary. 

\textit{Number of gapless surface mode as a topological invariant.}
If the number of boundary modes were to exceed 1, we would be able to gap them out in pairs without breaking $\mathcal{T}$. For example, if we have two boundary modes $\sigma_x k_x + \sigma_y k_y \pm m \sigma_z $ are both gapped and time reversal partners. Therefore the number of gapless boundary modes is either 1 or 0,  which is called the $Z_2$ invariant.

\textit{Chern insulator.}
Breaking $\mathcal{T}$ at the boundary, say by depositing magnetic impurities on the surface, can gap out the only gapless mode on the surface of a topological insulator. Now the boundary is described by the 2D massive Dirac Hamiltonian
\begin{equation}
	H_{2D}  = \sigma_x k_x + \sigma_y k_y + m\sigma_z
\end{equation}
called the \textit{Chern insulator}. Just as in the 3D case, there will be a gapless mode at the boundary where the mass switches sign, say at $x=0$ from $-|m|$ to $|m|$. Since $k_y$ is a good quantum number, the ansatz $|\psi\rangle  = e^{-\int_0^x m(x') dx'} e^{i k_y} |\sigma_y = 1 \rangle$ solves the eigenvalue equation. This mode has the dispersion $k_y$, hence is chiral, meaning it only propagates in $+y$ direction and can not backscatter. Indeed it is the boundary mode of the quantum Hall insulator. 
The above discussion implies that we can obtain  quantum Hall effect if  we take a spherical topological insulator and pierce it with a magnetic field. The gapless surface modes on the upper and lower hemispheres will acquire gaps with opposite signs due to the magnetic field. A chiral mode will develop on the equator. If we put terminals on two antipodal points on the equator and pass current between them, being chiral, only half of the equator will carry the current. The excess electrons compared to the other half, leads to a transverse voltage. We can compute the ensuing Hall conductivity by counting the excess charge in a 1D channel.
If the current is $I$, then $N  =I/(ev_F L)$ additional electrons populate one side of the equator with length $L$. The difference in transverse chemical potential due to this many excess electrons is found by using the dispersion relation $\hbar v_F \delta k =  h v_F N L = h I/e  = e \Delta V $. This means the Hall conductivity is $\sigma_{xy} = e^2/h$.  This is the quantum anomalous Hall effect (QAHE).  The time reversal breaking is usually achieved by magnetic impurities deposited on the surface of real topological insulators. 

\textit{Topological state of graphene, Haldane model.}
We also have a strictly 2D realization of the 2D Dirac Hamiltonian, that is graphene. Graphene comes with two copies of Dirac fermions due to having two high symmetry points in the BZ, excluding the spin degeneracy. These can acquire equal and opposite mass gaps when $\mathcal{T}$ is intact but $\mathcal{I}$ is broken, say due to sublattices being at different energy. The resulting system is an ordinary insulator. If on the other hand, $\mathcal{T}$ is broken, as in Haldane model~\cite{Haldane_ParAnom_PRL88}, the two fermions acquire the same mass, the resulting becomes a Hall insulator, just like the surface of a TI with broken $\mathcal{T}$. However, in actual graphene, there is an additional spin degeneracy. If a spin-orbit coupling exists, and $\mathcal{T}$ is intact, we have two copies of the Haldane model that are time-reversals of each other. The chiral edge modes then counter propagate and therefore the Hall conductivity vanishes. However, since these modes have opposite momentum, and spin is locked to momentum, in the presence of electric field, they create a net spin  current and produce the quantum spin Hall effect (QSHE)~\cite{KaneMele_QSHE_PRL05}

\textit{Fermi arcs.}
The relation between the bulk band structure and the boundary modes is one of the key points in the study of topological systems. These modes due to their protected structure has unusual transport signatures and lead to precise quantization of transport coefficients such as Hall conductivity. In addition to the gapless surface mode of a topological insulator and the chiral edge modes of a gapped surface mode, we also have chiral modes due to Weyl fermions called Fermi arcs. If the Weyl fermion comes incident to a boundary, say $x =  0$ from the left, it reflects with  momentum in $-x$-direction. However, since chirality of a Weyl fermion is fixed, the spin in $x$-direction must flip as well. This means the Weyl fermion at the boundary can not have spin in $x$, hence is an eigenstate of $\cos(\alpha) \sigma_y + \sin(\alpha) \sigma_z$. If we assume $\alpha = 0$, and that we have a positive chirality Weyl fermion at $\mathbf{k}=0$,  we get the same chiral solution as in the Chern insulator, $\psi_+ = \exp(-k_z x + i k_y )|\sigma_y =  +\rangle $ but this time the decay constant into the bulk is $k_z<0$. If we do the same analysis for the Weyl fermion with opposite chirality, we find that it is $\psi_- = \exp(k_z x + i k_y )|\sigma_y =  +\rangle $, so this time, $k_z>0$ so that the wave decays into the bulk in $x<0$. As $k_z \to 0$, the decay constant approaches zero and the surface state becomes a bulk mode.   Since Weyl points come in chiral partners located at $k^*_\pm$, the surface mode has $k_z< k_+^*$ and $k_z > k_-^*$, hence a `Fermi arc' in the BZ. Just as in the Chern insulator, these chiral modes contribute to Hall transport. Since now there are $a_z (k_+^*- k_-^*)/2\pi$ modes available,  if the sample has size $a_z$ in $z$-direction, the 3D quantum anomalous Hall conductivity is $ \sigma_{xy} = \Delta k e^2/(2\pi h) $, where $\Delta k $ is the separation of nodes in BZ.
 
All this looks quite curious,  but why exactly is it that the boundary and bulk properties of certain modes are related and in any way relevant to transport properties of actual materials with so much other complications such as disorder and interactions?  

\textit{Berry Curvature.}
To answer this question, we simply write down the equation of motion for a wave packet of electrons moving in the lattice. The Lorentz force due to external electromagnetic field steers this packet around the crystal, more generally in the phase space. However, the Hamiltonian of the electron depends on the position of the packet in the phase space. This means we are dealing with a situation where the Hamiltonian is adiabatically changed through a parameter, in this case momentum,  as the wavefunction evolves, which was considered by Berry. The wavefunction acquires a phase at every point in parameter space to  account for the changes in the Hamiltonian, called the Berry phase. This immediately produces a gauge field defined on the parameter space, which is BZ in a crystal.  After all, what is electromagnetic field? It is a manifestation of wavefunctions acquiring different phases at every point of space-time. Now it also acquires local phases in momentum space, hence there must be an analogous term in the equation of motion
\begin{eqnarray}
		\dot{\mathbf{x}} &=& \nabla_k \varepsilon - \dot{\mathbf{k}}\times \pmb{\Omega},\\
		\dot{\mathbf{k}} &=& -e(\mathbf{E} + \dot{\mathbf{x}} \times \mathbf{B})
\end{eqnarray}
The electric-like component is due to the phase that is local in time, that is the dynamical phase $e^{-i \varepsilon t}$ a wave-function acquires during its evolution. The spatial part of phase is 
\begin{equation}
	\mathcal{A}_i = -i \langle u |  \partial_{k_i} | u \rangle 
\end{equation}
where $u$ is the periodic part of the Bloch wave. Therefore the arising magnetic-like field is
\begin{equation}
	\pmb{\Omega} = \nabla_k \times \mathbf{\mathcal{A}}
\end{equation}

\textit{Berry Curvature due to Weyl points.}
Now we see that the BZ structure directly influences the motion of electron, we can calculate the Berry curvature $\Omega$ due to a Weyl node. We find that it satisfies
\begin{equation}
	\frac{1}{2\pi} \oint \mathbf{\Omega}\cdot d \mathbf{S}  = \pm 1
\end{equation}
when the surface surrounds a $+/-$ chirality Weyl point respectively. The fact that Weyl points must come in chiral partners imply that the if we take a surface that contains all Weyl points, or the BZ itself, the result is zero. This means that  $+/-$ chirality Weyl nodes are sources/sinks of Berry curvature. 

\textit{Berry Curvature and Chern invariant.}
The integral of Berry curvature over the BZ is called the Chern number and turns out to be an integer topological invariant. In the two band case, it measures how many times the pseudo-spin vector wraps around the unit sphere as we traverse across the whole BZ.   At an interface between two crystals with different Chern numbers, the net chirality of boundary modes must match the difference in Chern numbers, a fact commonly referred to as the `bulk boundary correspondence'~\cite{kane2013topological,bernevig2013topological}.

\textit{Topologically protected transport properties.}
Moreover, transport properties can be expressed in terms of the Chern number, because computation involves a summation over occupied states. In an insulator, this means integrating over the completely filled band over the BZ. Specifically in the Chern insulator, this integral turns out to be the Chern number. Disorder and interactions can not destroy this property as it is due to filled states~\cite{kane2013topological,bernevig2013topological}.

\section{Hamiltonians and Kinetics}

The surface states of topological insulators are described by the following Dirac-Rashba Hamiltonian:
\begin{equation}
H_{TI} = A\, (s_x k_y - s_y k_x) + \lambda s_z (k_x^3 - 3k_xk_y^2),
\end{equation}
where $A$ and $\lambda$ are material-specific parameters, ${\bm s}$ is the vector of Pauli spin matrices, and ${\bm k}$ is the two-dimensional wave vector. The hexagonal warping term is particularly strong in Bi$_2$Te$_3$. These states reside on opposite surfaces of a three-dimensional slab, yet usually all surfaces have topological states, which in Hall transport in particular mean that current can flow around the edges. The limitations of effective ${\bm k}\cdot{\bm p}$ methods for thin films are discussed in \cite{Nechaev_TI_Ham_2016}, where it is pointed out that open boundary conditions may yield different conclusions regarding the edge states than \textit{ab-initio} calculations.

Dirac and Weyl semimetals are described as different cases of the following Hamiltonian \cite{Armitage_Weyl_RMP_2018},
\begin{equation}
H_{DWSM} = \hbar v \tau_x {\bm \sigma}\cdot{\bm k} + m 	\, \tau_z + b \, \sigma_z + b' \, \tau_z \sigma_x,
\end{equation}
where ${\bm \sigma}$ represents a pseudospin degree of freedom, ${\bm \tau}$ the valleys or nodes, m is a mass parameter, and b and $b_0$ are effective internal Zeeman fields. Dirac semimetals have a single node and correspond to the case $m = b = b' = 0$, while the simplest model of a Weyl semimetal has two nodes that arise in the case $|b|>|m|$ and $b'=0$, that act as source and sink of Berry curvature.  In known materials, there are typically many more pairs of nodes. If the Weyl points arise due to broken inversion symmetry  with respect to $M_x, M_y$ mirror planes, there are 4 pairs of Weyl nodes,  ($\pm k_x, \pm k_y, \pm k_z$) including time reversal partners. There are as many as eight pairs in TaAs family due to the additional $C_4$ rotational symmetry ($\pm k_y, \pm k_x, \pm k_z$)  \cite{Binghai_Felser}). 

The separation between a Weyl node and its partner with opposite chirality determines the electromagnetic response of the material that contains the so called 'axion term'. Such a term is the source of chiral magnetic effect (CME) and the quantum anomalous Hall effect (QAHE) in Weyl semimetals \cite{Goswami_Tewari,Si_Burkov,Zyuzin_Burkov}.

Transition metal dichalcogenides are described by the following generalised Hamiltonian \cite{Dixiao_MoS2_2012}:
\begin{equation}
H_{TMD} = A \, (\tau \sigma_x k_x + \sigma_y k_y) + \frac{\Delta}{2} \, \sigma_z - \lambda \, \tau \, \frac{\sigma_z - \mathbb{I}}{2}\, s_z,
\end{equation}
where ${\bm s}$ represents spin, $\tau = \pm$ is the valley, and ${\bm \sigma}$ is an orbital pseudospin index, analogous to the sublattice pseudospin encountered in graphene. The spin splitting at the valence band top caused by the spin-orbit coupling is denoted by $2\lambda$. Additional terms encapsulate the spin splitting of the conduction band, the electron-hole asymmetry and the trigonal warping of the spectrum \cite{Kormanyos_PRB2013, Kormanyos_2DM_2015}. The trigonal warping term has the form:
\begin{equation}
H_{TW} = \frac{\kappa}{2} \, (\sigma_+ k_+^2 + \sigma_- k_-^2),
\end{equation}
where $\sigma_\pm = \sigma_x \pm i \, \sigma_y$. There is always a gap between the valence and conduction bands, which is manifested in the mass appearing in each of the copies of the Dirac Hamiltonian. Consequently, to satisfy time reversal invariance, the materials have two valleys, which are related by time reversal. Interaction terms coupling TMDs to external electromagnetic fields are covered in \cite{Gong_NC2013}.

Linear response calculations traditionally employ standard, well-established transport theory techniques such as the Kubo and Keldysh formalisms or the semiclassical wave-packet approach combined with the Boltzmann equation. Yet the presence of spin- and pseudospin-charge coupling in topological material Hamiltonians causes electromagnetic fields to induce inter-band dynamics, which have a subtle interplay with disorder. Such effects are most clearly seen in the density-matrix theory \cite{Interband_Coherence_PRB2017}. The single-particle density matrix $\rho$ obeys the quantum Liouville equation 
\begin{equation}
\td{\rho}{t} + \frac{i}{\hbar} \, [H, \rho] = 0,
\end{equation}
where $H$ is the total Hamiltonian. The density matrix is decomposed into two parts: one part, denoted by $\bkt{\rho}$, is averaged over impurity configurations, while the remainder, which is eventually integrated over, is denoted by $g$: $\rho = \bkt{\rho} + g$, with $\bkt{g}=0$. In linear response to an electric field ${\bm E}$ the density matrix comprises equilibrium and non-equilibrium components $\bkt{\rho} = \bkt{\rho_0} + \bkt{\rho_E}$, where $\bkt{\rho_E}$ is the correction to the equilibrium density matrix $\bkt{\rho_0}$ to first order in ${\bm E}$. The kinetic equation is linearised with respect to ${\bm E}$:

\begin{eqnarray}
&&
\displaystyle \td{\bkt{\rho_E}^{mm'}}{t} + \frac{i}{\hbar} \, [H_0, \bkt{\rho_E}]^{mm'} + J(\bkt{\rho_E})^{mm'} = 
\label{kineq}\\
&&
\frac{e{\bm E}}{\hbar} \cdot \bigg\{\delta^{mm'} \frac{\partial f_0(\varepsilon^m_{\bm k})}{\partial {\bm k}} 
+ i \mathcal{\bm R}^{mm'}_{\bm k} [f_0(\varepsilon^m_{\bm k}) - f_0(\varepsilon^{m'}_{\bm k})]\bigg\},
\nonumber
\end{eqnarray}
where $H_0$ is the band Hamiltonian, $J(\bkt{\rho})$ is a generalised scattering term, $f_0(\varepsilon^m_{\bm k})$ the Fermi-Dirac distribution, and $\mathcal{\bm R}^{mm'}_{\bm k} = \dbkt{u^m_{\bm k}}{i \pd{u^{m'}_{\bm k}}{\bm k}}$ is the Berry connection. Its appearance in the driving term gives rise to the Berry curvature intrinsic contribution to the Hall conductivity of systems with broken time reversal symmetry, and to other response properties. The Fermi occupation number difference factor makes it evident that this term drives off-diagonal response and therefore inter-band coherence contributions to the electrical response. 

This formulation is exactly equivalent to the quantum Boltzmann equation and has shed light on the physical origin of the chiral anomaly of Weyl semimetals \cite{Akihiko_PRB2017}. It shares a similar philosophy with the Keldysh method but without requiring an \textit{Ansatz} for the Keldysh component and integration over an additional energy variable, which become opaque in complex, multi-band systems. In this approach one can immediately separate intrinsic effects, extrinsic effects, and effects that combine interband coherence and disorder. A similar, Boltzmann equation-based theory gives a full account of phonon physics \cite{Qiuzi_PRB2012}, while an related approach being developed at present \cite{Xiao_JPCM2018} relies on the semiclassical transport framework. It will be interesting to see what insight can be gained from the density matrix theory in describing localisation physics, which received tremendous attention in the first half of this decade \cite{He_Bi2Te3_Film_WAL_ImpEff_PRL11, Competition_WL_WAL_2011, Shan_TIF_MassiveDirac_PRB2012, Hai-Zhou_Conductivity_2014_prl, Weizhe_TITF_2014_prb, Adroguer_WAL_2015, Weizhe_Weyl_2017, Weizhe_Materials_2017} and interaction physics in topological materials \cite{Culcer_TI_Int_PRB11}. The theory is being generalised to second order in the electric field \cite{Nandy_Tilt2019} where, in addition to the Berry curvature dipole term to be discussed below, additional disorder-mediated corrections to the non-linear Hall tensor were identified that have the same scaling in the impurity density.

\section{2D Topological Insulators}

In this section, we review the experimental work on two-dimensional (2D) topological insulators (TIs) that are protected by the time reversal symmetry and can be characterized by topological invariant $Z_2$. First identified by Kane and Mele, this type of 2D TIs can be manifested as the quantum spin Hall effect (QSHE) in transport\cite{Kane2005,Kane2005b}, and thus often referred to as quantum spin Hall (QSH) insulators. 

\subsection{CdTe/HgTe/CdTe quantum wells}

\begin{figure*}[h]
	\centering
	\centering
	\includegraphics[width=15cm]{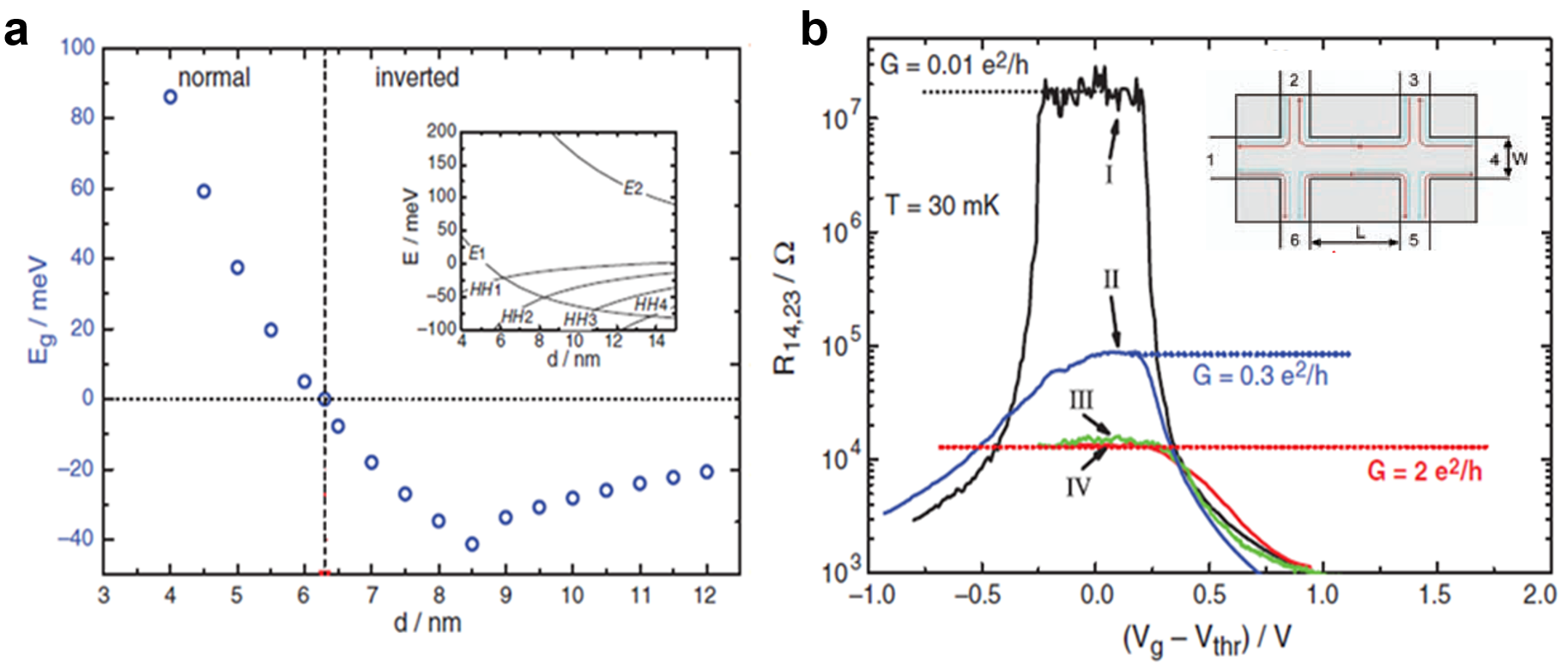}
	\caption{Observation of the quantum spin Hall effect in HgTe/CdTe quantum wells. (a) Band gap $E_g$ of HgTe/CdTe quantum well as a function of $d$, the thickness of the HgTe layer. Band inversion (negative $E_g$) takes place when $d>d_c=6.3$\,nm. The inset shows the evolution of several subbands of the quantum well. (b) Four-terminal resistances of four Hall-bar shaped HgTe/CdTe quantum well samples (I-IV). Sample I is in the normal regime ($d=5.5$\,nm), whereas samples II-IV are in the inverted regime ($d=7.3$\,nm). The sizes of the samples II, III and IV, given in the formats of $L\times W$, are $20\times13.3$\,$\mu$m$^2$, $1\times1$\,$\mu$m$^2$, and $1\times0.5$\,$\mu$m$^2$, respectively. The definitions of $L$ and $W$ are given in the upper-right inset. Adapted from K\"onig \textit{et al.}, Science \textbf{318}, 766 (2007)~\cite{Koenig2007}.}
	\label{fig:QSHE_HgTe}
\end{figure*}

Following the prediction of Bernevig \textit{et al.}~\cite{Bernevig2006} that the HgTe/CdTe quantum well with inverted band structure could be a 2D TI, K\"onig \textit{et al.} reported the evidence of QSHE in this system in 2007~\cite{Koenig2007}. As shown in Fig.~\ref{fig:QSHE_HgTe}, the four-terminal resistances of micrometer-scale Hall-bar shaped HgTe/CdTe samples in the inverted regime is approximately $2h/e^{2}$, a value expected for the helical edge transport in the ballistic regime. In contrast, for a sample with the thickness of the HgTe layer less than 6.3\,nm, the critical thickness for band inversion, the resistance increases to the order of mega-ohm, consistent with the insulating behavior expected for the topologically trivial regime. Further evidence for the QSHE in HgTe quantum wells was obtained with non-local transport in micrometer-sized devices of various geometries~\cite{Roth2009}. The experimental results are in agreement with the Landauer-B\"utikker theory adapted for the edge transport in 2D TIs. However, even for micrometer-sized QSH samples, the quantization of edge conductance is much less precise than the quantum Hall effect~\cite{Schopfer2007}. This was attributed to spin dephasing~\cite{JiangH2009} or inelastic backscattering processes related to charge puddles in the bulk \cite{Vayrynen2013,Vayrynen2014}. For a sufficiently long QSH edge channel, the transport is no longer in the ballistic regime, and the longitudinal resistance increases linearly with channel length $L$, namely $R \sim \frac{h}{2 e^{2}} \frac{L}{L_{\varphi}}$, where $L_{\varphi}$ is the spin dephasing/inelastic backscattering length, usually on the order of micrometer.

The edge channels in the inverted HgTe quantum wells were imaged with a scanning microwave impedance probe technique, and the edge conduction was however found to vary very little for magnetic fields up to 9\,T \cite{Ma2015}. The robustness of edge transport was also observed in a nonlocal electron transport experiment. Using samples with voltage probes separated up to 1\,mm, Gusev \textit{et al.} found that the nonlocal resistance is of the order 100\,k$\Omega$ and insensitive to the strength of the strength of in-plane magnetic field for $B <5 $T \cite{Gusev2011}. Such low resistances are quite surprising, given that the inelastic scattering length is on the order of micrometer in HgTe/CdTe edge channels \cite{Koenig2007,Roth2009}.

\subsection{\label{sec:InAsGaSb}InAs/GaSb}

InAs/GaSb quantum wells were also predicted to be a 2D TI candidate~\cite{LiuCX2008}. In 2011, Knez \textit{et al.} reported the evidence for edge transport in this system, despite that the parallel bulk conduction had to be subtracted from the total conductance~\cite{Knez2011}. Du \textit{et al.}~\cite{DuLJ2015} subsequently managed to use Si-doping to suppress the bulk conductivity and observed quantized edge channel conductance with a precision level of $\sim 1 \%$. The edge channel conductance in this sample was found to remain quantized for in-plane magnetic fields up to 12\,T and in a wide range of temperatures (20\,mK to 4\,K).

The insensitivity of the edge channel conductance to strong magnetic fields is puzzling at the first sight, since the Zeeman energy is expected to open a gap, leading to a destruction of the helical edge states. This phenomenon was explained recently by two groups, who pointed out that the small energy gap generated by the Zeeman energy is buried in the bulk valence band, and hence the helical edge transport is barely influenced by the magnetic field~\cite{LiCA2018,Skolasinski2018}. Similar argument was also used to explain the robustness of edge transport in HgTe quantum wells~\cite{Skolasinski2018}. It should be noted, however, if the Dirac point is indeed buried in the bulk bands, this would be are detrimental to pursue the Majorana physics (see Secs.~\ref{CS} and \ref{TWS} for a review of topological superconductivity), despite that pronounced superconducting proximity effect has been demonstrated in a HgTe/CdTe QSH system~\cite{Hart2014}.

Nichele \textit{et al.} reported that the transport properties of micrometer-sized InAs/GaSb wells in the non-inverted regime are phenomenologically similar to those observed in the inverted regime~\cite{Nichele2016}. The downward band bending of the InAs conduction near the sample edge was proposed as a possible origin of the topologically trivial edge states~\cite{Nichele2016}. The existence of such trivial edge states was further manifested in counterflowing edge transport in the quantum Hall regime~\cite{Akiho2019}. Shojaei \textit{et al.} recently measured the temperature and magnetic field dependences of a dual-gated InAs/GaSb quantum well, and concluded that the small hybridization gap (a few meV) in the inverted regime is overwhelmed by the disorder effect, and the transport is thus similar to a disordered two-dimensional  metal of symplectic class~\cite{Shojaei2018}.

\subsection{Other 2D TI candidates}

In addition to aforementioned semiconductor heterostructures, many 2D materials have been identified theoretically to be 2D TIs, such as monolayers of Si~\cite{LiuCC2012}, Ge~\cite{LiuCC2012}, Sn~\cite{XuY2013}, Bi~\cite{ReisF2017}, ZrTe5~\cite{WengHM2014} and WTe2~\cite{QianXF2014}, Bi bilayers~\cite{YangF2012}, as well as many hybrid structures based on graphene~\cite{RenYF2016}. A lot of experimental efforts were devoted to single-element TI candidates (e.g. silicene, germanene, stanine~\cite{ZhuFF2015}), which were prepared mostly with molecular beam epitaxy (MBE) and characterized in situ with scanning tunneling microscopy (STM) or angle resolved photoemission spectroscopy (ARPES). Evidence for the existence of edge channels has been obtained with STM measurements of Bi bilayers grown on Bi2Te3 thin films~\cite{YangF2012} and on Bi crystals~\cite{Drozdov2014}. In a subsequent ARPES experiment on Bi bilayers, however, very large Rashba spin-splitting was observed in the edge states, suggesting a topologically-trivial origin~\cite{Takayama2015}.

In spite of a lot of progress, most of these single-element films have so far been grown on conducting substrates, and thus are unsuitable for elucidating their topological nature with transport studies. Recently, Reis \textit{et al.}~\cite{ReisF2017} reported that honeycomb-structured Bi monolayers can be epitaxially grown on insulating SiC substrates. By using scanning tunneling spectroscopy (STS) measurements, they observed an energy gap of 0.8 eV, and also obtained evidence for conducting edge states. With help of the first principles calculations, Reis \textit{et al.} claimed that the large gap arises from covalent bonding between the Bi orbitals and the substrate and it is topologically nontrivial~\cite{ReisF2017}. If the QSHE is confirmed in the future, this system will be exceptionally attractive for exploiting the helical edge transport at high temperatures.

An alternative approach to obtain 2D TIs is to exfoliate layered compounds with strong spin-orbit interaction, such as 
$\mathrm{WTe}_{2}$~\cite{QianXF2014,TangSJ2017,PengL2017,FeiZY2017} and $\mathrm{ZrTe}_{5}$~\cite{WengHM2014,WuR2016,LiXB2016}. 
Following a theoretical prediction of the 1T' phase of  transition metal dichalcogenide (TMDC) monolayers being 2D TIs~\cite{QianXF2014}, much work has been done on $\mathrm{WTe}_{2}$ monolayers, the only member of the $\mathrm{MX}_{2}$ family (M=W, Mo, and X=Te, Se, S) in which the 1T' phase is energetically favored~\cite{TangSJ2017}. An ARPES experiment performed by Tang \textit{et al.} showed that $\mathrm{WTe}_{2}$ monolayer has a bulk energy gap of 55 meV, much larger than those in HgTe/CdTe and InAs/GaSb quantum wells. The existence of edge channels in $\mathrm{WTe}_{2}$ monolayers has been confirmed with the STM~\cite{TangSJ2017,PengL2017}, transport~\cite{FeiZY2017}, and scanning microwave impedance probe measurements~\cite{ShiYM2019}. More recently, Wu \textit{et al.}~\cite{WuSF2018} reported the edge conductance values consistent with the QSHE. In this experiment, however, the observation of quantized edge transport requires a spacing of 100 nm or lower for the neighboring voltage probes in the devices based on $\mathrm{WTe}_{2}$ monolayers. On the other hand, the edge conductance varies very little for temperatures below 100 K. These features are quite different from those of the HgTe/CdTe~\cite{Roth2009} and InAs/GaSb~\cite{DuLJ2015} quantum wells, in which the quantized edge conductance is limited to liquid helium temperatures, but can survive for the channel lengths of the micrometer scale. It remains to be understood why the edge conductance in $\mathrm{WTe}_{2}$ monolayers is so robust against thermal agitation, but very demanding on the channel length. In addition to $\mathrm{WTe}_{2}$, the layered compound $\mathrm{ZrTe}_{5}$ has also attracted a lot of interest. This can be attributed to its controversial topological classification, very low carrier density, and high carrier mobility~\cite{WengHM2014,WuR2016,LiXB2016,TangFD2019}. The latter features make $\mathrm{ZrTe}_{5}$ particularly suitable for seeking exotic quantum transport properties. Recently, Tang \textit{et al.} observed a 3D quantum Hall effect in $\mathrm{ZrTe}_{5}$ bulk single crystals~\cite{TangFD2019}. Although $\mathrm{ZrTe}_{5}$ can be easily exfoliated down to nanometer thicknesses, transport measurement of $\mathrm{ZrTe}_{5}$ monolayer, a 2D TI candidate with a band gap of 0.1\,eV~\cite{WengHM2014}, has not been reported so far, probably due to the chemical instability of this compound.

\begin{figure*}[t]
	\centering
	\includegraphics[width=15cm]{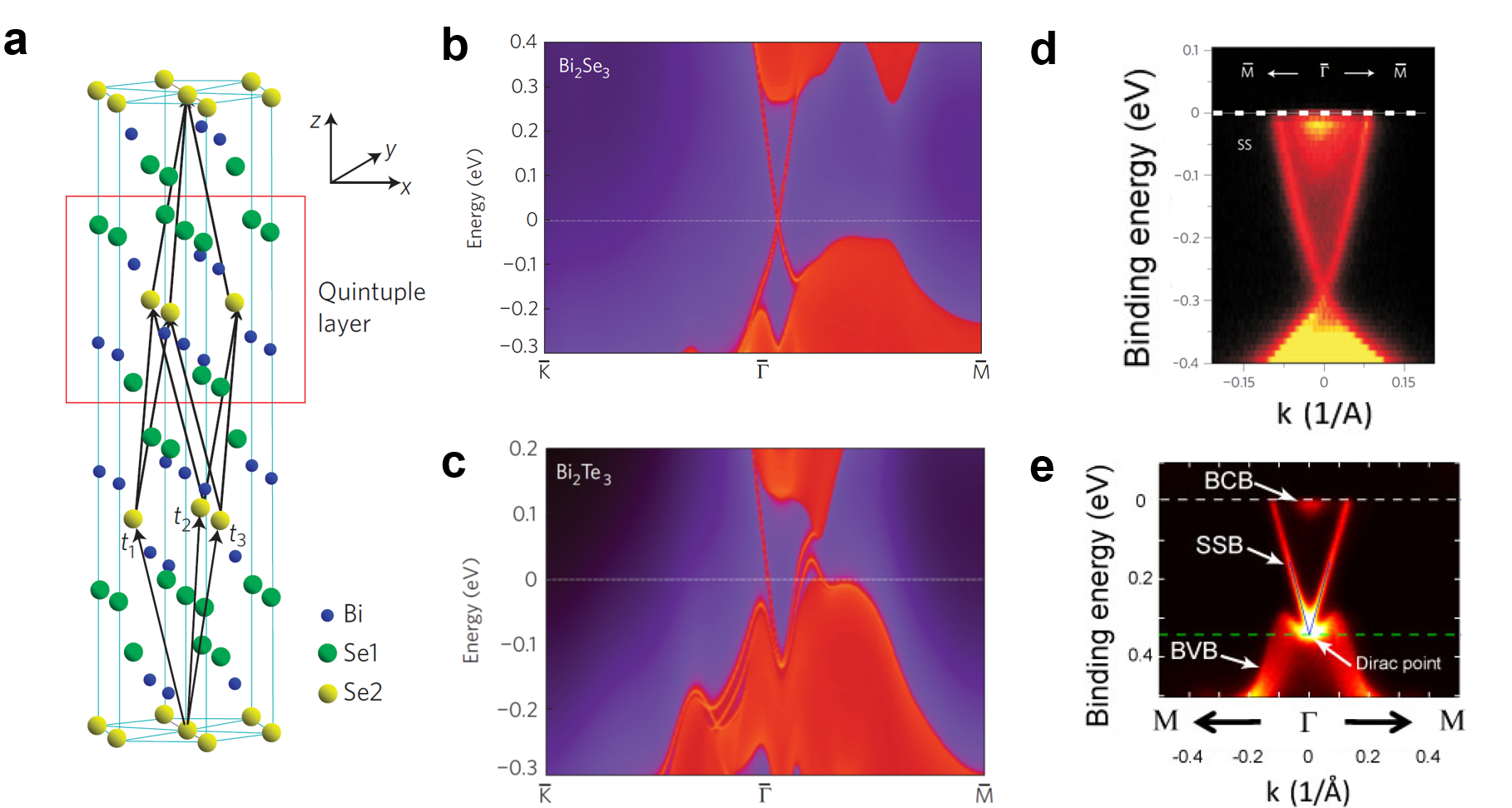}
	\caption{Crystalline and electronic structures of the Bi$_2$Se$_3$ family of 3D TIs. (a) Crystal structure of Bi$_2$Se$_3$, which has a layered structure of Van der Waals type, which can be separated in to quintuple layers (QLs) of Se-Bi-Se-Bi-Se. Each QL is about 1\,nm thick. (b,c) Band diagrams of Bi$_2$Se$_3$ (b) andBi$_2$Te$_3$ (c) given by first principles calculations. The gapless surface states are also shown. (d,e) Band structures of Bi$_2$Se$_3$ (d) and Bi$_2$Te$_3$ (e) obtained with ARPES measurements. Adapted from Zhang \textit{et al.}, Nat. Phys. \textbf{5}, 438 (2009)~\cite{ZhangHJ2009}; Xia \textit{et al.}, Nat. Phys.\textbf{ 5}, 398 (2009)~\cite{XiaY2009}; Chen \textit{et al.}, \textbf{325}, 178 (2009)~\cite{ChenYL2010}. }
	\label{fig:3DTIbandstructure}
\end{figure*}

\section{3D Topological Insulators}

The 3D counterpart of the 2D TIs can be characterized by four $Z_2$ indexes, $\left\{v_{0},\left(v_{1}, v_{2}, v_{3}\right)\right\}$. This type of 3D TIs are also protected by the time-reversal symmetry, and can be divided into strong and weak 3D TIs, corresponding to $v_{0}=1$ and 0, respectively~\cite{Hasan2010}. The latter can be regarded as a stack of 2D TIs and therefore have gapless states on the side surfaces, while the former does not have a 2D counterpart. The strong 3D TIs have so far received much more attention than the weak 3D TIs. In this review, we therefor solely focus on the strong 3D TIs, and refer to them as 3D TIs or TIs for brevity.

\subsection{Early experiments on 3D TIs}

Early experimental efforts on 3D TIs were mainly focused on confirming the existence of the helical surface states. ARPES measurements played a decisive role in identifying 3D TIs~\cite{Hasan2010,QiXL2011,Ando2013}. Among them, $\mathrm{Bi}_{1-\mathrm{x}} \mathrm{Sb}_{\mathrm{x}}$ is the first material confirmed to be a 3D TI~\cite{Hsieh2008}. 
The electronic structure of $\mathrm{Bi}_{1-\mathrm{x}} \mathrm{Sb}_{\mathrm{x}}$ is, however, very complicated due to the coexistence of multiple bands in the surface states. 
This makes $\mathrm{Bi}_{1-\mathrm{x}} \mathrm{Sb}_{\mathrm{x}}$ not suitable to be studied as a model 3D TI system. 
Experimental attention was quickly shifted to the compounds in tetradymite family ($\mathrm{Bi}_{2} \mathrm{Se}_{3}$, $\mathrm{Bi}_{2} \mathrm{Te}_{3}$, $\mathrm{Sb}_{2} \mathrm{Te}_{3}$ and their derivatives), 
in which the surface states are featured by a single Dirac cone, accompanied by a large bulk band gap ~\cite{Hsieh2008,ZhangHJ2009,RenZ2010,JiaS2011} (See Fig.~\ref{fig:3DTIbandstructure}). 
Among them, the binary compounds (e.g. $\mathrm{Bi}_{2} \mathrm{Se}_{3}$) can be prepared in the form of bulk single crystals, thin films, nanobelts or nanoplates, but they tend to have a high density of defects, and consequently the bulk remains conducting even at liquid helium temperatures~\cite{Ando2013,YangWM2013}. Synthesis of ternary or quaternary compounds turned out be an effective approach to suppress the bulk conductivity. One example is $\left(\mathrm{Bi}_{1-\mathrm{x}} \mathrm{Sb}_{\mathrm{x}}\right)_{2} \mathrm{Te}_{3}$, in which the chemical potential can be continuously tuned from the bulk conduction band to the valence band by increase the Sb concentration~\cite{ZhangJ2011,KongD2011}. This doping scheme is based on the fact that undoped $\mathrm{Bi}_{2} \mathrm{Te}_{3}$ is often n-type and $\mathrm{Sb}_{2} \mathrm{Te}_{3}$ is p-type. The other example exploiting the compensation effect is $\left(\mathrm{Bi}_{1-\mathrm{x}} \mathrm{Sb}_{\mathrm{x}}\right)_{2}\left(\mathrm{Te}_{1-\mathrm{y}} \mathrm{Se}_{\mathrm{y}}\right)_{3}$, in which the composition control can increase the bulk resistivity to the order of ohm$\cdot$cm for single crystal samples~\cite{RenZ2011}.

In addition to ARPES, STM was also used extensively in the 3D TI studies. Quasiparticle interference experiments provided the evidence for the suppression of backscattering~\cite{Roushan2009,ZhangT2009,Alpichshev2010}, a property arising from the spin-momentum locking (and $\pi$ Berry phase, see Section I, in the surface states. The Landau level spectra obtained with STS measurements are consistent with the linear dispersion of surface Dirac fermions~\cite{ChengP2010}. STM/STS measurements also produced atomic scale information of various defects in 3D TIs~\cite{DaiJX2016}. Such information is valuable for gaining in-depth understanding of the physics of defect formation, and thus offers new opportunities for further improving the sample quality.

Additional evidence for the $\pi$ Berry phase in the surface states was also gained from the Shubnikov-de Haas (SdH) oscillation measurements. The longitudinal conductivity in high magnetic fields satisfies
\begin{equation}
\Delta \sigma_{x x} \propto \cos \left[2 \pi\left(\frac{B_{F}}{B}+\frac{1}{2}-\frac{\gamma}{2 \pi}\right)\right],
\end{equation}
where $B_{F}=\frac{1}{\Delta\left(\frac{1}{B}\right)}=\frac{S_{F}}{4 \pi^{2}} \frac{h}{e}$ is the frequency of SdH oscillations. Here, $S_{F}$ is the extremal area of the Fermi spheroid perpendicular to the magnetic field $B$ for a 3D system (and in 2D, it is reduced to the area inside the Fermi circle), and $\gamma$ is the Berry's phase, which takes a value of $\pi$ for a TI surface and 0 for an ordinary 2D electron system. The $\gamma$ value can be extracted from the plot of Landau level index $n$ as a function of $\frac{1}{B_{n}}$, i.e. $n=\frac{F}{B_{n}}-\frac{\gamma}{2 \pi}$, where $B_{n}$ is the magnetic field corresponding to the nth conductivity minimum. SdH measurements performed on $\mathrm{Bi}_{2}\mathrm{Se}_{3}, \mathrm{Bi}_{2}\mathrm{Te}_{3}$ and $\mathrm{Bi}_{2}\mathrm{Te}_{2}\mathrm{Se}$ single crystals yielded nontrivial values of Berry's phase for the surface states~\cite{QuDX2010,Analytis2010,RenZ2010,JiaS2011,RenZ2011}. However, caution has to be taken in order to avoid inaccurate assignment of longitudinal conductivity minima~\cite{XiongJ2012}, and to account for nonlinearity in the energy-momentum relation due to broken particle-hole symmetry, as well as the Zeeman effect. Nevertheless, SdH measurements can provide valuable electronic parameters relevant to the device applications, such as the carrier concentration, effective mass, and quantum lifetime.

\subsection{Electronic properties of 3D TIs}

\subsubsection{Aharonov-Bohm effect}

\begin{figure*}[h]
	\centering
	\includegraphics[width=15cm]{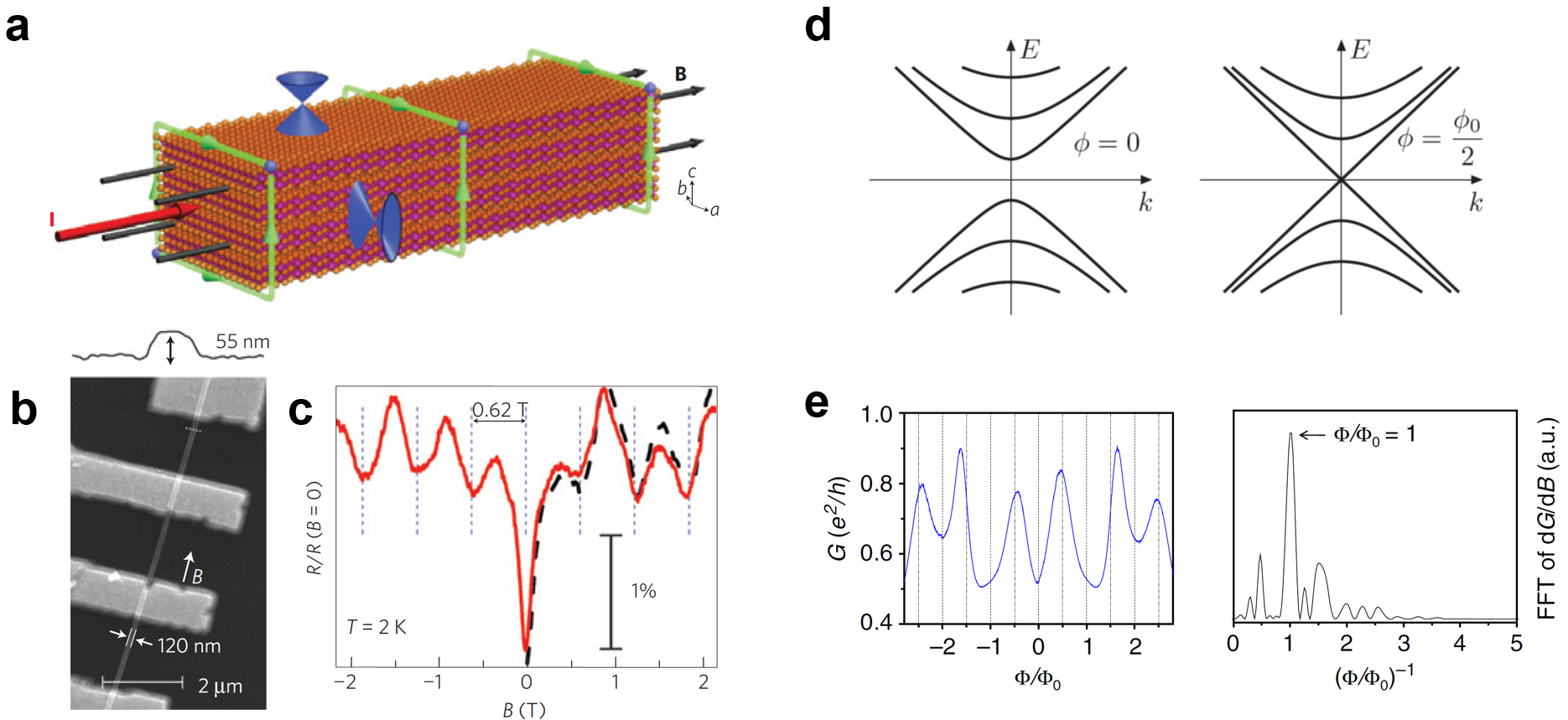}
	\caption{AB/AAS oscillations in TI nanowires. (a) Schematic sketch of the measurement geometry, in which the current and magnetic field are applied parallel to the long axis of the wire. (b) Scanning electron image of a Bi$_2$Se$_3$ nanowire with multiple electrical contacts. (c) Resistance fluctuations with $h/e$-period oscillations when a parallel magnetic field is applied. (d) Band diagrams of a TI nanowire with zero (left) and half (right) flux quantum applied (i.e. $\phi=0$ and $\phi_0$). (e) Conductance fluctuations in a Bi$_{1.33}$Sb$_{0.67}$Se$_3$ nanowire in which the chemical potential is lowed by surface coating of F4-TCNQ molecules (strong acceptors) and electrical back-gating. Panels (a-c) are taken from Peng \textit{et al.}, Nat. Mater. (2009)~\cite{PengHL2010}; panel (d) from Bardarson \textit{et al.}, PRL (2010)~\cite{Bardarson2010}, and panel (e) from Cho \textit{et al.}, Nat. Commun. (2015)~\cite{Cho2015}.}
	\label{fig:ABandAASoscillation}
\end{figure*}

The interplay between the Aharonov-Bohm (AB) phase, the Berry's phase and quantum confinement effect makes TI nanostructures very appealing for exploring novel mesoscopic physics. 
In 2009, Peng \textit{et al.} observed that when a magnetic field is applied parallel to the long axis of $\mathrm{Bi}_{2} \mathrm{Se}_{3}$ nanowires, the conductance exhibits oscillations with a period of $h/e$, with the maxima appearing at integer multiples of the flux quantum ($\Phi_{0}=h / e$)~\cite{PengHL2010}. These features are in contrast to the Altshuler-Aronov-Spivak oscillations~\cite{Altshuler1981} observed in hollow metallic cylinders~\cite{Sharvin1981} and carbon nanotubes~\cite{Bachtold1999}, in which the interference between various time-reversed paths leads to $h/2e$-periodicity oscillations for the diffusive transport. In the diffusive regime, the $h/e$-period Aharonov-Bohm oscillations, which would be most pronounced in a single ring-like structure, are expected to be suppressed in a nanowire due to the ensemble average of various electron trajectories, which carry random phase factors~\cite{Washburn1986}.

According to the theory reported in Refs.~\cite{ZhangY2010,Bardarson2010}, the $h/e$ oscillations observed in the TI nanowires~\cite{PengHL2010} can be regarded as a transport hallmark of the helical surface states in 3D TIs. The $\pi$ Berry's phase modifies the 1D quantization condition around the perimeter of nanowires, leading to a gap opening at the Dirac point when the magnetic flux through nanowire is even multiples of half-flux-quantum $\left(\frac{\Phi_{0}}{2}=\frac{h}{2 e}\right)$, and a gapless 1D mode for odd numbers of half-flux-quantum~\cite{ZhangY2010,Bardarson2010}. In the ballistic regime, this peculiar electronic structure results in the $h/e$ oscillations with conductance maxima (minima) located at odd (even) multiples of $\Phi_{0} / 2$. The experimental data reported in Ref.~\cite{PengHL2010}, however, exhibit the opposite behavior. This was explained by Bardarson \textit{et al.}~\cite{Bardarson2010}, who considered the joint effects of doping and disorder, and showed that for certain disorder strengths, the conductance maximum can oscillate between the zero flux and half flux quantum as the chemical potential varies. This phenomenon was subsequently observed in $\mathrm{Bi}_{2} \mathrm{Se}_{3}$ nanowires by Jauregui \textit{et al.}~\cite{Jauregui2016}. It was also predicted that both $h/e$ and $h/2e$ oscillations can appear in TI nanowires and their relative strength varies with increasing disorder strength~\cite{Bardarson2010}. This was confirmed by the transport measurement of Se-encapsulated $\mathrm{Bi}_{2} \mathrm{Se}_{3}$ nanowires, in which the disorder strength was controlled by intentional aging of the samples~\cite{Hong2014}.

A direct observation of the 1D helical mode in a TI nanowire, however, requires the chemical potential tuned close to the Dirac point. In Ref.~\cite{Cho2015}, the Fermi level in $(\mathrm{Bi}, \mathrm{Sb})_{2} \mathrm{Se}_{3}$ nanowires was controlled by electrical gating and the surface coating of acceptor-type molecules. The conductance maxima at half flux-quantum (and minima at zero-flux) appear to be consistent with the theory. Similar conductance oscillations were also observed in a back-gated $\mathrm{Bi}_{2} \mathrm{Se}_{3}$ nanowire sample~\cite{Hong2014}, but the amplitude of conductance oscillations was, however, one order of magnitude lower than that observed in Ref.~\cite{Cho2015}. Therefore, it is desirable to further explore the helical 1D mode at half flux quantum in the samples with weak disorder and low chemical potential. In this regime, the variation in the longitudinal conductance is expected to be close to $e^{2}/h$ as the magnetic flux is increased from zero to $\Phi_{0}/2$~\cite{Bardarson2010}.

\subsubsection{Weak antilocalisation}

\begin{figure*}[h]
	\centering
	\includegraphics[width=15 cm]{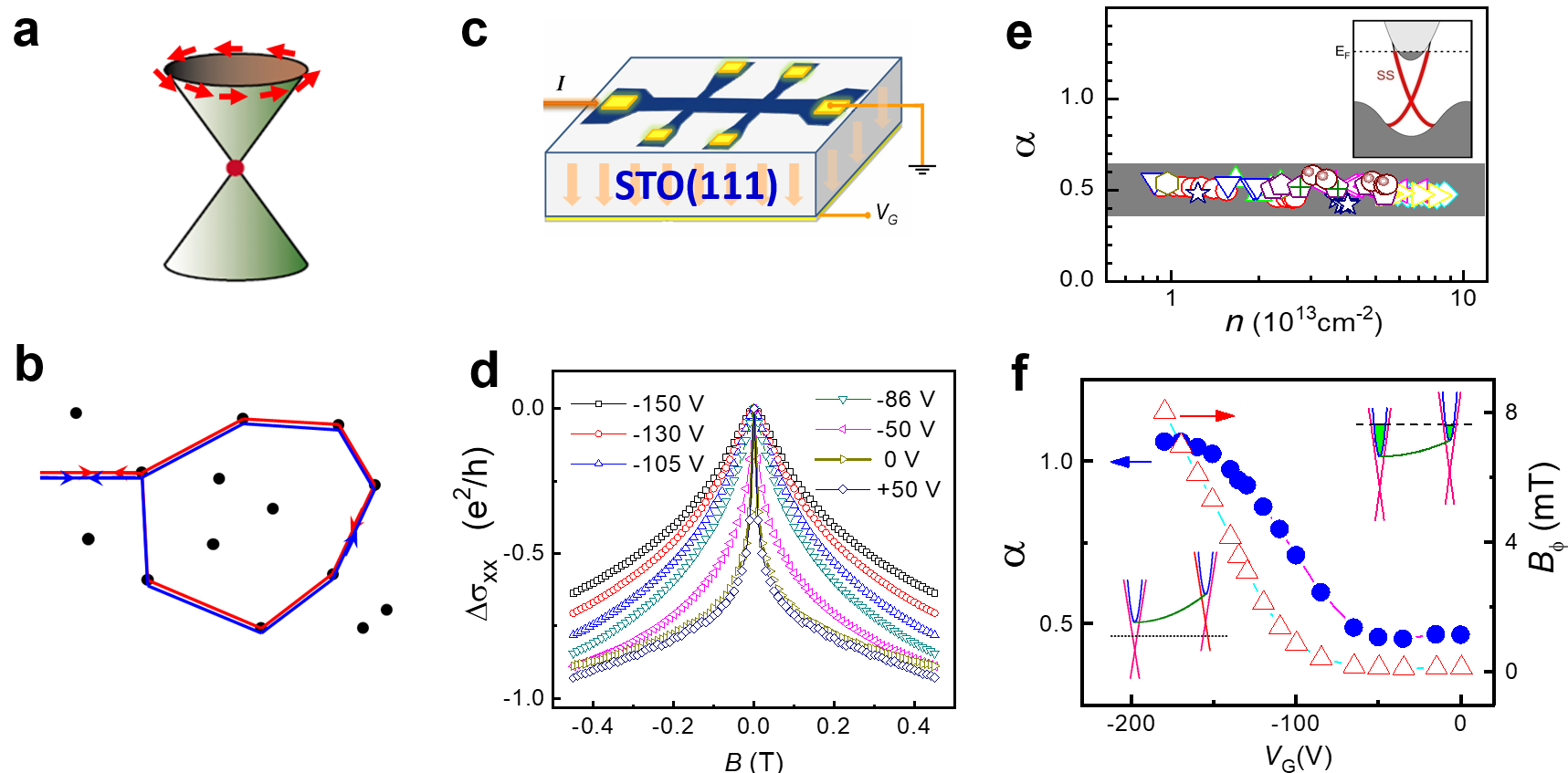}
	\caption{Weak antilocalization effect in TI thin films. (a) Schematic sketch of helical 3D TI surface states, in which the spin-momentum locking leads to a Berry's phase of $\pi$. (b) Destructive interference between a pair of time-reversed path due to the $\pi$ Berry's phase. (c) Sketch of the Hall-bar shaped Bi$_2$Se$_3$ thin film samples, in which the chemical potential is tuned with a back-gate via the STO substrate as a high-kappa dielectric. (d) Sheet carrier density dependence of the prefactor alpha extracted from the fits of magnetoconductivity data to the HLN equation. The $\alpha$ value remains close to $1/2$ for a wide range of carrier densities (0.8 - 8.6$\times10^{13}$\,cm$^{-2}$), which correspond to Fermi levels located in the bulk conduction band (upper inset). (f)  Gate-voltage dependence of a $\mathrm{Bi_2Se_3}$ sample, in which the $\alpha$ value can be tuned from about $1/2$ to nearly $1$, when the bulk carriers are depleted by applying negative gate voltage. The upper and lower insets show the band diagrams for $\alpha=1/2$ and $1$, respectively. Adapted from  Chen\textit{et al.}, Phys. Rev. Lett. \textbf{105}, 176602 (2010); \textit{et al.}, Phys. Rev. B \textbf{83}, 241304 (2011)~\cite{Chen2011Tuna}; Zhang \textit{et al.}, Adv. Func. Mater. \textbf{21}, 2351 (2011)~\cite{ZhangGH2011}.}
	\label{fig:WALeffect}
\end{figure*}

Another method for probing the surface transport is to utilize the weak antilocalisation (WAL) effect. It can be understood in the following semiclassical picture. At low temperatures, electrons can remain phase coherent after many scattering events, namely $l_{\varphi} \gg l_{e}$, where $l_{\varphi}$ the dephasing length, and $l_{e}$ is the mean free path. Since a pair of time-reversed paths for a closed electron trajectory has exactly the same length, constructive interference takes place for a particle with negligible spin-orbit coupling. This leads to an increase in the returning probability of the particle and hence decrease in conductivity. For a Dirac fermion on the TI surface, the spin-momentum locking introduces an additional phase $\pi$ between the pair of counterpropagating loops. The consequence is the destructive interference and the suppression of backscattering. This is known as weak antilocalisation, an effect opposite to the weak localisation~\cite{Hikami1980Spin,Bergmann1984Wea,Chakravarty1986Wea}.

The WAL effect has been observed frequently in 3D TIs~\cite{Ando2013}. It was, however, not easy to attribute it to the helical surface states exclusively, due to coexistence of parallel conduction channels, such as the bulk layer~\cite{Chen2010Gate,Kim2011Thic} and the surface accumulation layer arising from downward band bending~\cite{King2011Larg}. The spin-orbit coupling strengths are also very strong in these unwanted 2D channels, and they may belong to the same symmetry class (symplectic metal) as the helical surface states~\cite{Hikami1980Spin,Suzuura2002Cros}. Therefore, the WAL correction to longitudinal conductivity in these conducting channels can be described by the same equation, if they can be treated independently. In addition to the parallel conductivity, the coupling between these channels (e.g. the surface-bulk scatterings and inter-surface hybridization) further complicates the analysis of the WAL effect in many 3D TI samples~\cite{Garate2012Weak,LinCJ2013Para}. Nevertheless, valuable information on the surface states can still be obtained by systematically tuning the chemical potential  and quantitatively analyzing of the transport data.

A convenient method to unveil the WAL effect in TIs is to measure the longitudinal resistivity in perpendicular magnetic fields. The magnetoconductivity can be described by the Hikami-Larkin-Nagaoka (HLN) equation~\cite{Hikami1980Spin} simplified for the strong spin-orbit coupling limit:
\begin{equation}
\sigma(B)-\sigma(0) = -\frac{\alpha e^{2}}{\pi h}\left[\ln \left(\frac{B_{\varphi}}{B}\right)-\Psi\left(\frac{1}{2}+\frac{B_{\varphi}}{B}\right)\right]
\end{equation}
in which $\Psi$ is the digamma function, $B$ is the magnetic field, $B_{\varphi}$ is the dephasing field, $l_{\varphi}$ is the dephasing length. The value of prefactor $\alpha$ is dependent on the number of conduction channels and the coupling strengths between them. For instance, $\alpha=1/2$ if only one channel contributes, and $\alpha=n/2$ for $n$ independent and equivalent channels. In case of two channels with non-negligible inter-channel coupling or asymmetry in the dephasing lengths (i.e. $l_{\varphi, 1} \neq l_{\varphi, 2}$), $\alpha$ would take a value between $1/2$ and 1 and the $B_{\varphi}$ value would be determined by a complicated function of $l_{\varphi, 1}$, $l_{\varphi, 2}$, and other transport parameters~\cite{Garate2012Weak}. It is worth noting that the HLN formula needs to be modified to account for strong spin-orbit scattering expected in topological materials \cite{Adroguer_WAL_2015, Weizhe_Weyl_2017, Weizhe_Materials_2017}.

Measurements of TI thin films or microflakes with conducting bulk often yield $\alpha$ values close to $1/2$. This can be attributed to strong coupling between the surface and bulk states, which makes the sample behaving like a single channel system~\cite{Garate2012Weak}. When the bulk carriers are depleted by electrical gating or reduced by appropriate doping, crossover from $\alpha \sim 1 / 2$ to $\alpha \sim 1$ can be observed~\cite{Chen2011Tuna,Checkelsky2011Bulk,Steinberg2011Elec,Kim2013Cohe,Brahlek2014Emer} (See Fig.~\ref{fig:WALeffect} for an example). The $\alpha$ value close to 1 is a signature of decoupling between two conduction channels. If the film is sufficiently thin or the bulk carrier density is low enough, a full depletion of bulk carriers can be achieved with electrical gating. In this case, $\alpha=1$ would mean a pair of equivalent surfaces decoupled to each other. Such a transport regime is suitable for pursuing intrinsic quantum transport properties of the helical surface states. For instance, Liao \textit{et al.} found that the electron dephasing rate has an anomalous sublinear power-law temperature dependence in the $(\mathrm{Bi},\mathrm{Sb})_{2}\mathrm{Te}_{3}$ thin films in the decoupled surface transport regime. In contrast, when the same film is tuned to the bulk conducting regime, the dephasing rate returns to linear temperature dependence commonly seen in conventional 2D electron systems. The coupling of the surface states to charge puddles in the bulk was proposed as a possible origin of the sublinear power law~\cite{LiaoJ2017Enha}.

\subsubsection{Quantum Hall effect}

\begin{figure*}[h]
	\centering
	\includegraphics[width=12cm]{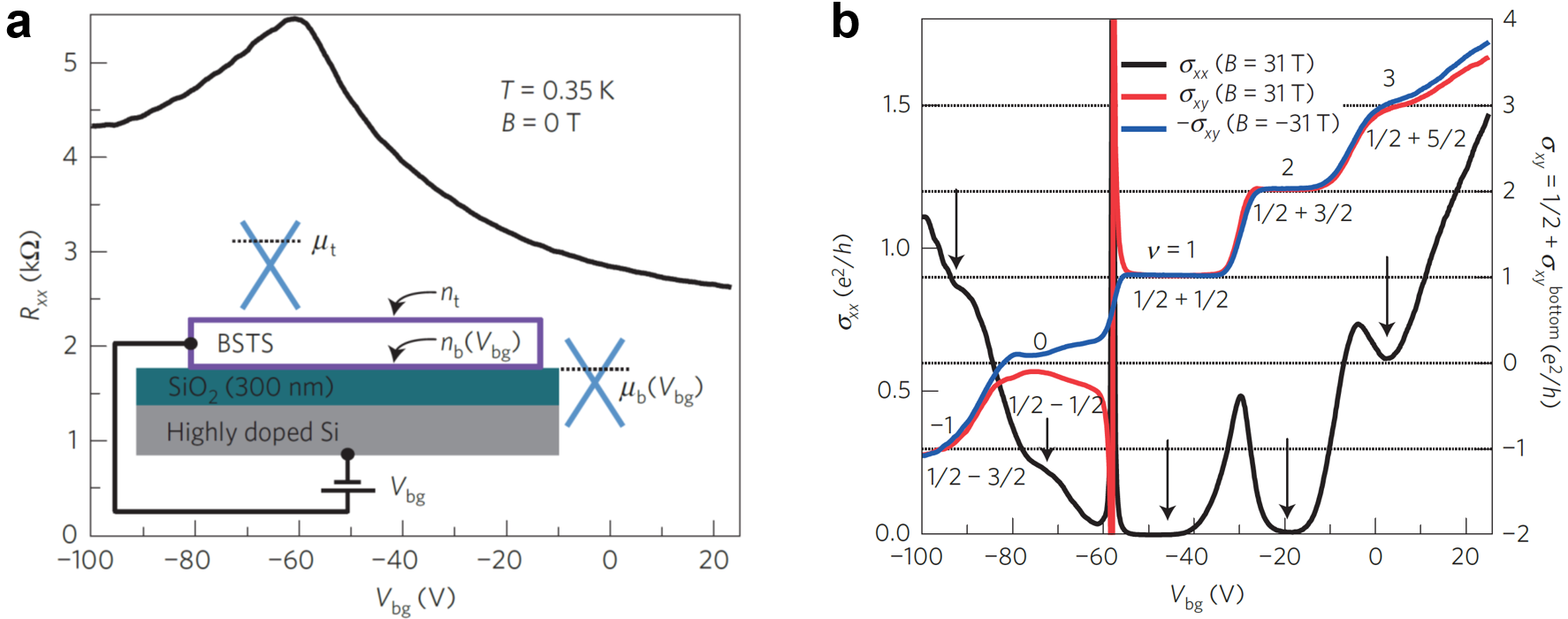}
	\caption{Observation of the quantum Hall effect in a 3D TI. (a) Gate-voltage dependence of longitudinal resistance of (Bi,Sb)$_2$(Te,Se)$_3$ (BSTS) microflake, in which the chemical potential is tuned with a back gate via a 300\,nm thick SiO$_2$ layer. (b) Longitudinal conductivity $\sigma_{xx}$ and $\sigma_{xy}$ plotted as functions of back-gate voltage in a magnetic field of $B=31$\,T applied perpendicularly to the BSTS flake. Here the total Hall conductivity contains contributions from the top and bottom surfaces. The top surface component remains quantized as $\frac{1}{2}e^2/h$, while the bottom surface contribution can be tuned with the back-gate voltage. Adapted from Xu \textit{et al.}, Nat. Phys. \textbf{10}, 958 (2014)~\cite{XuY2014Obse}}
	\label{fig:quantumhalleffectin3dtis}
\end{figure*}

The Dirac electrons on the TI surface states can in principle exhibit half-integer quantum Hall effect (QHE) due to the existence of a zero-energy Landau level~\cite{Hasan2010,QiXL2011}. Despite the absence of spin or valley degeneracies in many 3D TIs, direct observation of the half-integer quantum Hall plateau has not been possible because both the top and bottom surfaces contribute to transport. The quantised Hall conductivity in a slab geometry can be written as
\begin{equation}
\sigma_{x y} = \sigma_{t} +\sigma_{b}= (\nu_{t} + \nu_{b})\frac{e^2}{h} = (N_{t}+\frac{1}{2}+N_{b}+\frac{1}{2})\frac{e^2}{h},
\end{equation}
where $\sigma_{H}$ is the Hall conductivity, $\nu_{i}$ the filling factor, and $N_{i}$ the Landau level index of surface $i$, with $i=t$, $b$. Here $t$ and $b$ denote the top and bottom surfaces, respectively. In a magnetic field of the order of 10 T, the carrier mobility of several thousand cm$^{2}$/V$\cdot$s is usually sufficient for observing the QHE. Bismuth chalcogenides with such mobilities were not difficult to obtain even in early days of TI research~\cite{QuDX2010}. The failure in observing the QHE in TIs can be attributed to the conducting bulk, which provides backscattering paths for the chiral edge channels in the quantum Hall regime.

The first observation of QHE in TIs was realized in 2011 with a strained 70 nm thick HgTe epilayer with electron mobility $3.4\times10^{4}$\,cm$^{2}$/V$\cdot$s. Multiple quantum Hall plateaus in $\sigma_{xy}$ were observed but they were accompanied substantial longitudinal conductivities. In 2014, two groups managed to observe the QHE in bismuth chalcogenides. Xu \textit{et al.} observed well-defined quantum Hall plateau of the zeroth Landau level at $B >15 T$ and $T=0.35 K$ using bottom-gated microflakes exfoliated from $(\mathrm{Bi}, \mathrm{Sb})_{2}\left(\mathrm{Te},\mathrm{Se}\right)_{3}$ single crystals~\cite{XuY2014Obse}. Yoshimi \textit{et al.} utilized high quality $(\mathrm{Bi}, \mathrm{Sb})_{2}\mathrm{Te}_{3}$ thin films grown in InP(111) substrates and realized the $\sigma_{H}= \pm 1 e^{2}/h$ plateaus in a magnetic field of 14\,T and at $T=40$\,mK~\cite{Yoshimi2015Quan}. Using a sample with quality higher than those in Ref.~\cite{XuY2014Obse} and dual gating, Xu \textit{et al.} studied the zero Hall state (with $\left(\nu_{t},\nu_{b}\right)=(-1/2,1/2)$ or $\left(\frac{1}{2},-\frac{1}{2}\right)$) in detail. Their joint local and nonlocal measurements suggest the existence of a quasi-1D dissipative edge channel with nearly temperature independent conductance for $T<50$\,K~\cite{XuY2016Quan}.

The QHE has also been observed in ungated $\mathrm{Bi}_{2} \mathrm{Se}_{3}$ thin films of both n-type and p-type. Such observations represent a remarkable technical achievement in reducing the density of defects in $\mathrm{Bi}_{2} \mathrm{Se}_{3}$ by optimizing the MBE growth processes~\cite{Koirala2015Reco,Moon2018Solu}. It is also noteworthy that Zhang \textit{et al.} recently observed a well-defined quantum Hall $\nu_{tot}=1$ plateau in an exfoliated Sn-doped $\left(\mathrm{Bi}, \mathrm{Sb}\right)_{2}\left(\mathrm{Te}, \mathrm{S}\right)_{3}$ microflake in magnetic fields less than 4\,T at $T=6$\,K~\cite{XieFJ2019Phas}. Further improvement in the quality of TIs may enable observations of the fractional QHE and other interaction-induced quantum phenomena.

\begin{figure*}[t]
	\centering
	\includegraphics[width=15cm]{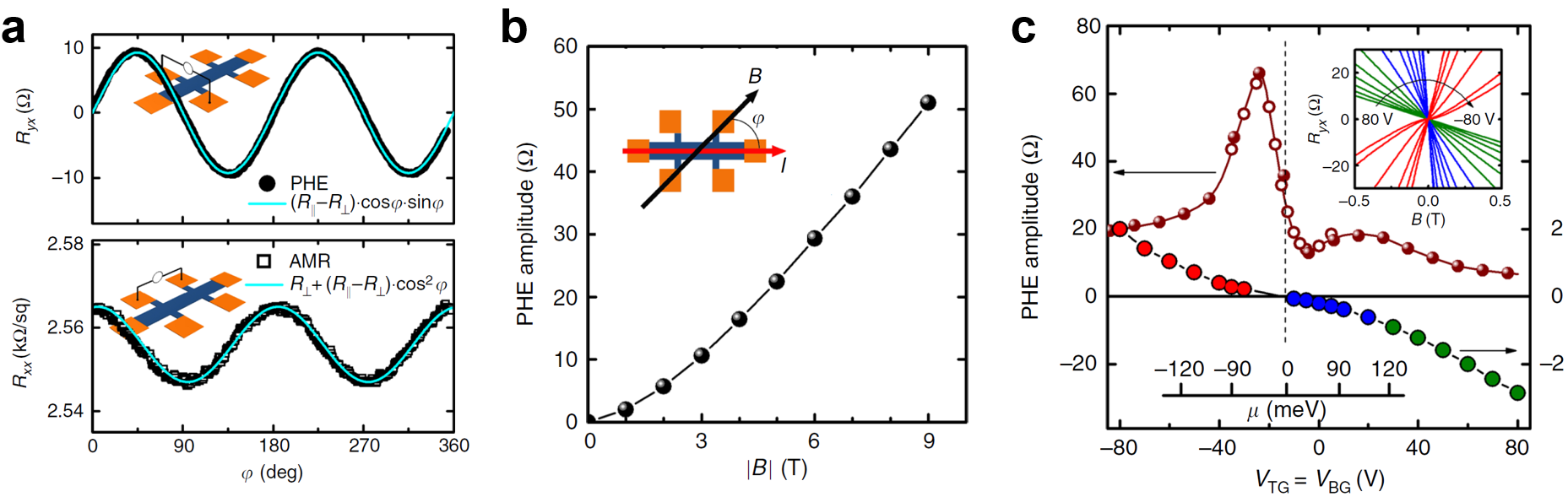}
	\caption{Planar Hall effect (PHE) and anisotropic magnetoresistance (AMR) in a dual-gated (Bi,Sb)$_2$Te$_3$ thin film. (a) Hall resistance $R_{yx}$ and longitudinal resistance $R_{xx}$ versus $\varphi$, the angle between the current and the in-plane magnetic field. (b) The amplitude of PHE as a function of the strength of magnetic field. (c) PHE amplitude as a function of gate voltages, which are symmetrically tuned (i.e. the top and bottom gate voltages remain equal). The upper right inset shows the Hall resistance curves under various gate voltages. The change in chemical potential can be estimated from the gate-voltage dependence of the Hall resistance (bottom curve and scale bar). Adapted from Taskin \textit{et al.}, Nat. Commun. \textbf{8}, 1340 (2017)~\cite{Taskin2017}.}
	\label{fig:planarhalleffectandanisotropicmrina3dtithinfilm}
\end{figure*}

Theoretical developments on exotic quantum Hall physics in topological materials merit a brief mention. In TIs, the half-quantised quantum Hall effect on a single surface was addressed in \cite{Konig_PRB2014}, while the Landau levels of thin films were recently discussed \cite{Song-Bo_SR2015}. The Hall plateaux fall into two patterns, in one pattern the filling number covers all integers while only odd integers in the other. The quantum spin Hall effect, normally protected by time reversal symmetry, is destroyed by the magnetic field together with structure inversion asymmetry. Li \textit{et al} \cite{LiX_PRL2013} showed that in a MoS$_2$ trilayer the Landau level energies grow linearly with B, rather than with $\sqrt{B}$, and display an unconventional Hall plateau sequence consisting of odd negative integers and even positive integers, with the crossover at $n = 0$. A spin-resolved $p-n$ junction also exhibits Landau levels at fractional filling factors. Later, it was shown that, at low doping, TMDs under shear strain develop spin-polarized Landau levels residing in different valleys, with Landau level gaps between 10 and 100 K \cite{Cazalilla_PRL2014}. A superlattice arising from a Moire pattern can lead to topologically nontrivial subbands, with edge transport quantised. Ref.~\cite{LiX_PRL2016} focused on the (111) surface of SnTe in a quantising magnetic field, predicting a nematic phase with spontaneously broken $C_3$ symmetry due to the competition between intravalley Coulomb interactions that favor a valley-polarised state and weaker intervalley scattering that increases in strength as the magnetic field is ramped up.

\subsubsection{Planar Hall effect}

When a magnetic field is applied parallel to the surface of a 3D TI, the transport properties were once considered to be unaffected because the Zeeman term can be gauged way by shifting the Dirac cone in the momentum space.  Taskin \textit{et al.}, however, observed that in dual-gated devices of bulk-insulating $(\mathrm{Bi}, \mathrm{Sb})_{2} \mathrm{Te}_{3}$ thin films with dominating surface transport the in-plane magnetic field can induce an anisotropic magnetoresistance (AMR) and the planar Hall effect (PHE)~\cite{Taskin_NC2017}. In contrast to the anomalous Hall effect, the planar Hall effect consists of a resistivity anisotropy induced by an \textit{in-plane} magnetic field, which anisotropically lifts the protection of surface Dirac fermions from backscattering, in a configuration in which the conventional Hall effect vanishes. Both AMR and PHE are phenomenologically similar those encountered in ferromagnetic materials. The former exhibits an angular dependence of $\cos ^{2} \varphi$ and the latter is proportional to $\sin \varphi \cos \varphi,$ where $\varphi$ is the angle between the magnetic field and the applied current. The AMR and PHE are two aspects of the same anisotropic scattering process in conventional magnetic materials, and consequently they have the same amplitude. Taskin \textit{et al.} proposed a theoretical model to account for these effects in 3D TIs. It is based on the assumption that the in-plane magnetic field breaks the time-reversal symmetry, and removes the protection from backscattering for spins perpendicular to the magnetic field, but at the same time maintains the protection for spins parallel to the magnetic field. This model can qualitatively reproduce the double-dip structure of the AMR/PHE amplitude as function of gate voltage. They estimated that about 10$\%$ of impurity scatterings are related to spin flips at $B=9$\,T~\cite{Taskin_NC2017}. The planar Hall effect can appear in the presence of magnetic disorder in a TI/ferromagnet structure when the external magnetic field is aligned with the magnetisation orientation \cite{Chiba_PRB2017}, as well as in tunneling across a single ferromagnetic barrier on a TI surface when the magnetisation has a component along the bias direction \cite{Scharf_PRL2016}. Recently, the PHE and AMR effects were observed in $\mathrm{Sr}_{0.06} \mathrm{Bi}_{2} \mathrm{Se} 3$~\cite{HuangH2018} and Sn-doped $(\mathrm{Bi}, \mathrm{Sb})_{2}\left(\mathrm{Te}, \mathrm{S}\right)_{3}$ samples~\cite{WuB2018}. Nandy \textit{et al.} developed a semiclassical Boltzmann theory of the transport in $\mathrm{Bi}_{2} \mathrm{Se}_{3}$ and showed that the nontrivial Berry phase of the bulk states can also lead to the PHE, as well as negative longitudinal magnetoresistance~\cite{Nandy2017}. Certain contributions to the planar Hall effect can be traced to the Berry phase and orbital magnetic moment familiar from semiclassical dynamics \cite{Nandy_SR2018}. Similar phenomena have been observed topological semimetals~\cite{HuangXC2015} and discussed in the context of chiral anomaly~\cite{Burkov2017,Nandy2017}. However, according to a recent theoretical study, the negative longitudinal magnetoresistance is not necessarily associated with the chiral anomaly~\cite{DaiX2017}.

\begin{figure*}[t]
	\centering
	\includegraphics[width=15 cm]{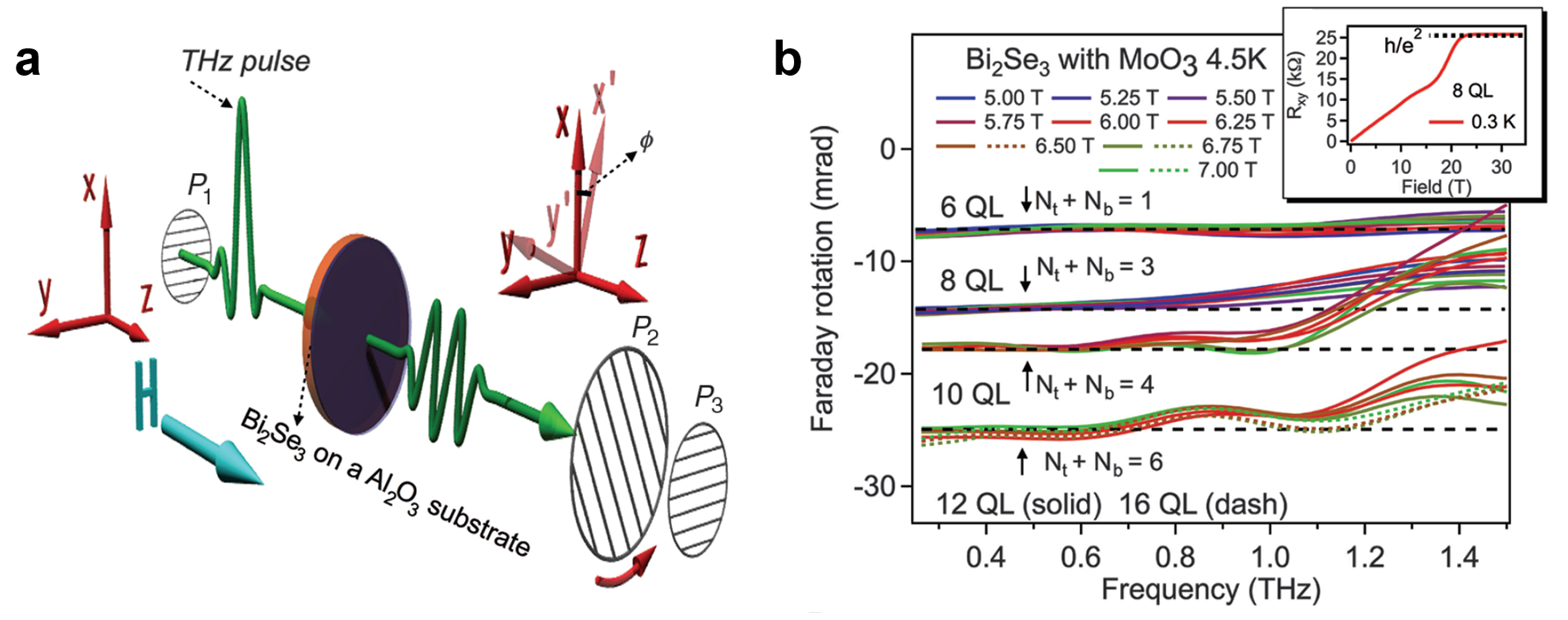}
	\caption{Topological magnetoelectric effect in Bi$_2$Se$_3$ thin films. (a) Sketch of the Faraday rotation measurement setup in the THz regime. (b) Faraday rotation angle as a function of the frequency for severalBi$_2$Se$_3$ thin films with thicknesses of 6 to 12 quintuple layers (QL). The dashed horizontal lines denote theoretical values expected from the magnetoelectric effect. The upper-right corner shows the quantum Hall effect of a 8 QL thin film prepared in a similar condition. Adapted from Wu \textit{et al.}, Science\textbf{ 354}, 1124 (2016)~\cite{WuL2016}.}
	\label{fig:TME}
\end{figure*}

\subsubsection{Topological magnetoelectric effect}

In the previous subsections we have been mainly concerned with dc transport properties. The difference between a 3D TI and an ordinary insulator can also be manifested in their response to the electromagnetic field. According to the topological field theory~\cite{Qi&Zhang2008}, the Maxwell Lagrangian has an axion term $\frac{\alpha}{4 \pi^{2}} \theta \mathbf{E} \cdot \mathbf{B}$, in which $\alpha=\frac{e^{2}}{\hbar c}$ is the fine structure constant, and $\theta=\pi$ for a 3D TI. In contrast, $\theta$ takes a value of 0 for an ordinary insulator. When the time-reversal symmetry (TRS) is broken (by applying external magnetic field, utilizing proximity effect of a magnetic insulator, or magnetic doping, see Section~\ref{magnetic_doping}), the axion term leads to some interesting modifications in the Maxwell equations, and hence the topological magnetoelectric effect~\cite{Qi&Zhang2008, Qi&Zhang2009}. This can be detected by optically techniques, such as Faraday and Kerr rotations. In 3D TIs, the Faraday and Kerr rotation angles are quantized in units of the fine structure constant, if the Fermi level can be placed between the Landau levels or inside the magnetic gap induced by exchange interaction. In the former case the Faraday rotation angle follows $\tan \theta_{F}=\alpha\left(N_{t}+\frac{1}{2}+N_{b}+\frac{1}{2}\right)$ in a free-standing TI film, while the latter is corresponding to a quantum anomalous Hall insulator in which $\tan \theta_{F}=\alpha$.

The quantized Faraday rotation has been observed with time-resolved terahertz (THz) spectroscopy by a few groups~\cite{WuL2016,Dziom2017,Okada2016}. Wu \textit{et al.} used high quality Bi2Se3 thin films grown on sapphire substrates and capped with MoO3, in which the chemical potential is as low as 30 meV even without gating. Fig.\,\ref{fig:TME} shows the Faraday rotation angles are close to the quantized values expected for total Landau level filling factor $N_{t}+N_{b}=1,3,4,$ and 6. It is noteworthy that the quantization of $\theta_{F}$ can be observed in magnetic fields down to 5\,T, much lower than $B$ = 20 T required for observation the QHE in similar films~\cite{WuL2016}. The quantized Faraday rotation was demonstrated in a strained HgTe thin film tuned with a Ru top gate in high magnetic fields~\cite{Dziom2017}. The THz spectroscopy measurements of Cr-doped $(\mathrm{Bi}, \mathrm{Sb}) \mathrm{Te}_{3}$ thin films showed that a material-independent scaling function of $\theta_{F}$ and Kerr rotation angle $\theta_{K}$ approaches to the fine structure constant as the dc Hall conductance increases toward 1 $e^{2}/h$~\cite{Okada2016}.

\subsection{Novel magnetism with 3D TIs}

\subsubsection{Breaking time-reversal symmetry in 3D TIs}

Breaking TRS in 3D TIs are predicted to result in a plethora of novel quantum phenomena, and many of them have been realized in experiment. The surface states with broken TRS can be described with the following Hamiltonian
\begin{equation}
H=\hbar v_{F}\left(k_{x} \sigma_{y}-k_{y} \sigma_{x}\right)+\Delta \sigma_{z},
\end{equation}
in which the mass term $\Delta \sigma_{z}$ is introduced by the exchange interaction. The mass term turns the gapless surface states into a system with a gap of $2|\Delta|$. The broken TRS also modifies the spin texture of the surface states. The electron spins are no longer locked perpendicularly to the momentum. They rather form a hedgehog-like spin texture in which the spin direction is perpendicular to the TI surface at $\mu=\pm\Delta$ and slowly evolves into the in-plane helical structure for $|\mu| \gg |\Delta|$. As a consequence, the Berry phase becomes $\varphi=\pi(1-|\Delta/\mu|)$ if the Fermi level lies outside the mass gap $(|\mu|>|\Delta|)$. When the Fermi level is located inside the gap, a topological invariant, known as the Chern number $C$, can be defined by integration over the Brillouin zone. For a slab-shaped sample, $C=\pm\left(\frac{1}{2}+\frac{1}{2}\right)=\pm1$, in which both the top and bottom surfaces contribute 1/2 to the total value, and the sign of $C$ is identical to that of $\Delta$.

The topologically nontrivial electronic structure described above can be manifested in electron transport. If the Fermi level lies in the mass gap, the quantum anomalous Hall effect (QAHE) occurs. The Hall conductivity follows $\sigma_{x y}=C \frac{e^{2}}{h}$, accompanied by vanishing longitudinal conductivity $\left(\sigma_{x x}=0\right)$. Both properties are the hallmarks of a Chern insulator, in which either QAHE or QHE can be observed. In the latter, the Chern number is equal to the integer filling factor of the Landau level, also known as the TKNN number~\cite{Thouless1982}. For a Chern insulator with boundary, electron transport is carried by 1D ballistic chiral edge states surrounding an insulating bulk. Such transport is often called dissipationless because the edge transport can be free from backscattering for macroscopic distances. It should be noted, however, that energy dissipation cannot be completely avoided in a Chern insulator. It takes place at the so-called hot spots at the corners of source and drain contacts~\cite{Klass1992}, and the contact resistance is of the order $\frac{1}{C} \frac{h}{e^{2}}$.  

When the Fermi level lies outside the mass gap, the bulk becomes conducting, and the Hall conductivity is no longer quantized. Ado \textit{et al.}~\cite{Ado_EPL2015} considered all three sources of the anomalous Hall (AH) conductivity, namely the intrinsic contribution, skew scattering and side jump, and predicted that $\sigma_{x y} \propto\left(\frac{\Delta}{\mu}\right)^{3}$ for $\mu \gg \Delta$. Lu \textit{et al.} predicted that the variation of the Berry phase from $\varphi=\pi$ at $|\mu| \gg |\Delta|$ to $\varphi=0$ at the edge of the mass gap (i.e. $\mu=\pm \Delta$) can induce a crossover from the WAL to the weak localisation~\cite{LuHZ2011}.

Two approaches have been widely explored to open the mass gap in 3D TIs~\cite{HeK2018,Tokura2019}. One is to introduce magnetic order by doping with transition metal elements. The other is to utilize interfacial exchange interactions in TI/magnetic insulator heterostructures. In the following two subsections, we shall review the experimental progresses in these two directions, while the subsequent subsection covers theoretical developments in this field.

\subsubsection{\label{magnetic_doping}Magnetically doped 3D TIs}

\begin{figure}[t]
	\centering
	\includegraphics[width=1\linewidth]{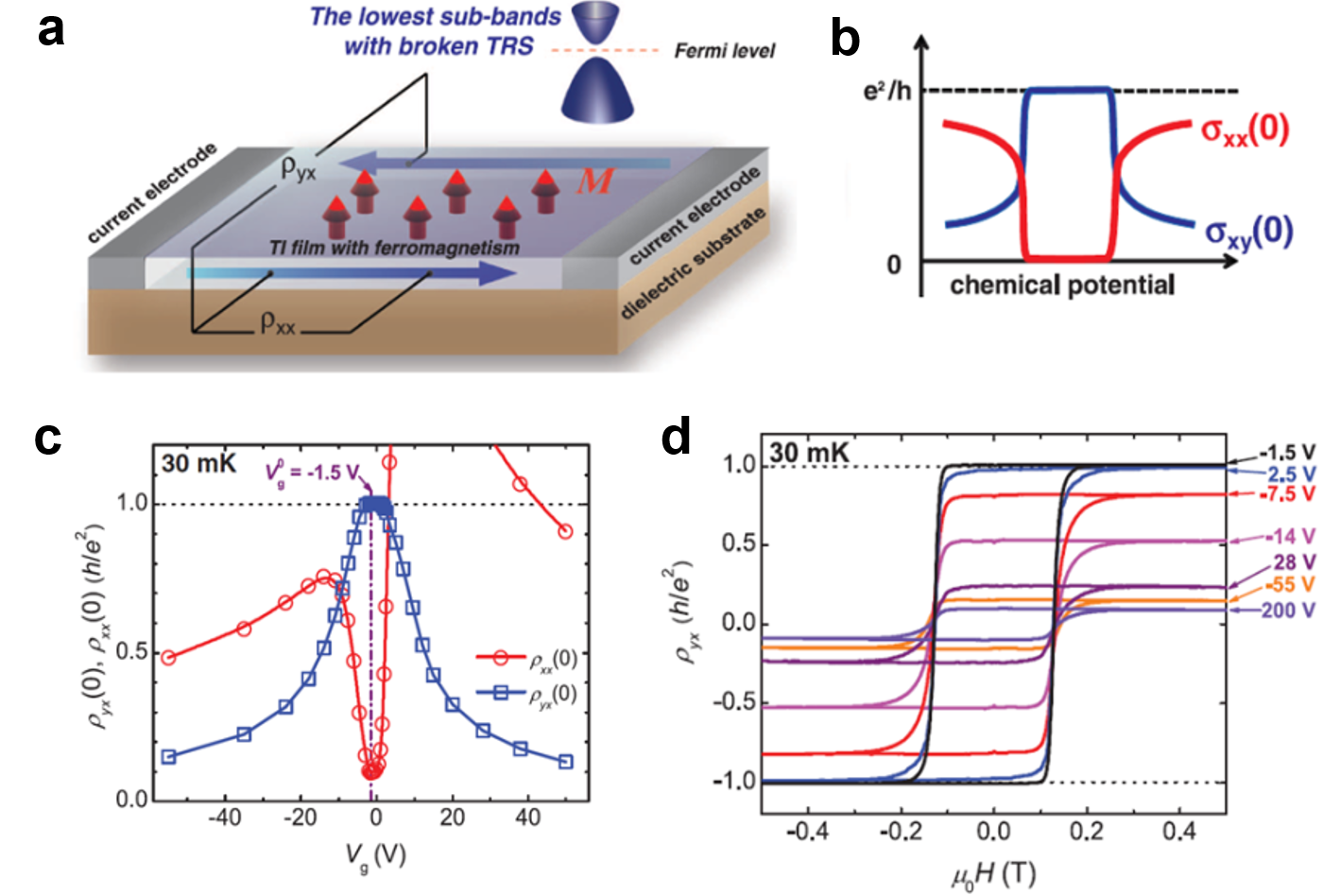}
	\caption{Observation of the quantum anomalous Hall effect. (a) Sketch of the backed-gated Cr-doped (Bi,Sb)$_2$Te$_3$ thin film. The perpendicular magnetization induced by the Cr doping turns the massless surface states into a massive, gapped 2D system. (b) Qualitative chemical potential of longitudinal conductivity ($\sigma_{xx}$) and Hall conductivity ($\sigma_{xy}$). Quantization of $\sigma_{xy}$ can be realized when the Fermi level is tuned into the exchange gap. (c) Experimentally observed longitudinal resistivity ($\rho_{xx}$) and Hall resistivity ($\rho_{xy}$) plotted as a function of the back-gate voltage. (d) Hall resistivity as a function of applied magnetic field at several gate voltages. Taken from Chang \textit{et al.}, Science \textbf{340}, 167 (2013)~\cite{ChangCZ2013Thin}.}
	\label{fig:quantumanomaloushalleffectinmagneticallydopedtis}
\end{figure}

Many transition metal and rare earth elements have been used as magnetic dopants to introduce ferromagnetic order in 3D TIs. Among them, Cr, V, Mn and Fe have received most attention~\cite{YuR2010,ChenYL2010,XuSY2012,ChangCZ2013Thin,ChangCZ2015}. In particular, Cr-doped $(\mathrm{Bi}, \mathrm{Sb})_{2} \mathrm{Te}_{3}$ (Cr-BST) thin films allowed for the first experimental observation of the QAHE~\cite{ChangCZ2013Thin}, as shown in Fig.\,\ref{fig:quantumanomaloushalleffectinmagneticallydopedtis}. 
Enormous experimental efforts have since been made to optimize the MBE growth Cr-BST thin films. However, observation of the QAHE in Cr-BST thin films is still limited to temperatures of a few hundred mK or lower, despite that the Curie temperature is on the order of 10\,K~\cite{HeK2018,Tokura2019}. This can be attributed partially to the spatial fluctuations in the Dirac mass gap. According to an STM study, the local exchange gap in a Cr-BST thin film varies from 9 meV to 51 meV due to fluctuations in the local density of magnetic dopants~\cite{Lee2015Imag}. In the first observation of QAHE~\cite{ChangCZ2013}, the zero-field longitudinal conductivity was about $0.1 \frac{e^{2}}{h}$ even at $T$=30 mK and the backscattering through the bulk states restricted the accuracy of quantum anomalous Hall (QAH) resistance to about 1$\%$~\cite{ChangCZ2013}. More precise QAHE was later realized in V-doped $\mathrm{Sb_2Te_3}$ thin films, in which the QAH plateau reached a precision level of 0.02$\%$ at $T$=25\,mK~\cite{ChangCZ2015}. Recently, two groups have managed to observe QAH resistances with precision levels of 1 part in $10^{6}$ and 0.1 part in $10^{6}$ in Cr-BST~\cite{Fox2018} and V-doped $(\mathrm{Bi}, \mathrm{Sb})_{2} \mathrm{Te}_{3}$ (V-BST)~\cite{Goetz2018} thin films, respectively.

A lot of efforts have also been devoted to increasing the observation temperature of QAHE. One method is to utilize the modulation doping technique developed in semiconductor heterostructures~\cite{Mogi2015}. Two ultrathin $(\mathrm{Bi}, \mathrm{Sb})_{2} \mathrm{Te}_{3}$ layers heavily doped with Cr are placed 1 nm from the top and bottom surfaces while the majority of the film is not magnetically doped. The Hall resistance in this multilayer structure reaches $0.97 \frac{h}{e^{2}}$ at $T$= 2 K. The other method is to use a co-doping scheme. The QAHE can be realized at $T$=1.5 K in a Cr and V co-doped $(\mathrm{Bi}, \mathrm{Sb})_{2} \mathrm{Te}_{3}$ sample. It is believed that the increased homogeneity is responsible for the realization of QAHE at higher temperatures~\cite{OuYB2018}.

It was also shown that there is a large difference in the coercive fields of Cr- and V-doped BST thin films. This property was employed to realize the axion insulator state~\cite{Mogi2017Amag,Mogi2017Tail,XiaoD2018}, in which the magnetization directions of Cr-BST and V-BST layers are opposite to each other, and the longitudinal conductivity vanishes like an QAH insulator, but the Hall conductivity displays a zero plateau. These features are nearly perfectly borne out in the transport data of a V-BST/BST/Cr-BST trilayer structure~\cite{Mogi2017Amag,Mogi2017Tail}. The axion insulator phase can exist in a wide range of magnetic fields $\left(\mu_{0} H=0.2-0.8 \mathrm{T}\right)$ at low temperatures. It is remarkable that the longitudinal resistance can be as high as $1 \mathrm{G} \Omega$ in this insulating phase. Similar results are also reported in Ref.~\cite{XiaoD2018}, which shows the zero-plateau of the Hall resistance, in addition to those of the longitudinal and Hall conductivities.

In contrast to the Cr and V doping, experiments with other magnetic dopants have been futile in producing a well-defined QAH phase to date~\cite{ChenYL2010,XuSY2012,Checkelsky2012,ZhangDM2012,ZhangM2014}. This was somewhat surprising considering that sizable gap opening in the surface states was observed in Fe-doped and Mn-doped $\mathrm{Bi}_{2} \mathrm{Se}_3$ thin films with ARPES measurements~\cite{ChenYL2010,XuSY2012}. In a subsequent experiment, however, S{\'a}nchez-Barriga \textit{et al.} showed that in Mn-doped $\mathrm{Bi}_{2} \mathrm{Se} 3$ thin films the gap in the surface states, which can be as large as 200\,meV, survives at room temperature, in stark contrast with the low Curie temperatures in the sample (less than 10 K for both bulk and surface magnetism)~\cite{Sanchez-Barriga2016}. They suggested that this kind of energy gap originates from $nonmagnetic$ resonant scatterings related to the impurities in the bulk, instead of the magnetic exchange interaction~\cite{Sanchez-Barriga2016}. Such nonmagnetic scatterings complicate the electronic structure and are detrimental to realizing the QHAE. Recently, Liu \textit{et al.} further showed that the AH resistance in $\left(\mathrm{Bi}_{1-\mathrm{x}} \mathrm{Mn}_{\mathrm{x}}\right)_{2} \mathrm{Se}_{3}$ thin film can have too components with different signs, and the surface component has a sign opposite to that of the QAH phases in Cr- or V-doped BST films~\cite{LiuN2018}. Similar sign in the AH resistance was also observed in Mn-doped $\mathrm{Bi}_{2}\left(\mathrm{Te}_{2} \mathrm{Se}\right)_{3}$~\cite{Checkelsky2012}, Mn-doped $\mathrm{Bi}_{2}\mathrm{Te}_{3}$~\cite{Lee2014}, and Cr-doped  $\mathrm{Bi}_{2}\mathrm{Te}_{3}$~\cite{ZhangJS2013,ZhangZC2014}. It remains to be investigated whether such 'anomalous' sign in $\sigma_{x y}$ is related to the non-magnetic scatterings \cite{Aydin_2019}. Nevertheless, the results described above suggest that the influence of magnetic dopants is far more complicated than the mean field exchange interaction considered in the massive Dirac fermion model. A deep understanding of the impurity effect, both magnetic and non-magnetic, are crucial to finding magnetic TIs with improved quality.

\subsubsection{Magnetic heterostructures}

Another approach to open a gap in the TI surface states is to make use of the proximity effect in TI/magnetic insulator (MI) heterostructures. Both ferromagnetic and antiferromagnetic insulators with spin aligned perpendicular to the surface are capable of generating a mass gap via the exchange interaction across the interface. One advantage of the heterostructure approach is that it may overcome the difficulty of magnetic inhomogeneity encountered in magnetically doped TIs~\cite{Lee2015Imag}. This not only allows for the observation of QAHE at high temperatures, but also facilitates the study of exotic quasiparticles, such as chiral Majorana zero modes~\cite{FuL2008Supe,FuL2009Prob,Akhmerov2009Elec} and magnetic monopoles~\cite{Qi&Zhang2008}.

Many ferromagnetic or ferrimagnetic insulators, including EuS~\cite{WeiP2013Exch,YangQ2013Emer}, GdN~\cite{Kandala2013}, $\mathrm{Y}_{3} \mathrm{Fe}_{5} \mathrm{O}_{12}$ (YIG)~\cite{LangMR2014Prox}, $\mathrm{BaFe}_{12} \mathrm{O}_{19}$~\cite{YangWM2014Prox}, $\mathrm{Cr}_{2} \mathrm{Ge}_{2} \mathrm{Te}_{6}$~\cite{Alegria2014Larg}, $\mathrm{Tm}_{3} \mathrm{Fe}_{5} \mathrm{O}_{12}$ (TIG)~\cite{TangC2017Abov},  $\mathrm{Fe}_{3} \mathrm{O}_{4}$~\cite{Buchenau2017Symm}, and $\mathrm{CoFe}_{2} \mathrm{O}_{4}$~\cite{HuangSY2017Prox}, have been used to fabricate the MI/TI heterostructures. Evidence for interfacial magnetic interactions has been obtained by the measurements of the WAL effect in perpendicular or parallel magnetic fields~\cite{WeiP2013Exch,YangWM2014Prox}, Kerr spectroscopy~\cite{LangMR2014Prox}, and polarized neutron reflectometry~\cite{Katmis2016Ahig,LiMD2017Dira}. However, the gap opening effect are very weak in most of these heterostructures, as evidenced by the fact that the reported AH resistances did not exceed several Ohms. Nevertheless, AH resistances up to 120\,$\Omega$ have been observed recently in a $(\mathrm{Bi}, \mathrm{Sb})_{2}(\mathrm{Te}, \mathrm{Se})_{3} /(\mathrm{Ga}, \mathrm{Mn}) \mathrm{As}$ heterostructure, in which the magnetic semiconductor layer is in an insulating regime and has a perpendicular easy axis~\cite{LeeJS2018Engi}. More recently, in a $\mathrm{Cr}_{2} \mathrm{Ge}_{2} \mathrm{T} \mathrm{e}_{6} / \mathrm{TI} / \mathrm{Cr}_{2} \mathrm{Ge}_{2} \mathrm{T}_{6}$ sandwich structure AH resistances of the order k$\Omega$ has been achieved~\cite{Mogi2019Larg}. It is noteworthy that in these two studies, both the TI and MI layers were fabricated with MBE in ultrahigh vacuum. This suggests that high interface quality is crucial in obtaining strong magnetic proximity effect in the TI surface states.

The exchange interaction between antiferromagnetic insulators and TIs may also open a sizable gap in the surface states. Based on the first principles calculations, the magnetic proximity effect in $\mathrm{MnSe} / \mathrm{Bi}_{2} \mathrm{Se}_{3}$ heterostructure can induce a gap of about 54 meV~\cite{LuoWD2013}. Another theoretical study of the same heterostructure, however, suggested that the gap associated with the Dirac surface states is very small (8.5 meV), and also coexists with trivial metallic states~\cite{Eremeev2013}. An ARPES study of MnSe ultrathin layers grown on $\mathrm{Bi}_{2} \mathrm{Se}_{3}$ however, revealed an energy gap of ~100 meV, which is free of any other electronic states. This surprising result is attributed to direct interaction of the Dirac surface states with a $\mathrm{Bi}_{2} \mathrm{Mn} \mathrm{Se}_{4}$ septuple (SL) layer (1 SL=Se-Bi-Se-Mn-Se-Bi-Se), which is formed from the intercalation of a MnSe layer into the quintuple layers of $\mathrm{Bi}_{2} \mathrm{Se}_{3}$. The calculation reported in Ref.~\cite{Hirahara2017Larg} also produces a Chern number $C=-1$, indicating high temperature QAHE can in principle be observed in this system. Very recently, exciting results have been reported on a antiferromagnetic compound in the same family, $\mathrm{Bi}_{2} \mathrm{Mn} \mathrm{Te}_{4}$. The first principles calculations suggested that when $\mathrm{Bi}_{2} \mathrm{Mn} \mathrm{Te}_{4}$, which has a layered structure, is exfoliated into ultrathin layers, the ground state can oscillate between a QAH insulator and an axion insulator phase, depending on whether the number of $\mathrm{Bi}_{2} \mathrm{Mn} \mathrm{Te}_{4}$ septuple layers is odd or even~\cite{Otrokov2019Uniq,LiJ2019Intr}. It is remarkable that a quantized Hall resistance was observed lately in a 5 SL microflake of $\mathrm{Bi}_{2} \mathrm{Mn} \mathrm{Te}_{4}$ at $T$=4 K, although a high magnetic field had to be applied to align the $\mathrm{Mn}^{2+}$ spins~\cite{DengYJ2019Magn}. Further work along this line may lead to the observation of both QAHE and the axion insulator state in zero magnetic field in stoichiometric materials. This might pave a way for discovering novel quantum phenomena as well as manipulate exotic quasiparticles for potential applications.

\subsubsection{Magnetic ordering in TIs: theory}

Magnetic order has a profound effect on charge transport through a material. When a current passes through a magnetised material topological mechanisms and scattering processes predominantly deflect electrons in one direction. This results in an additional current perpendicular to the driving current, which depends on the magnetisation, and vanishes if the material is non-magnetic. Whereas the Hall effect of classical physics requires an external magnetic field, this effect, termed the anomalous Hall effect, requires only a magnetisation. It is often the smoking gun for detecting magnetic order and remains one of the biggest focus areas for research on TIs. In ultra-thin films of magnetic topological insulator a quantised anomalous Hall effect (QAHE) exists, carried by edge states believed to be topologically protected, with a transverse conductivity of $e^2/h$. The quantum anomalous Hall effect is believed to be dissipationless and is being investigated for electronics applications \cite{He_ARCMP2018}. 

Prior to discussing anomalous Hall effects, we examine the critical aspect of magnetic ordering in TI, which has been the subject of a number of papers. In undoped TI the ordering was originally believed to be due to the van Vleck mechanism \cite{Yu_TIF_QAHE_Science2010} and to point out of the plane, in contrast to thin film ferromagnets, in which the magnetisation typically points in the plane. This interpretation is being challenged, as we shall see below. Interestingly, Rosenberg and Franz \cite{Rosenberg_PRB2012} showed that in a magnetically doped metallic TI the surface magnetic ordering can persist up to a higher critical temperature than the bulk, so that a regime exists where the surface is magnetically ordered but the bulk is not. This is believed to be because the metallic surface state is more susceptible to magnetic ordering than the insulating bulk. Magnetic impurities were also shown to give rise to a Kondo effect \cite{WangCulcer_TI_Kondo_PRB2013} and non-Fermi liquid behavior \cite{Principi_PRB2015}. Even in the absence of magnetism distinctive features appear in the static spin response in 2D TIs for the topologically nontrivial band-inverted structure \cite{Kernreiter_PRX2016}. Electronic correlations such as the Hubbard U affect the ordering in ferromagnetically doped TI thin films, as was seen in V-doped Sb$_2$Te$_3$ \cite{Kim_Correl_PRB2018}. The on-site Coulomb interaction can turn the TI thin film into a Mott insulator and facilitate it entering the quantum anomalous Hall phase, discussed below. Ferromagnetic order is determined by $p$-orbital-assisted long-range superexchange as well as short-range double-exchange between the partially filled $d$-bands, which enhances it relative to Cr doping.

In determining the highest possible observation temperature of the QAHE the two important energy scales are the band gap of the magnetic TI film, given by the size of the magnetisation, and the ferromagnetic Curie temperature. The smaller of the two defines the observation temperature of the QAHE. Two interesting publications have shed light on co-doping mechanisms that can be used to enhance the temperature at which the QAHE is observed, which are currently a few tens of mK. Qi \textit{et al} \cite{Zhenyu_QAHE_PRL2016} determined that the QAHE can occur at high temperatures in n-p co-doped TIs, taking as an example vanadium-iodine co-doped Sb$_2$Te$_3$. The chosen dopants have a preference for forming n-p pairs due to mutual electrostatic attraction, thereby enhancing their solubility. While doping with V alone would shrink the bulk gap, co-doping with I restores it to its original value. Even at 2$\%$ V and 1$\%$ I, the QAHE persists until 50K. The authors ascribe this enhancement to the fact that compensated $n-p$ codoping preserves the intrinsic band gap of the host material. In a similar vein, Kim \textit{et al} showed that the QAHE temperature can be enhanced significantly by Mo-Cr co-doping Sb$_2$Te$_3$ \cite{Kim_Order_PRB2017}. At the same time, these authors discovered that the ferromagnetic order in Cr-doped Sb$_2$Te$_3$ survives when the spin-orbit interaction is turned off, implying that the magnetic order is not governed by the van Vleck-type mechanism, which relies on nontrivial band topology. Since the system is an insulator for Cr doping $\le 10 \%$, the RKKY mechanism can also be discounted, while the surface states were found to be of secondary importance in magnetic ordering. DFT calculations reveal that magnetic ordering arises from long-range exchange interactions within quintuple layers, mediated by directional bonds with Te by certain sets of orbitals on the Cr and Sb atoms, in a similar way to hydrogen local moments in graphene.

A Zeeman field can also be induced in a TI via a magnetic proximity effect. Two studies from the same group shed light on this process for a TI on a ferromagnetic insulator \cite{Menshov_PRB2013} as well as for a TI on an antiferromagnetic insulator \cite{Eremeev_PRB2013}, the latter using analytics as well as density functional theory in Bi$_2$Se$_3$/MnSe(111). A unified qualitative picture emerges. Charge redistribution and mixing of orbitals of the two materials cause drastic modifications of the electronic structure near the interface. In addition to the topological bound state an ordinary bound state is present, which is gapped and spin polarised due to hybridisation with the magnet. The two overlap in space in such a way that the ordinary state mediates indirect exchange coupling between the magnet and the topological state, and the latter acquires a gap at the Dirac point.

\subsubsection{Spin-orbit torque}

\begin{figure}[t]
	\centering
	\includegraphics[width=1\linewidth]{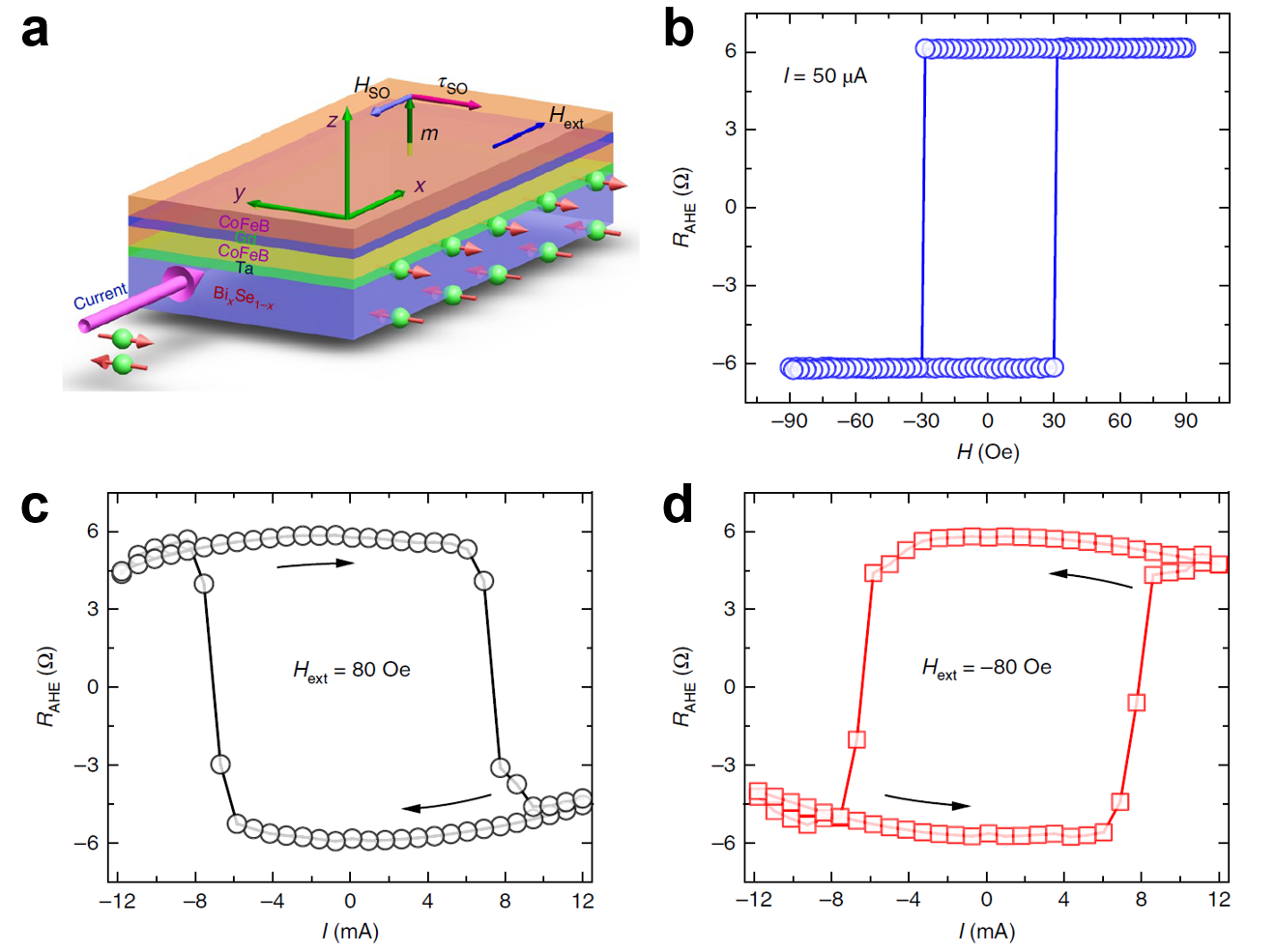}
	\caption{Room temperature current-induced magnetization switching in ferromagnetic metal/TI heterostructure. 
(a) Schematic of a Bi$_x$Se$_{1-x}$ (4\,nm)/Ta (0.5\,nm)/CoFeB (0.6\,nm)/Gd (1.2\,nm)/CoFeB (1.1\,nm) stack used in the measurements. 
(b) Anomalous Hall resistance $R_\mathrm{AH}$ as a function of applied magnetic field. 
(c,d) Current-induced switching of the magnetization due to the spin-orbit torque generated by the in-plane current in the Bi$_x$Se$_{1-x}$ underlayer in the presence of 80\,Oe (c) and $-80$\,Oe (d) in-plane bias $H$-fields. From Mahendra \textit{et al.}, Nat. Mater. 17, 800 (2018)\cite{Dc2018Room}.}
	\label{fig:currentinducedmagnetizationreversalintifmheterostructuresatroomtemperature}
\end{figure}

Much of the recent interest in topological insulators is centred around the phenomenon of spin-orbit torque \cite{Nikolic_SOT, Manchon_SOT}, which is the Onsager reciprocal of charge pumping. The spin-momentum locking in the TI surface states offers a convenient and efficient means to electrically generate a spin polarisation, which can be detected in dc transport by utilising ferromagnetic electrodes as the probe~\cite{LiCH2014Elec}. In a typical experiment a TI is placed on top of a ferromagnet, a current is driven through the TI, and the effect of the current on the magnetisation of the ferromagnet is observed. For a large enough current density the magnetisation may switch, and for technological applications it is desirable to make this critical current density as low as possible, with operation ideally possible at room temperature. 

The efficiency in the generation of SOT can be characterized by a dimensionless parameter, $\eta=\frac{2 e}{\hbar} \frac{J_{S}}{J_{c}}$, where $J_{s}$ is the spin current density and $J_{c}$ is the charge current density. ST-FMR measurements have been carried out on many TI/ferromagnet heterostructures, including $\mathrm{Bi}_{2} \mathrm{Se}_{3} / \mathrm{NiFe}$~\cite{Mellnik_Nature2014,WangY2015Topo}, $\mathrm{Bi}_{2} \mathrm{Se}_{3} / \mathrm{CoFeB}$~\cite{Jamali2015Gian}, $(\mathrm{Bi}, \mathrm{Sb})_{2} \mathrm{Te}_{3} / \mathrm{NiFe}$~\cite{Kondou2016Ferm}, and $\mathrm{Bi}_{2} \mathrm{Se}_{3} / \mathrm{YIG}$~\cite{Fanchiang2018Stron}. The extracted maximum value of the charge-spin conversion efficiency $\eta$ spreads from about 0.4 to 3.5 for each of these systems. Even though the $\eta$ values are in general larger than those of heavy metal/ferromagnet heterostructures~\cite{Mellnik_Nature2014}, they are two orders of magnitude smaller than that the values ($\eta=140-425$) obtained by the second harmonic measurements of the Hall voltages in $(\mathrm{Bi}, \mathrm{Sb})_{2} \mathrm{Te}_{3} /(\mathrm{Cr}, \mathrm{Bi}, \mathrm{Sb})_{2} \mathrm{Te}_{3}$ heterostructures~\cite{FanYB2014Magn}. Such a large discrepancy was resolved in Ref.~\cite{Yasuda_PRL2017}, in which the second harmonic Hall voltage is identified to mainly originate from asymmetric magnon scattering, instead of the contribution from the damping-like SOT due to the current induced spin polarization.

Recently, SOT-induced magnetisation switching has been demonstrated at room temperature in several ferromagnet/TI heterostructures, such as $\mathrm{CoTb} / \mathrm{Bi}_{2} \mathrm{Se}_{3}$~\cite{HanJH2017Room}, a TI-ferrimagnet heterostructure with perpendicular magnetic anisotropy, accompanied by a large spin-Hall effect, $\mathrm{NiFe}/\mathrm{Bi}_{2} \mathrm{Se}_{3}$~\cite{WangY2017Room} due to the spin polarisation induced in the topological insulator, $\mathrm{Bi}_{1-\mathrm{x}} \mathrm{Sb}_{\mathrm{x}} / \mathrm{Mn} \mathrm{Ga}$~\cite{Khang2018Acon} with a sizable spin-Hall effect, and $\mathrm{Bi}_{\mathrm{x}} \mathrm{Se}_{1-\mathrm{x}} / \mathrm{CoFeB}$~\cite{Dc2018Room}. The current density required for magnetic switching is often on the order of $10^{5} \mathrm{A} / \mathrm{cm}^{2}$, which is considerably lower than those for heavy metals (e.g. $5.5 \times 10^{6} \mathrm{A} / \mathrm{cm}^{2}$ for $\beta-\mathrm{Ta}$~\cite{LiuLQ2012}). These experiments demonstrated the potential of the TI-based magnetic random-access memory. However, it should be noted that the ferromagnetic layer used in these experiments are metallic and often have a conductivity much larger (or least comparable to) that of the TI layer. Substantial amount of energy is wasted due to the current shunting by the ferromagnetic layer. It would be very appealing to achieve efficient current induced magnetization reversal in a magnetic insulator. The 2D materials will offer excellent opportunities for fabricating TI/MI heterostructures with higher efficiency in generating SOT, since a number of 2D materials have been identified to be magnetic insulators(semiconductors)~\cite{Mounet2018}, and atomically sharped interface can be obtained handily with van der Waals epitaxy.

Two fundamental effects underpin spin-orbit torques. One is the generation of a spin polarisation of the surface states by an electrical current, or magneto-electric effect \cite{YLG_JETPL1989, YLG_JETP1991}. In a TI/ferromagnet heterostructure the current-induced spin polarisation in the surface states exerts a torque on the ferromagnetic layer, which can be regarded as a TI counterpart of the Rashba spin-orbit torque (SOT) in heavy metal/ferromagnet heterostructures~\cite{Ramaswamy2018}. This effect requires, at the very least, spatial inversion symmetry breaking, hence cannot originate from the bulk of the TI. The second effect is the spin-Hall effect, which, conversely, stems from the bulk of the TI and is zero for the surface states. This refers to a non-equilibrium spin current driven by an electric field. The exact origin of the strong torque observed in any one experiment, in particular whether it stems from the spin-momentum locking of the surface states or from spin-Hall currents in the bulk, tends to be unclear. In this context Ref.~\cite{Ghosh_PRB2018} considered a TI/ferromagnet heterostructure in which the Fermi energy was varied so that at one extreme transport is entirely surface-dominated while at the other it is entirely bulk-dominated, and showed that the spin Hall torque remains small even in the bulk-dominated regime. A current-induced spin polarisation was observed in Bi$_2$Se$_3$ by performing spin torque ferromagnetic magnetic resonance (ST-FMR) experiments \cite{Mellnik_Nature2014}, where the spin polarisation is in the plane. In a series of more recent experiments, the current-induced spin polarisation in WTe$_2$ was shown to be out of the plane \cite{MacNeill_NP2017, Li_WTe2_NC2018}.

\begin{figure}
	\begin{center}
		\includegraphics[width = \columnwidth]{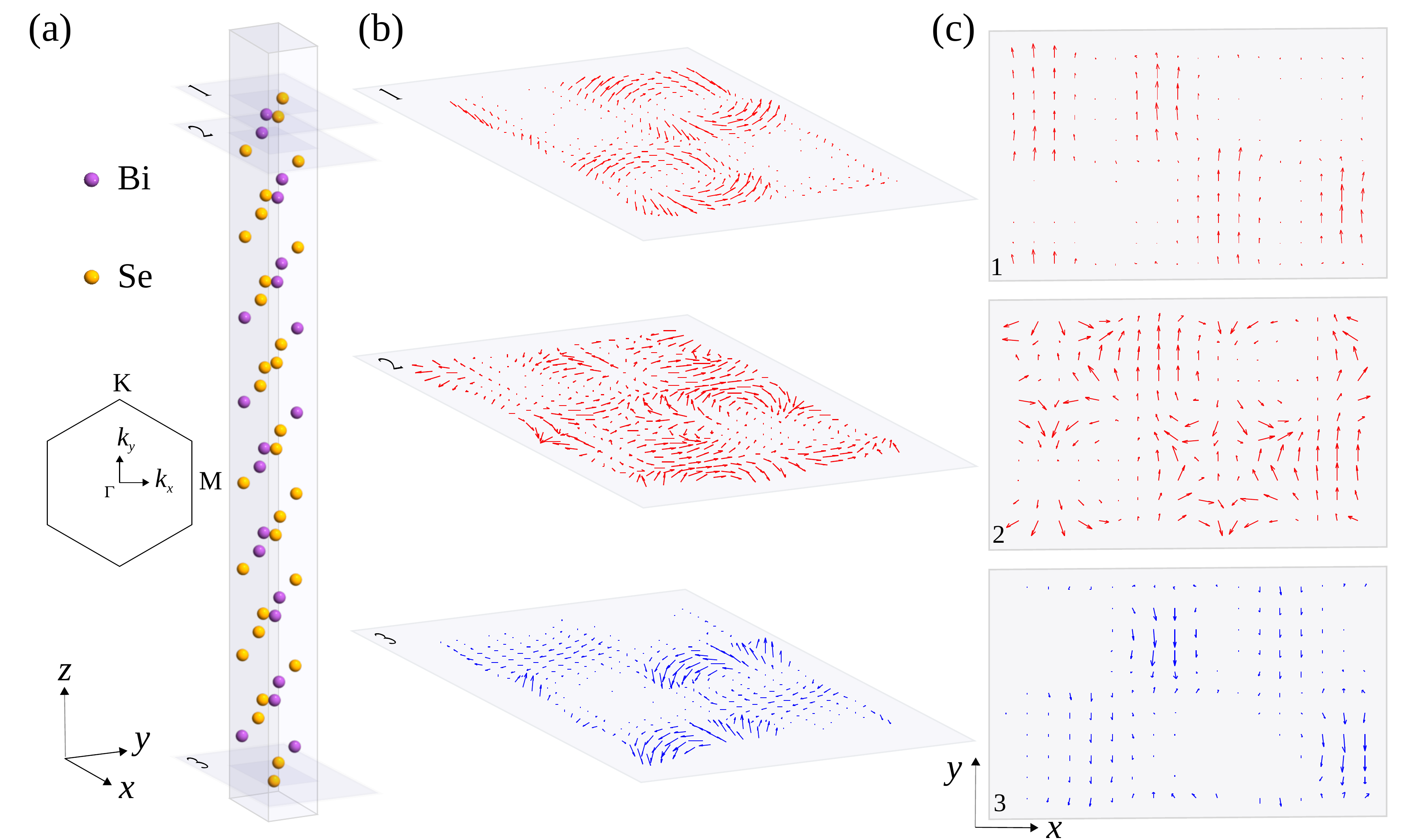}
		\caption{\label{Nikolic} Current-induced spin polarisation and spin texture, adapted from \cite{Chang_PRB2015}. (a) The arrangement of Bi and Se atoms in a supercell of a Bi$_2$Se$_3$ thin film with a thickness of 5 quintuple layers. The inset in panel (a) shows the Brillouin zone in the $k_x-k_y$ plane at $k_z = 0$. (b) The vector field of the non-equilibrium spin polarisation ${\bm S}({\bm r})$ within selected planes shown in (a), generated by injection of an unpolarised charge current along the $x$-axis. The planes 1 and 3 correspond to the top and bottom metallic surfaces of the film, while plane 2 resides in the bulk at a distance $d \approx 0.164$ nm away from plane 1. (c) The vector fields in (b) projected onto each of the selected planes in (a). The real space grid of r points in panels (b) and (c) has spacing $\approx 0.4 \AA$.}
	\end{center}
\end{figure}

The current-induced spin polarisation has a simple explanation. A steady-state current corresponds to a net momentum, and since the spin of the surface states is locked to the momentum, this automatically ensures there is a net spin polarisation. This is quite general in a topological insulator, as long as no warping terms are present, and is valid for currents both longitudinal and transverse to the applied electric field. Yet research is beginning to emerge demonstrating that the dynamics in the vicinity of the TI/ferromagnet interface are non-trivial, and may be vital in understanding what is measured experimentally. A numerical study of the spin-orbit magneto-electric effect \cite{Chang_PRB2015} showed that the stead-state surface spin polarisation extends $\approx$ 2nm into the bulk of the TI as a result of wave function penetration into the bulk, as in Fig.~\ref{Nikolic}. When the hexagonal warping term $\propto \lambda$ is taken into account, in addition to modifications to the conductivity \cite{Akzyanov_PRB2018}, an out of plane spin polarisation also emerges. A recent computational work considered a Bi$_2$Se$_3$/Co bilayer \cite{Nikolic_Co_NL2017} and demonstrated that the Co layer is substantially modified to acquire spin-orbit properties of  Bi$_2$Se$_3$, so when current flows through the Co, a non-equilibrium electronic spin density will be generated that is noncollinear to the Co magnetisation.

An additional contribution to the spin polarisation, and therefore the spin-orbit torque, exists when a TI or a 2D spin-orbit coupled semiconductor is placed on a ferromagnet with magnetisation perpendicular to the 2D plane. It has been termed \textit{intrinsic}, meaning it is due to the band structure, and it stems from the fact that the magnetisation opens a gap in the spectrum of the TI/semiconductor. This contribution to the spin polarisation is associated with the same term in the charge conductivity that leads to the anomalous Hall effect. In a 2D semiconductor with spin-orbit coupling linear in momentum it is known that the intrinsic anomalous Hall conductivity is cancelled out by scalar disorder when the vertex corrections are taken into account \cite{AHE_vertex_PRL_2006}, in analogy with cancellations occurring in the spin-Hall effect \cite{Inoue_RashbaSHE_Vertex_PRB04}. Ado \textit{et al} \cite{Ado_SOT_PRB2017} revealed that, as expected, for spin-independent disorder the same is true for the intrinsic spin-orbit torque, and the only remaining torque is due to the current-induced spin polarisation. We note that this cancellation requires both spin-orbit split sub-bands to cross the Fermi surface and does not apply to TI. Ref.~\cite{Ado_SOT_PRB2017} also introduced a more physical representation which allows the decomposition of the torques into a dissipationless component (field-like) invariant under time reversal, and a dissipative component (damping-like) that changes sign under time reversal. In a subsequent work, Xiao and Niu \cite{Xiao_SOT_PRB2017} showed that the net result for the intrinsic torque depends on the structure of the disorder potential.

For a non-equilibrium spin polarisation ${\bm S}$ acting on a magnetisation ${\bm M}$ the torque ${\bm T} \propto {\bm M} \times {\bm S}$. In the magnetism literature spin torques are customarily broken up into two components, one referred to as field-like, and the other as damping-like or antidamping-like. This nomenclature stems from the Landau-Lifshitz-Gilbert equation, in which the magnetic field term ${\bm M} \times {\bm H}$ is responsible for precession and the Gilbert term ${\bm M} \times ({\bm M} \times {\bm H})$ for damping of the precessional motion. In the context of spin transfer torques this nomenclature is justified, since the damping-like torque is dissipative, while the field-like contribution is dissipationless. This nomenclature has also permeated the literature on spin-orbit torques, where the field-like and anti-damping-like torques $\propto {\bm M} \times {\bm S}$ and $\propto {\bm M} \times ({\bm M} \times {\bm S})$ respectively, where ${\bm S}$ is the current-induced spin polarisation. As explained in \cite{Ado_SOT_PRB2017} the analogy is not exact and can be misleading. The current-induced spin polarisation, for example, has different directions for Rashba and Dresselhaus spin orbit coupling, yet is always dissipative, as it depends on the scattering time $\tau$. For a realistic structure both the Rashba and Dresselhaus interactions tend to be present and the direction of the current-induced spin polarisation is unknown a-priori. 

Fischer \textit{et al} \cite{Fischer_PRB2016} concentrated on heterostructures comprising either a TI on a ferromagnet or a TI on a magnetically-doped TI. In both cases the magnetisation was taken to be in the plane. In addition to the current-induced spin polarisation, an additional out-of-plane torque was found, arising from spin diffusion across the interface combined with spin precession of the current-induced spin polarisation around the in-plane magnetisation. The two have different transformation properties under magnetisation reversal. For a TI on a ferromagnet both torques have comparable efficiencies, while for a TI on a magnetically doped TI the spin transfer-like torque is found to dominate. 
Within a similar setup, differences between the single Fermi surface Rashba-Dirac Hamiltonian and two Fermi surface Rashba Hamiltonian are considered \cite{Ndiaye_PRB2017}. 

A series of papers have reported drastic enhancements of spin-orbit related effects in hybrid graphene-TI structures. In Ref.~\cite{Jin_PRB2013} it is reported that epitaxial graphene on the TI Sb$_2$Te$_3$ evolves into the quantum spin-Hall phase and develops a spin-orbit gap of 20 meV. Another paper \cite{RV_PRB2017} reports the possibility that the addition of graphene monolayers or bilayers to a TI-based magnetic structure greatly enhances the current-induced spin polarisation by a factor of up to 100, due to the high mobility of graphene and to the fact that graphene very effectively screens charge impurities, which are the dominant source of disorder in topological insulators. Zhang \textit{et al} computed the spin-transfer torque in graphene-based TI heterostructures \cite{Zhang_SR2018}, induced by the helical spin-polarized current in the TI, which acts as a quantum spin Hall insulator. The torque was found to have a similar magnitude to ferromagnetic/normal/ferromagnetic graphene junctions, and to be immune to changes in geometry. A more recent work studied the spin proximity effect in graphene/TI heterostructures \cite{Song_NL2018}, predicting a sizable anisotropy in the spin lifetime in the graphene layer. Finally, when the spin-orbit coupling is sizable, the current-induced spin polarisation and spin-Hall effect drastically alter the non-local resistance of graphene \cite{Huang_NLR_PRL2017}, which can become negative and oscillate with distance, even in the absence of a magnetic field. Even though the results were derived for adatom-functionalised graphene, they are expected to apply generally to 2D systems exhibiting both current-induced spin polarisations and spin-Hall transport.

\subsubsection{Anomalous Hall effects}

The controversy surrounding the origins of the anomalous Hall effect go back nearly seventy years \cite{Nagaosa-AHE-2010}. Three main contributions have been identified, one of which is intrinsic and is now known to be related to the Berry curvature of Bloch electrons. It is associated with a deflection of particle trajectories under the action of the spin-orbit interaction in the band structure of the material, and was shown to be important in TI and TI thin films \cite{Culcer_TI_AHE_PRB11, Haizhou_PRL2013} as well as in longitudinal transport in the presence of a magnetic field \cite{Dai_PRL2017}. The other two are termed skew scattering and side jump, and are extrinsic, meaning they depend on the disorder configuration. Skew scattering refers to asymmetric scattering of up and down spins, while side jump represents a transverse shift in the wave-packet centre of mass in the course of scattering, also asymmetric between spin up and spin down. Skew scattering and side-jump were originally introduced for an electron with a scalar dispersion scattering off a spin-dependent potential, and were much later generalised to spin-dependent dispersions in the presence of scalar potentials. The complex interplay of spin-dependent dispersions and spin-dependent scattering potentials has been addressed in a small number of papers \cite{Tse_SDS_IESHE_PRB06, Xintao_PRB2013, Adroguer_WAL_2015, Weizhe_Materials_2017, Weizhe_Weyl_2017}, while the subtle debate surrounding the side-jump in particular is covered extensively in Ref.~\cite{Nagaosa-AHE-2010}. It should be noted that the intrinsic contribution also has a disorder correction, which in the Kubo formalism includes the ladder diagrams \cite{AHE_vertex_PRL_2006} and in the density matrix formalism involves an additional driving term \cite{Interband_Coherence_PRB2017}. This is simply a reflection of the fact that (i) the non-equilibrium correction to the density matrix is an expansion in powers of the disorder strength $n_i$ and (ii) the leading term in the expansion is $\propto n_i^{-1}$, since it is linear in the transport scattering time that is needed to keep the Fermi surface near equilibrium. The next-to-leading term is thus of order $n_i^{(0)}$.

Efforts to identify topological systems exhibiting a quantised anomalous Hall effect continue. Thin-film topological crystalline insulators with ferromagnetically ordered dopants can support quantum anomalous Hall phases with Chern numbers between -4 and 4 \cite{Fang_PRL2014}. The QAHE can be induced by an in-plane magnetisation in atomic crystal layers of group-V elements with a buckled honeycomb lattice according to \cite{Zhong_QAHE_PRB2017}. For weak and strong spin-orbit couplings, the systems harbor QAHEs with Chern numbers of C=$\pm 1$ and $\pm 2$, respectively, which could be observable at room temperature. Very recently, a new material entered the topological insulator stage, when a topological phase transition tuned by an electric field was demonstrated in ultrathin Na$_3$Bi \cite{Collins_Nature2018}, with an accompanying quantum spin-Hall effect. Very recently a QAHE at a relatively high temperature was reported in  a flat-band twisted bilayer graphene sample \cite{Serlin_MoireBLG_QAHE_2019} in which strong correlations result in the system to choosing a single valley and well-defined spin orientation.

The conductivity of TIs doped with magnetic impurities has been the subject of several papers. In TI the situation is somewhat different depending on whether the surface states are metallic or not, due to the presence of Berry curvature monopoles of opposite polarities at the top of the valence band and the bottom of the conduction band. When the chemical potential is in the gap one expects one TI surface to contribute $e^2/h$ to the anomalous Hall conductivity. As soon as the chemical potential passes the bottom of the conduction band this topological contribution is cancelled by the monopole in the conduction band. What remains is the Fermi surface contribution, which depends on the magnetisation and on the disorder profile. Ref.~\cite{Sabz_JPCM2015} showed that the surface conductivity of magnetic TIs is anisotropic, and strongly depends both on the direction of the spins of magnetic impurities and on the magnitude of the bulk magnetization. It reaches a minimum when the spin of surface impurities are aligned perpendicular to the surface of TI, because the backscattering probability is enhanced due to the magnetic torque exerted by impurities on the surface electrons. Moreover, Zarezad \textit{et al} demonstrated numerically that randomly distributed magnetic clusters with temperature-dependent mean sizes are liable to form on the surface of the TI \cite{Zarezad_PRB2018}. The anisotropic magnetoresistance depends strongly on the spin directions of the magnetic clusters, in a very different way from the case of non-interacting impurities.

\begin{figure}
	\begin{center}
		\includegraphics[width = \columnwidth]{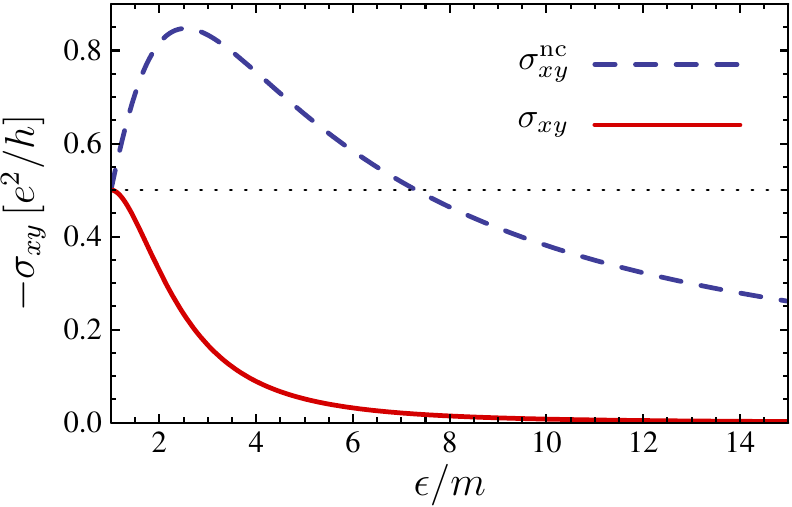}
		\caption{\label{Ado} Anomalous conductivity calculated in the standard non-crossing approximation ($\sigma_{xy}^{nc}$, dashed line), and including the contribution of the crossed diagrams ($\sigma_{xy}$, solid line), adapted from \cite{Ado_EPL2015}. 
}
	\end{center}
\end{figure}

Undoubtedly one of the biggest revelations in this field has been the fact that diagrams with two intersecting disorder lines, an inherent part of skew scattering on pairs of closely located defects, influences the anomalous Hall effect substantially and reduces the Fermi surface contribution at high densities \cite{Ado_EPL2015}, as shown in Fig.~\ref{Ado}. Going beyond the ladder approximation is therefore imperative, and subsequent works also showed that the disorder potential correlation length modifies the result \cite{Ado_PRB2017}, as well as spin-charge correlations in the disorder profile \cite{Aydin_2019}. 

Hall transport has also been investigated under optical driving fields. A dynamical Hall effect, for example, can be driven by a strong \textit{a.c.} electromagnetic field as seen in \cite{Tse_PRB2016} for light of subgap frequency near the absorption edge in a magnetically doped TI. Although the light is off-resonance, in the strong-field regime there is always a finite electron population in the conduction band due to nonlinear effects. A similar analysis was performed on the \textit{quantum} anomalous Hall effect in intense fields \cite{Lee_PRB2017}, where the Hall conductivity was shown to remain close to $e^2/(2h)$ at low fields and low frequencies. At strong fields, the half quantisation is destroyed and the dynamical Hall conductivity displays coherent oscillations as a function of field strength due to the formation of Floquet sub-bands and associated transitions between them.

A topic gaining currency at the interface between topological materials and magnetism is the interplay of skyrmions with TI surface physics. Skyrmions are topological magnetic excitations with particle-like properties, in which the spin at the core and the spin at the perimeter point in opposite directions. They result from the competition between the Dzyaloshinskii-Moriya interactions and exchange interactions, and give rise to a nonzero Berry curvature in real space. This is associated with an \textit{emergent} magnetic field, which deflects conduction electrons and causes a \textit{topological} Hall effect. Skyrmions on TI surfaces provide the opportunity to examine the interplay of real-space and momentum-space Berry curvatures. In general skyrmion effects cannot be treated perturbatively as topological features are missed in such approaches. Araki and Nomura combined a non-perturbative solution to the scattering of massless Dirac fermions with the Boltzmann equation, demonstrating analytically that skyrmions contribute to the anomalous Hall conductivity \cite{Araki_PRB2017} because at the skyrmion boundary Dirac electrons acquire a phase factor that is absent in the dynamics of Schrodinger electrons. The essential ingredient is the sign change in the out-of-plane magnetic texture between the centre of the skyrmion and the boundary, since in the skyrmion model considered the Berry curvature is zero. Scattering of massive Dirac fermions by a skyrmion in a TI/ferromagnet structure was the subject of a concomitant numerical investigation \cite{Andrikopoulos_SR2017}, which concentrated on its effect on the longitudinal conductance. Under certain circumstances the electrical signal due to the skyrmion may be distinguishable from the uniform ferromagnetic background.  

A concerted effort is being directed towards the understanding of magnonic effects in magnetic TI heterostructures. These include the spin-Seebeck effect, a voltage signal induced in a metallic system by thermally driven spin currents in adjacent magnetic systems, which was examined in the vicinity of a TI/ferromagnetic insulator interface \cite{Okuma_PRB2017}. In this system the spin-Seebeck effect is induced by surface electrons scattering off the nonequilibrium magnon population at the surface of the thermally driven ferromagnetic insulator. Similarly, \cite{Imai_JPSJ2018} identified magnon-drag thermoelectric effects stemming from the electromotive force induced by magnons and a thermoelectric analogue of the anisotropic magnetoresistance. Yasuda \textit{et al} report a large unidirectional magnetoresistance in TI heterostructures \cite{Yasuda_PRL2016}, which is attributed to asymmetric scattering of electrons by magnons. Its large magnitude is due to spin-momentum locking and a small Fermi wave number at the TI surface, and is expected to be maximized around the Dirac point.

\subsubsection{Topological protection}

Topological protection in TI has preoccupied researchers for over a decade. It is variously understood to mean: (i) that the topological surface states are protected against Anderson localisation due to the fact that backscattering is forbidden; (ii) that the topological surface states cannot be eliminated by any time-reversal preserving perturbation; (iii) that the edge states in TIs are topologically protected. Whereas (i) and (ii) are widely accepted we stress that (i) does not imply high mobilities can be achieved in TIs, as was believed for some time, since carriers can scatter through any angle other than $\pi$, and in fact TIs continue to have very poor mobilities. The controversy surrounding (iii) affects the QAHE and QSHE and is especially relevant in light of the experimental observation that quantised conductance occurs in very short channels, of a few hundred nm. 

\begin{figure}
	\begin{center}
		\includegraphics[width = \columnwidth]{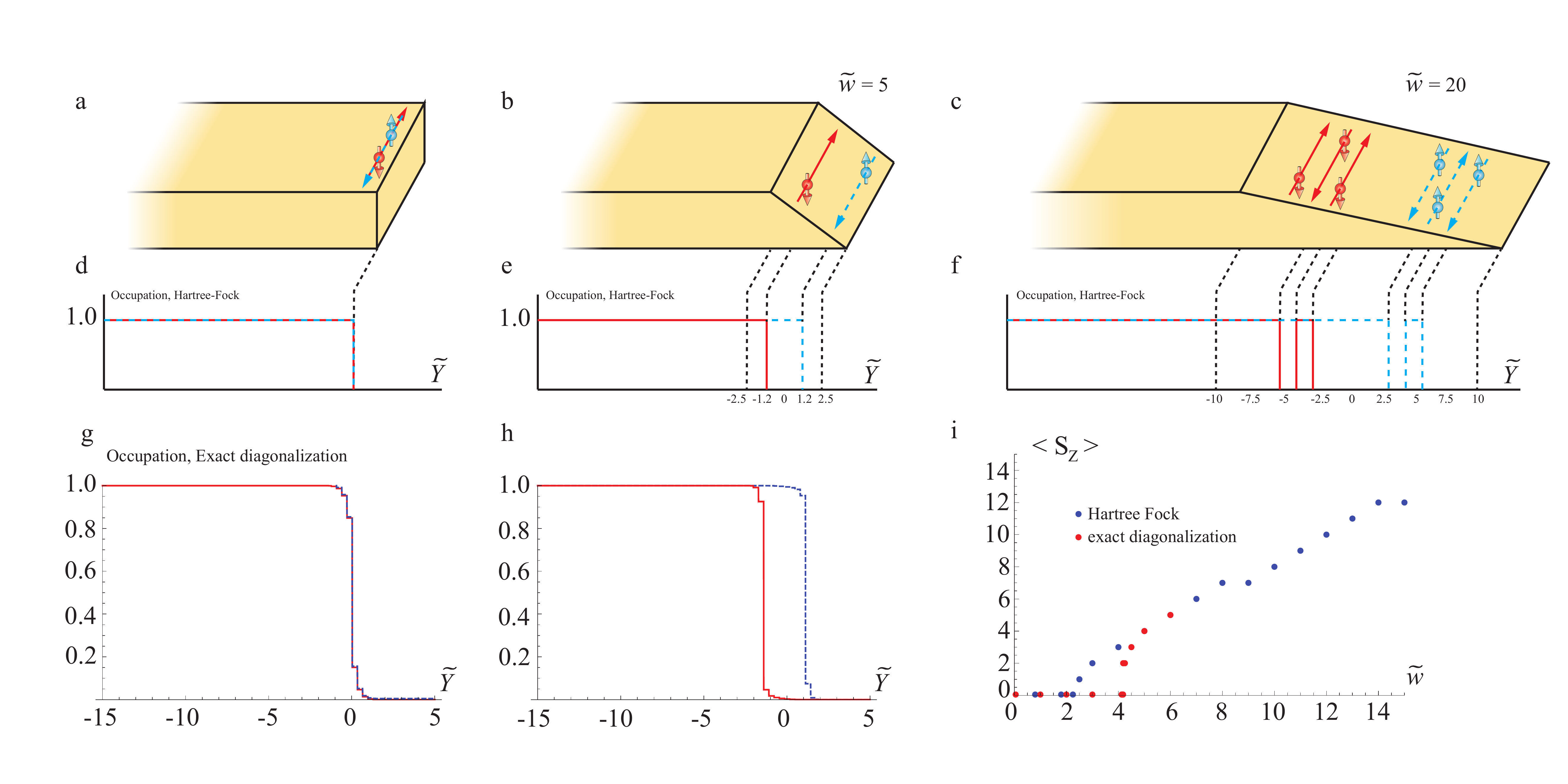}
		\caption{\label{Gefen}
Edge reconstruction and topological protection, adapted from from \cite{Wang_PRL2017}. Panels (a)--(c) describe the schematics of the results for three different distributions of the confining positive charge (light orange), characterised by $\tilde{w}$, the length scale over which it decays to zero. The edge modes are marked by broken blue (spin up) and solid red (spin down) lines. Panels (d)--(f) depict the occupations of the electronic states, using the Hartree-Fock approximation, demonstrating a single drop in density for a sharp edge ($\tilde{w}=0$) in (d), spin separation for smoother edge ($\tilde{w} = 5$) in (e), and an even smoother edge ($\tilde{w}=20$) in (f). $\tilde{Y}$ denotes the position of state, in units of the effective magnetic length; $\tilde{Y} = 0$ is the center of the density drop. Panels (g) and (h) depict the same distributions as in (d) and (f), respectively, using exact diagonalisation. Panel (i) depicts the edge spin magnetisation as a function of the slope of the positive-charge density, suggesting a continuous phase transition.}
	\end{center}
\end{figure}

A series of studies have challenged the universality of topological protection. To begin with, the topological surface states themselves are sensitive to the type of metallic contacts placed on the TI, as demonstrated by ab initio calculations of Bi$_2$Se$_3$ \cite{Spataru_PRB2014}. These reveal that Au and graphene leave the spin-momentum locking mostly unaltered, while Ni, Pd, and Pt strongly hybridise with the TI and relax spin-momentum locking. Size quantisation effects also influence the topological properties \cite{Kotulla_NJP2017}. Interestingly, the edge itself may experience spontaneous time-reversal symmetry breaking due to edge reconstruction when a smooth potential is considered, rather than the infinitely sharp theoretical approximation \cite{Wang_PRL2017}, as in Fig.~\ref{Gefen}. The electron density seeks to mimic the positive-charge distribution on the gate. If this falls smoothly to zero near the edge, the electron density will mimic this by separating the edge modes, each giving rise to a decrease in density. Since the edge modes have opposite chiralities, time reversal symmetry is spontaneously broken. In this case backscattering is enabled and the conductance quantisation of the quantum spin-Hall effect is consequently destroyed, while a spontaneous anomalous Hall effect appears at zero magnetic field. 

Much of the discussion centres on the effect of impurities on edge state transport. Tanaka \textit{et al} examined the effect on the conductance of a quantum spin-Hall device of a magnetic impurity, which can backscatter an electron from one edge state to the other \cite{Tanaka_PRL2011}. If the Kondo exchange is taken to be isotropic, so that the total spin of the electrons and impurities is conserved, and all electrons moving in the same direction are taken to have the same spin, the correction to the conductance due to this impurity vanishes in the dc limit, in contradiction to an earlier paper \cite{Maciejko_2DTI_Kondo_PRL09}. Anisotropic exchange introduces corrections that can be sizable above the Kondo temperature, but are suppressed as $T \rightarrow 0$. The treatment was generalised by Altshuler \textit{et al}, who considered scattering by a disordered chain of Kondo impurities \cite{Altshuler_PRL2013}. As the authors point out, in disordered systems with strong spin-orbit interactions it is unlikely that any component of the total spin is conserved. When this is taken into account, backscattering processes emerge that persist down to absolute zero. The edge electrons experience Anderson localisation for an arbitrarily weak anisotropy in the coupling to the spin impurities provided the sample is long enough. Interactions between electrons cannot destroy the localisation, unless they happen to be strongly attractive.

Black-Schaffer and Nazarov focused on the coupling between surface and bulk states in the presence of strong non-magnetic potential impurities, which create localised resonances appearing at ever lower energies as the impurity strength is increased \cite{Schaffer_PRB2012}. At large strengths the resonance goes through the Dirac point, causing two Dirac points to emerge on both sides of the resonance. Both the surface states and the resonances penetrate approximately 10 layers into the sample, enabling second-order bulk-assisted scattering processes, which act to destroy the topological protection. Coupling between the edges and the bulk was investigated from a different angle in a subsequent paper \cite{Zhang_Coexist_PRB2014}, where edge and bulk states were considered to be at the same Fermi energy. In this case backscattering between the two leads to Anderson localisation of both edge and bulk states. Finally, in the presence of even weak electron-electron interactions, short-range nonmagnetic impurities act as noncollinear magnetic scatterers, which enable strong backscattering and cause deviations from quantisation even at zero temperature \cite{Novelli_PRL2019}. 

Due to fluctuations in the donor density, doping TIs tends to create a non-uniform potential landscape consisting of electron and hole puddles, especially dangerous at small energies. It is natural to ask what effect these have on edge transport. In \cite{Vayrynen_PRL2013} a puddle is modeled by a quantum dot, coupled to the helical edge states by tunnelling, and quantum dots with even numbers of electrons are considered. Elastic processes involving electron dwelling in the dot do not lead to any backscattering. Yet inelastic backscattering processes are enhanced by electron dwelling in the dot by increasing the time electrons interact with each other. At temperatures lower than the quantum dot energy spacings, the conductance correction depends strongly on the position of the Fermi level with respect to the dot energy levels. The enhancement of the resistance is much stronger for dots with an odd number of electrons, due to the Kondo effect \cite{Vayrynen_PRB2014}, but the corresponding temperature dependence is relatively weak. Indeed, current experiments see no signature of Kondo transport \cite{Fuhrer_Na3Bi_Kondo_2019}.

\section{Valley-dependent phenomena}

Whereas TMDs have a massive Dirac spectrum, unlike TIs, this describes a lattice pseudospin degree of freedom rather than an angular momentum stemming from the real spin. Nevertheless, TMDs have two degenerate valleys, which are related by time reversal, so that the masses in the two valleys have opposite signs. Unlike graphene they typically have sizable spin-orbit interactions. Large spin splittings were identified in the conduction band of TMD monolayers in \cite{Kosmider_PRB2013}, while the MoS$_2$-WS$_2$ heterojunction was shown to have an optically active band gap \cite{Kosmider_Hetero_PRB2013} with the lowest energy electron-hole pairs spatially separated and living in different layers. The spin-orbit interaction has a noticeable effect on the optical conductivity \cite{Gilbertini_PRB2014, Xiao_Rashba_PRB2016}.

Given that the spin is essentially locked to the valleys it is not clear that the two can be manipulated \textit{independently}, yet accessing the elusive valley degree of freedom is fascinating from a fundamental standpoint. The massive Dirac spectrum gives rise to the possibility of manipulating the valley degree of freedom. For example, the analogue of the anomalous Hall effect in doped TMD monolayers is the \textit{valley} Hall effect, which has been detected in MoS$_2$ \cite{Mak_Science2014} and has generated considerable excitement. Physically it corresponds to an anomalous Hall effect with different signs for different valleys. There is no net charge current, because the anomalous Hall currents from the two valleys cancel each other out, but electrons from different valleys flow to different sides of the sample generating a valley polarisation, which can be detected using circularly polarized light. In bilayer MoS$_2$ this was shown to be controllable electrically using a top gate voltage, which breaks the inversion symmetry of the bilayer \cite{Lee_NN2016}. The valley Hall effect generates a non-local resistance, which has a non-trivial dependence on the longitudinal resistivity, weakening at large valley Hall angles \cite{Beconcini_PRB2016}, where the valley Hall angle is determined by the ratio of the valley Hall conductivity and the longitudinal charge conductivity.

Because the valleys are spin polarised the valley Hall effect is typically accompanied by a spin-Hall effect \cite{Feng_PRB2012, Wenyu_PRB2013}, valley currents tend to be spin-polarised \cite{Zhang_TMD_PRB2014}, and spin and valley polarisations are generated concomitantly, although they can be distinguished at high magnetic fields \cite{Tahir_PRB2016}. Likewise, spin and valley noise are coupled \cite{Tse_PRL2014}, with fluctuations in the Faraday rotation signal being connected to the valley degree of freedom as well as to the spin, a fact that is ascribed to intervalley scattering processes. Again, given that spin noise is sensitive to an applied magnetic field, spin and valley dynamics may be distinguished in certain regimes. On a similar note, TMDs also exhibit strong hyperfine interactions, which can create a feedback mechanism in which spin-valley currents generate significant dynamical nuclear polarization which in turn Zeeman shifts excitonic transitions out of resonance with an optical driving field, saturating the production of spin-valley polarisation \cite{Sharma_PRB2017}. 

Similar findings have been reported theoretically in Bi monolayers Bi$_2$XY, where X,Y $\in \{$H, F, Cl, Br, or I$\}$ \cite{Zhou_NPJQM2018}, where a staggered exchange field is introduced by doping with transition-metal atoms or by magnetic substrates. Unsurprisingly these findings have stimulated a lot of activity in optics \cite{Mak_NPhot2016, Mak_NPhot2018}, and various schemes have also been proposed for controlling and enhancing the spin and valley polarisations and currents by employing electric and magnetic fields \cite{Tahir_PRB2016, Tahir_EPL2017} and optical techniques \cite{Kovalev_NJP2018}.

A half-quantised valley Hall effect has been predicted when the chemical potential lies in the mass gap, driven by the same Berry curvature mechanism as the quantised anomalous Hall effect in TIs \cite{Lensky_PRL2015}. By mapping the system onto the Landau-Zener problem, the authors argue this response is dominated by bulk currents arising from states just beneath the gap rather than by edge modes, as the latter are not topologically protected, as in TI, and may be absent. The potential gradient due to the external electric field divides the system into three regions, of which the middle region is gapped and the surrounding regions have carriers in the conduction and valence bands respectively. Upon reflection from the central gapped region carriers are believed to experience side jumps leading to a Hall effect. A dissipationless response is found even when topologically protected edge modes are absent, and is independent of the gap size. To date this claim remains unverified experimentally. In TMD bilayers the situation is somewhat different \cite{Kormanyos_PRB2018}. The Berry curvature has sizable contributions from both the intralayer and the interlayer couplings, the latter leading to a dependence on the stacking configuration and enabling tunability in double gated devices. The valley and spin Hall conductivities are not quantised, but can change sign as a function of the gate electric field. Structures that may host topologically protected states supporting persistent spin or valley currents include interfaces where the Dirac mass changes sign in 2D Dirac materials with spin-orbit coupling \cite{Abergel_NJP2014}. In this case the topologically protected states are so-called Jackiw-Rebbi modes with a linear dispersion, supporting spin and valley currents parallel to the interface.

Whereas the valley Hall effect has been responsible for the bulk of transport research on TMDs, magneto-transport has begun to attract attention. Seeing as the valley pseudospin and the magnetic field are both odd under time reversal one may expect magneto-transport phenomena without a counterpart in single-valley systems. Sekine and MacDonald demonstrated that the interplay between the Berry curvature, a perpendicular magnetic field, and disorder scattering in TMD monolayers gives rise to a longitudinal magnetoconductivity contribution that is odd in the valley pseudospin and odd in the magnetic field \cite{Akihiko_PRB2018}. For this contribution to be visible a valley polarisation must exist in the system. The dependence of the magnetoresistance on the magnetic field changes from quadratic to linear when a finite valley polarisation is induced by optical pumping. 

\section{Probing Fermi arcs in Weyl semimetals}

\begin{figure}
	\begin{center}
		\includegraphics[width = \columnwidth]{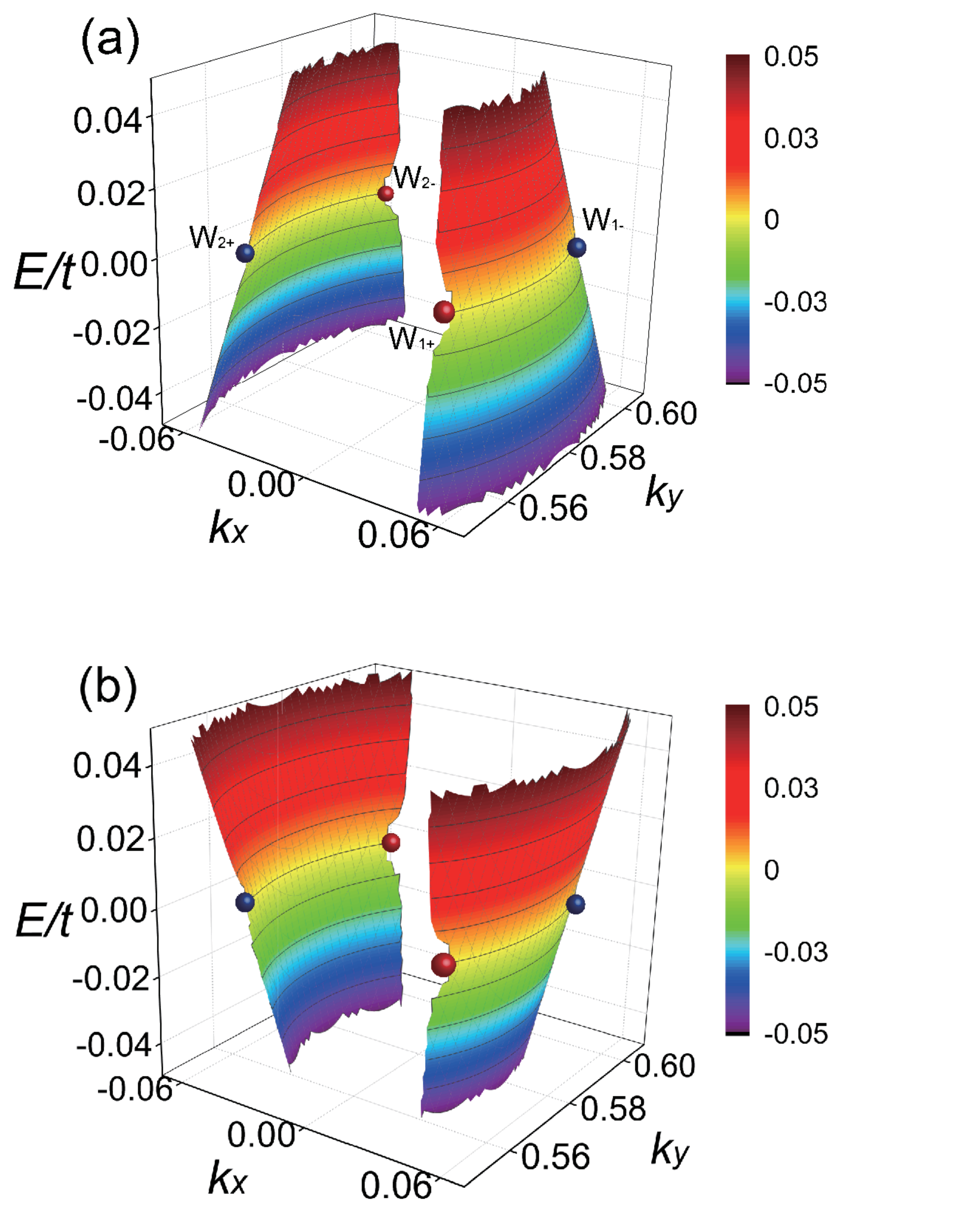}
		\caption{\label{Okugawa} Schematic of the Fermi arcs on opposite WSM surfaces, adapted from \cite{Okugawa_PRB2014}. (a) Top surface (b) Bottom surface. The red (blue) points are the gapless points, which have positive (negative) monopole charges for the Berry curvature.}
	\end{center}
\end{figure}

Weyl semimetals harbor unusual surface states known as Fermi arcs, with each arc connecting the surface projections of two Weyl nodes of opposite chirality, as shown in Fig~\ref{Okugawa}. For a basic understanding their dispersion can be approximated as linear and, as shown in \cite{Okugawa_PRB2014}, for a surface in the $xy$-plane may be written as $\varepsilon_{\bm k} = \pm \hbar v k_y$, where $v$ is the magnitude of the velocity, and the line connecting the two Weyl nodes is $\parallel k_x$. The Fermi arc states exist for $-k_0 \le k_x \le k_0$, where $\pm k_0$ represent the locations of the Weyl nodes. Note that: (i) the two Fermi arcs are only degenerate at $\varepsilon_{\bm k} = 0$; (ii) Fermi arcs on opposite surfaces have opposite velocities; and (iii) in contrast to TI surface states, Fermi arc states are open, effectively disjoint segments of a 2D Fermi surface, connecting the Fermi surfaces for carriers with opposite chiralities that otherwise appear to be disconnected. As the system evolves from a Weyl semimetal to a TI with decreasing thickness, the arcs on opposite surfaces merge into a surface Dirac cone \cite{Okugawa_PRB2014}. In a similar manner to TI surface states, Fermi arc wave functions in real space extend into the material. If the sample is very thin, in analogy with a TI thin film, the wave functions corresponding to the Fermi arcs on the top and bottom surfaces overlap significantly, which enables back scattering between certain pockets of the Brillouin zone. This is expected to enhances Friedel oscillations, as seen in the local density of states \cite{Hosur_PRB2012}. 

Since Fermi arcs couple points in the Brillouin zone with opposite Berry curvatures, it has been believed that their existence is topologically protected. Recent studies however cast doubt on this belief, and to date it remains an open question. For example, Wilson \textit{et al} \cite{Wilson_PRB2018} argue that Fermi arc states in WSMs hybridise with Lifshitz rare states in the bulk to a much greater degree than TI surface states, since the latter exist in the bulk gap and are localised whereas WSMs are gapless and the rare states are only quasilocalised. This coupling is non-perturbative and gives the arcs spectral weight in the bulk so that they are no longer bound to the surface and may no longer be said to be topologically protected even for weak disorder. Interestingly, the surface chiral velocity persists even for strong disorder, making the Fermi arc visible in spectroscopic measurements. Experimentally, quasiparticle scattering and interference was imaged on the surface of the WSM TaAs \cite{Inoue_Sci2016}, providing spectroscopic evidence of Fermi arc states. Nevertheless, the scattering wave vectors observed experimentally are consistent with theoretical predictions that assume particle propagation through the bulk of the sample in addition to propagation on the surfaces. Indeed, aside from hybridisation with rare states, the connection between surface and bulk provided by the Fermi arcs has highly non-trivial consequences not only for scattering and interference, but also for transport. 

To begin with, scattering of electrons between the surface and bulk states caused by inhomogeneities introduces dissipation in Fermi arc transport \cite{Gorbar_PRB2016}. In a 1D description that neglects surface-bulk coupling, quenched disorder effects result in a single phase factor that is odd under exchange of spatial variables. Consequently, disorder effects disappear from the current-current correlation function, which would imply a dissipationless longitudinal conductivity. Disorder averaging, however, is still responsible for the finite width of the surface states, without which the longitudinal conductivity would diverge. 
The authors of \cite{Gorbar_PRB2016} show rigorously, however, that an impurity scatters surface waves into the bulk, resulting in dephasing of the Fermi arc states and dissipation. The deeper conclusion is that generically an effective theory of surface states in the presence of disorder is not well-defined for gapless systems. 

The geometry of the Fermi arc dispersion is also important \cite{Resta_PRB2018}. A straight arc eliminates the impact of electron-phonon scattering on surface transport because scattering only occurs between states with the same velocity along the direction of the current. Scattering off surface disorder is also suppressed in this geometry, such that for strong disorder a straight arc yields a surface conductivity 1-2 orders of magnitude larger than a TI. This is traced to the different hybridisation strengths between surface and bulk in WSMs and TIs. 

In the absence of a magnetic field Fermi arcs may alter the anomalous Hall response of WSMs by introducing a \textit{residual} contribution, with the result depending on the degree of tilting of the arcs on opposite surfaces \cite{Zhao_EPL2018}. In an external magnetic field it emerges \cite{Potter_NC2014} that closed orbits can be formed in which Fermi arc states on opposite surfaces are connected by trajectories that traverse the bulk of the sample. As a result quantum oscillations in e.g. the density of states can be observed in magnetic fields up to a critical value, given by the thickness of the sample. Following the prediction, Shubnikov-de Haas oscillations involving Fermi arc states were indeed detected in Cd$_3$As$_2$ \cite{Moll_Nat2016}. In strong magnetic fields, Fermi arcs can also give rise to a 3D quantum Hall effect, with complete loops formed via \textit{wormhole} tunneling assisted by the Weyl nodes \cite{Wang_3DQHE_PRL2017}. A peculiar phenomenon present in WSMs in magnetic fields is the chiral magnetic effect, the generation of an electrical current parallel to a magnetic field when the Weyl nodes are offset in energy by a finite amount. Fermi arcs are expected to make a finite contribution to the chiral magnetic effect opposite in sign to the bulk contribution, which change the sign of the overall result \cite{Baireuther_NJP2016}. This holds even in the infinite-system limit because, even though the number of surface modes decreases, the remaining modes are more sensitive to magnetic flux.

Research is burgeoning on the unusual optical properties of Fermi arc states. They interact strongly with light and have a large optical conductivity for light polarised transversely to the arc \cite{Shi_PRB2017}. They affect the plasmon dispersion, which is already unconventional in bulk WSMs \cite{Pellegrino_PRB2015, Hofmann_PRB2016}. There the axion term lifts the degeneracy of the three gapped plasmon modes at ${\bm q} = 0$ \cite{Pellegrino_PRB2015}. This because WSMs are gyrotropic, meaning their dielectric tensor is asymmetric, and the degree of asymmetry is proportional to the separation between the nodes, which acts as an effective applied magnetic field in momentum space. In magnetic systems with broken time reversal symmetry Fermi arc plasmons are chiral, with constant frequency contours that are open and hyperbolic \cite{Song_PRB2017}. Their dynamics is also strongly linked to quantum non-local effects, for example they decay by emitting electron-hole pairs in the bulk \cite{Andolina_PRB2018}.

\section{Non-linear electrical response}

Undoubtedly the most exciting novel development in topological materials transport has been the take-off of non-linear electrical effects, which can be enabled by the lack of inversion symmetry, the application of a magnetic field, or the presence of a valley polarisation. An early work illustrated the role of mirror planes in determining the charge and spin response \cite{Misawa_2011}, yet the community is still only beginning to appreciate the richness and variety of physical phenomena that emerge when higher-order responses in the applied fields are allowed. Geometrical phase effects in first order response, as well as their interplay with disorder, are relatively well understood. However in the second order response most of the ground work remains to be done. The quantity of interest is the next term in perturbation theory beyond linear response to an electric field, which can be formulated diagrammatically, semiclassically, or in terms of the density matrix. The non-linear optical response, particularly strong in TMDs, encompasses conceptually new phenomena including second harmonic generation and nonreciprocal, rectification and shift currents. Non-reciprocal refers to phenomena that have a built-in bias direction, such current flow in a $p-n$ junction. These relatives of the  photogalvanic and photovoltaic effects are frequently encoded by the Berry phase, the toroidal moment, and the magnetoelectric monopole, or may have extrinsic origins such as magnon scattering. The relevant concepts are beautifully summarised in Ref.~\cite{Nagaosa_NRR_NC2018}. Growth in this field has been spurred by spectacular experimental developments, motivated by the possibility of detecting the relevant response by scanning higher harmonics of the applied frequency. One of the grand aims is to find a Hall effect in time-reversal symmetric systems: two papers have reported a nonlinear Hall effect in bilayer/few-layer WTe$_2$ \cite{NLH_2019, Kang_NM2019}. In \cite{NLH_2019} the nonlinear Hall effect results in a much larger transverse than longitudinal voltage, with a nonlinear Hall angle of nearly $\pi/2$, which may have topological origins \cite{BCD_2018}. The non-linear Hall effect is generally extrinsic in time-reversal invariant systems, but can be intrinsic if time-reversal symmetry is broken \cite{Gao_Shift_PRL2014, Gaoyang_PRB2018}.

The static non-linear Hall conductivity contains an intrinsic contribution proportional to the Berry curvature dipole in reciprocal space, that is, the term $k{\bm \Omega}_{\bm k}$, where ${\bm \Omega}_{\bm k}$ is the Berry curvature \cite{Inti_PRL2015, Facio_PRL2018, Konig_PRB2019}. The nonlinear Hall effect arising from the Berry curvature dipole in TMDs with time-reversal symmetry was examined in Ref.~\cite{You_BCD_TMD_PRB2018}, which showed that such a current is present when only one mirror line exists, while in certain TMD phases a finite Berry curvature dipole emerges when strain or electrical displacement fields are applied. WSMs are also expected to be excellent candidates for nonlinear effects because of their large Berry curvature concentrated near the Weyl points, and in a related study, the Berry curvature dipole was investigated in WSMs \cite{Binghai_BCD_PRB18}, concluding that type-II Weyl points, having a strong tilt, were preferable to type-I. In this vein, Nandy and Sodemann \cite{Nandy_Tilt2019} calculated the non-linear Hall conductivity of two-dimensional tilted Dirac fermions using a multi-band quantum Boltzmann equation, identifying disorder contributions in addition to the intrinsic Berry curvature dipole. Recent work by the same group \cite{Inti_Acceleration_2019} reveals that the rectification current obeys a sum rule controlled by the Berry connection. Rectification at relatively high frequencies has also been studied in Ref.~\cite{Isobe_2018}. Non-linear Hall effects can also be induced by disorder \cite{Du_2018}. In the context of tilted Dirac cones we note the revealing renormalisation group work of Ref.~\cite{Yang_Tilt_PRB2018} on the Coulomb interaction and quenched disorder in tilted TIs. Along the tilting direction a random scalar or vector potential dynamically generates a new type of disorder, dominant at low energies, which turns the system into a compressible diffusive metal, with the fermions acquiring a finite scattering rate. The band-touching point is replaced by a bulk Fermi arc in the Brillouin zone. The consequences of these features for transport remain to be determined.

\begin{figure}
	\begin{center}
		\includegraphics[width = \columnwidth]{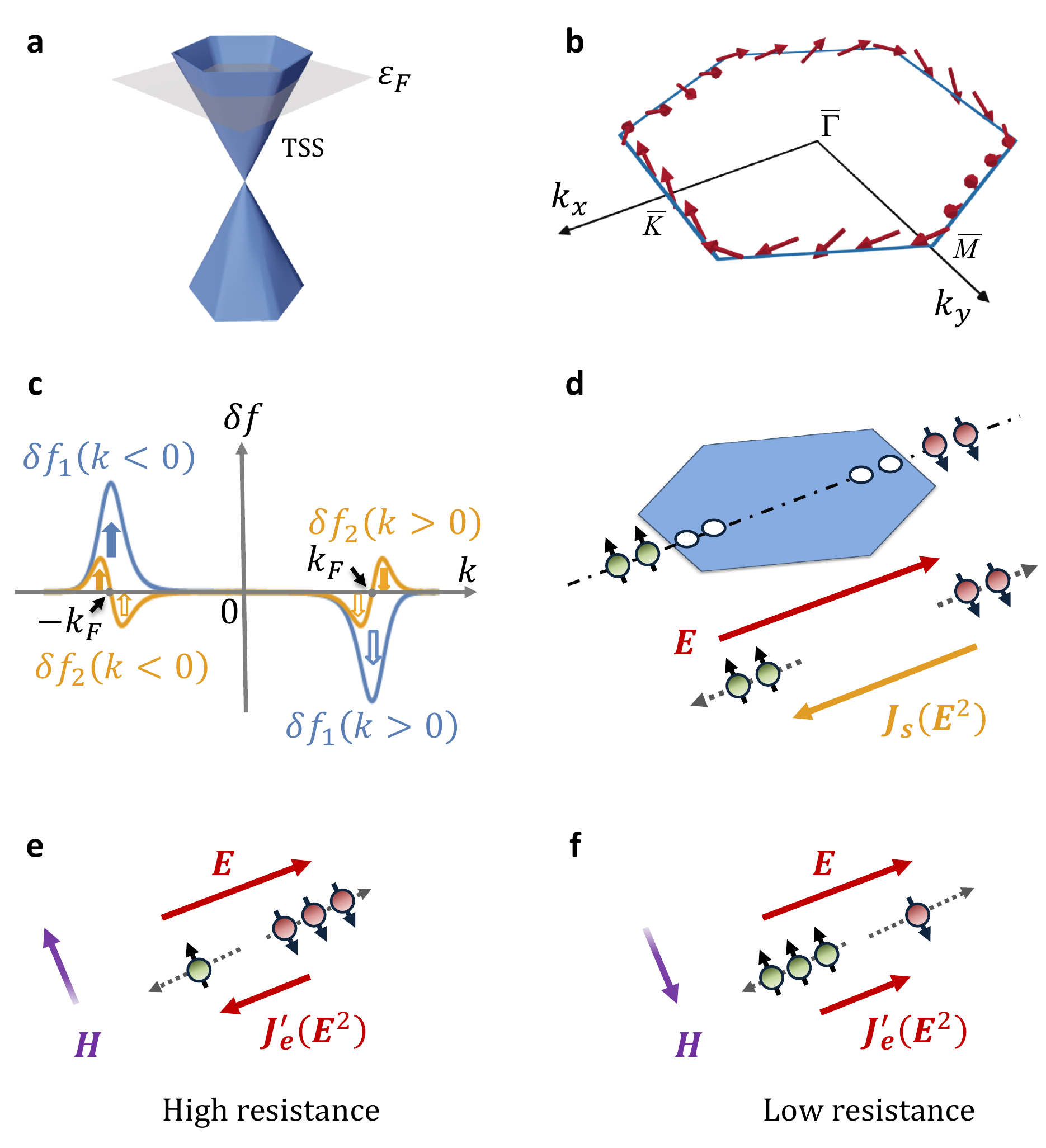}
		\caption{\label{Pan} Bilinear magnetoelectric resistance, adapted from \cite{Hyunsoo_NP2018}. (a) Hexagonally warped energy dispersion for the surface states with Fermi surface lying in the conduction band. (b) Hexagonally warped spin texture at the Fermi contour of the surface states. (c) Variation of the electron distribution along the ${\bm k}$-axis parallel to the applied electric field ${\bm E}$: $\delta f_1$ (blue curve) and $\delta f_2$ (yellow curve) are the corrections to the equilibrium distribution of first and second order in the electric field, respectively. Solid arrows represent the excess of electrons with spins along the arrow direction, and hollow arrows represent depletion of the same. (d) When an electric field ${\bm E}$ is applied along a certain direction in ${\bm k}$-space (dash-dotted line), a non-equilibrium spin current ${\bm J}_s(E^2)$ is generated at the second order of the electric field, due to spin-momentum locking. (e) and (f) When an external magnetic field is applied, the nonlinear spin current is partially converted into a charge current ${\bm J}_e'(E^2)$: a high-resistance state is reached (e) when the magnetic field is antiparallel to the spin direction of the electronic states with ${\bm k} \parallel {\bm E}$, while a low-resistance state is reached (f) when the magnetic field is parallel to that spin direction.}
	\end{center}
\end{figure}

An in-plane magnetic field has a subtle effect on the non-linear electrical response of a hexagonally warped TI at frequencies very close to the DC limit \cite{BMER_2019}. Whereas an in-plane magnetic field merely shifts the origin of a Dirac cone with no physically measurable effect, when the Hamiltonian contains a warping term the effect on the spectrum is non-trivial, and a strong non-linear response results, termed bilinear electromagnetic response. Its sign and the magnitude depends sensitively on the orientation of the current with respect to the magnetic field as well as the crystallographic axes, so that the spin texture of the topological surface states could be mapped via a transport measurement. The bilinear magnetoelectric resistance was measured in hexagonally warped TIs in \cite{Hyunsoo_NP2018}, cf. Fig.~\ref{Pan}. On the other hand, in Ref.~\cite{Yasuda_PRL2017} a second harmonic Hall voltage was detected in the presence of in-plane magnetic field and magnetisation in TI heterostructures, believed to be due to asymmetric magnon scattering. In a similar manner to a magnetic field, a time-reversal breaking valley polarisation allows second harmonic generation even in centrosymmetric crystals, and this in turn can provide a direct measure of the valley polarisation \cite{Hipolito_2DM2017}.

Noncentrosymmetric crystals are anticipated to exhibit a dc photocurrent in the nonlinear optical response \cite{Sipe_PRB2000}. This so-called \textit{shift current} has attracted intensive attention as part of the bulk photovoltaic effect in ferroelectric materials, in the quest for efficient solar cell paradigms. Its fundamentally origin is the fact that, as an electron is excited by light from the valence band into the conduction band, its centre of mass changes due to the difference in the value of the Berry connection in the two bands. The final expression can be formulated in terms of the Berry curvature dipole in momentum space discussed above and is gauge invariant. Since the centre of mass is shifted the effect has been termed a shift current.  It is being pursued in topological materials as well. Kim \textit{et al} \cite{Kim_Shift_PRB2017} identified a sizable shift current generated in hexagonally warped TIs by linearly polarised light. A related nonlinear \textit{spin} current also exists in TIs, which can be excited by THz light \cite{Hamamoto_NLSC_PRB2017}. When inversion symmetry is broken but time-reversal symmetry is preserved a nonlinear anomalous Hall effect emerges in certain TMDs \cite{NLH_Tune_2DM2018}, while a dissipationless nonlinear anomalous node conductivity is also expected in WSMs \cite{Rostami_NLAH_PRB18}.


\section{Chiral superconductivity, Majorana edge modes and related phenomena}
\label{CS}

In the preceding sections, we have dealt with normal topological materials in which no pairing interaction among electrons takes place.
However, pairing interactions bring about new topological phenomena. 
Some earlier theoretical examples pertain to superfluid helium 3 ($^3$He) \cite{Salomaa87,Salomaa88}, anyon superfluids \cite{Wen91}, 
$d$-wave superconductors \cite{Tanaka95,Tanaka96,Kashiwaya00}, 
and the fractional quantum Hall effect \cite{Read00}. 
The theory \cite{Read00} has illustrated the concept of a topological superconductor as a system with a bulk pairing gap and a gapless  
Majorana mode at the boundary or on a topological defect. 
Such emergent Majorana modes obey the quantum statistics of non-Abelian anyons \cite{Ivanov01, Kitaev01,Kitaev03}, 
offering a route to braiding-based topological quantum computation \cite{Nayak08}.
Research into topological superconductivity has intensified with the discovery of topological insulators and related topological materials, 
with a flurry of theoretical as well as experimental activity that followed. 
Distinct classes of topological superconductors have been identified (see, e.g., Refs. \cite{Schnyder08,Morimoto15,Chiu16}).
Their properties have been a subject of several review articles 
\cite{Tanaka12_Rev,Alicea12_Rev,Beenakker13_Rev,Elliot15,DasSarma15,Sato17,Aguado17,Haim18}. 

Still, the field of topological superconductivity is growing fast, having in some areas advanced well beyond the existing review literature. This pertains, in particular, to the latest developments in chiral superconductivity, Majorana edge modes, the fractional Josephson effect as well as unconventional Cooper pairing in topological materials. These topics define the primary scope of our review article. We have endeavoured to mention key ideas and developments in the overlapping areas such as specific models and realisations of Majorana zero modes, their tunneling spectroscopy and related transport phenomena. These topics have been covered in several review articles. Further details of the underlying physics and a survey of the results can be found easily in the cited references.

\subsection{Chiral superconductivity. A primer}

Superconductivity originates from attractive pairwise interactions between electrons in a metal.
As is common in many-body physics, the pairing interaction can be treated in the mean-field approximation, 
allowing the description of a superconducting state by the Bogoliubov- de Gennes (BdG) Hamiltonian
\begin{equation}
\mathcal{H}_{\bm k} =
\Biggl[
\begin{array}{cc}
H_{\bm k} &  \Delta_{\bm k} \\
\Delta^{\dagger}_{\bm k} & - H^*_{-{\bm k}}
\end{array}
\Biggr], 
\qquad 
\Psi_{\bm k} = 
\Biggl[
\begin{array}{c}
u_{\bm k} \\
v_{\bm k}
\end{array}
\Biggr].
\label{BdG}
\end{equation}
It is a $2 \times 2$ matrix in which $H_{\bm k}$ is a single-particle Hamiltonian of the normal system, while the off-diagonal entry 
$\Delta_{\bm k}$ and its hermitian conjugate $\Delta^{\dagger}_{\bm k}$ account for the pairing interaction.  
The state $\Psi_{\bm k}$  has two (Nambu) components, $u_{\bm k}$ and $v_{\bm k}$,
being a particle- and a hole-like wave functions of the normal system, respectively.  

\begin{figure}[t]
	\begin{center}
		\includegraphics[width=70mm]{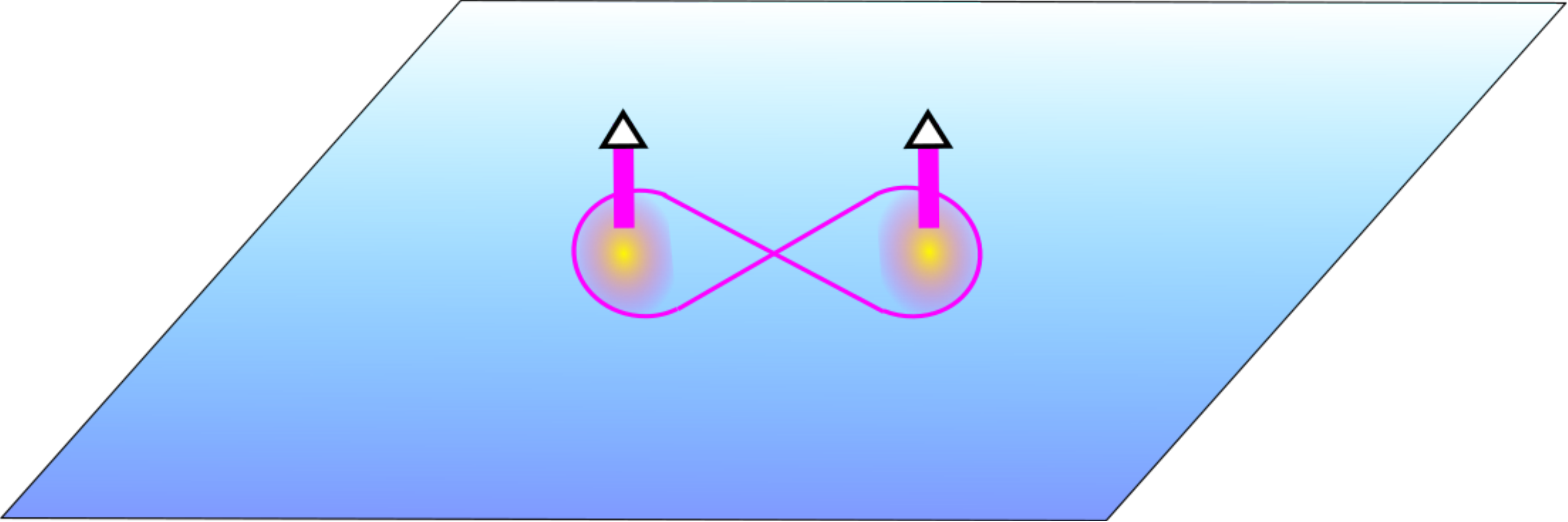}
	\end{center}
	\caption{
		Schematic of an equal-spin $p$ - wave electron pairing assumed in Eqs. (\ref{Delta_ESP}) and (\ref{BdG_ESP}).
	}
	\label{ESP_fig}
\end{figure}

\begin{figure}[t]
	\begin{center}
		\includegraphics[width=85mm]{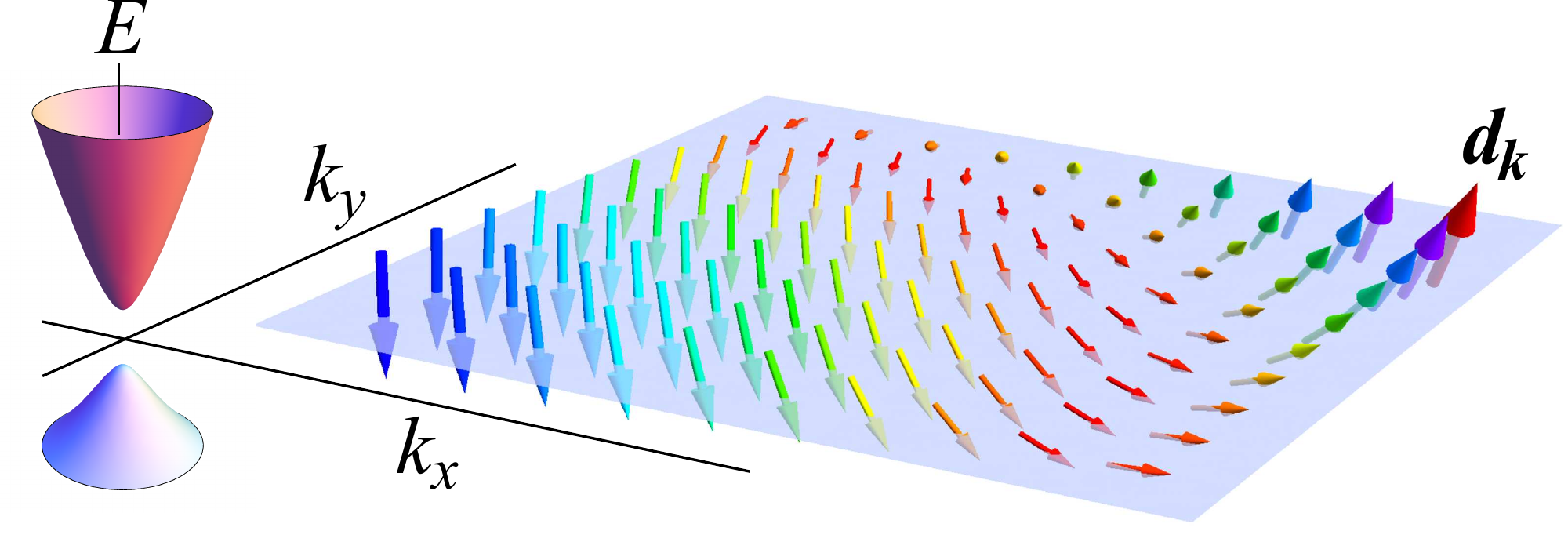}
	\end{center}
	\caption{
		Schematic of the Nambu pseudospin texture in a topological superconducting state for $\mu m> 0$ in Eq. (\ref{BdG_d}).
		The texture has the skyrmion topology with a nontrivial Chern number $C= \pm 1$.
	}
	\label{Top_fig}
\end{figure}

\begin{figure}[t]
	\begin{center}
		\includegraphics[width=85mm]{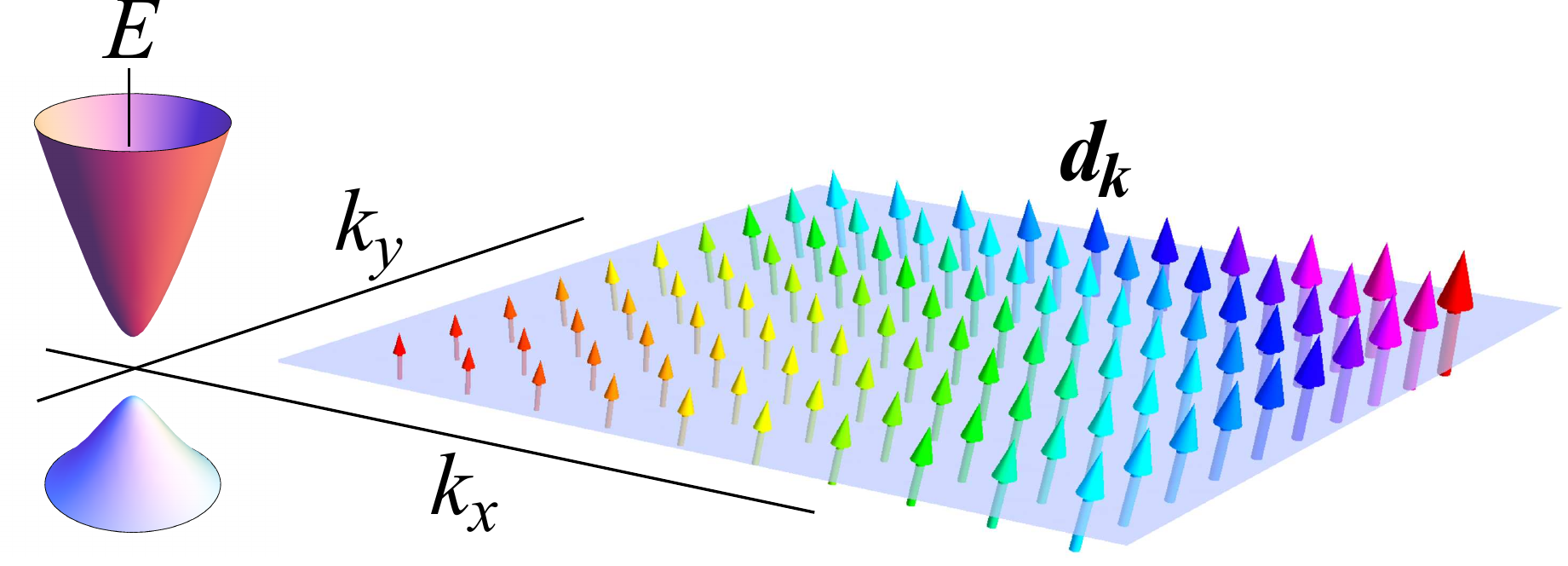}
	\end{center}
	\caption{
		Schematic of the Nambu pseudospin texture in an ordinary superconducting state for $\mu m < 0$ in Eq. (\ref{BdG_d}).
		In this case, the Chern number is trivial, $C=0$.
	}
	\label{Ord_fig}
\end{figure}

Coming from the pairing interaction, the matrix structure of the BdG Hamiltonian is crucial for emergent topological superconductivity.
A paradigmatic example is a 2D superconductor with an odd ${\bm k}$ - parity gap function \cite{Read00,Ivanov01}:

\begin{equation}
\Delta_{\bm k} = \Delta^\prime (k_x + ik_y),
\label{Delta_ESP}
\end{equation}
where the constant $\Delta^\prime$ is taken real.
Equation (\ref{Delta_ESP}) describes equal-spin pairs in the orbital $p$ - wave state with the quantum number $m_\ell =1$ (see also Fig. \ref{ESP_fig}). 
It has a partner with the quantum number $m_\ell =-1$. 
These are the 2D analogues of the $A$-phase of superfluid $^3$He \cite{Leggett75,Vollhardt90}. 
We shall see later that the $k_x + ik_y$ - pairing (\ref{Delta_ESP}) is also an effective model for the hybrid structures of quantum anomalous Hall insulators 
and conventional superconductors. As for the normal-state Hamiltonian, here we choose the simplest, parabolic band conductor 
with $H_{\bm k} = \frac{{\bm k}^2}{2m} - \mu$, where $\mu$ is the chemical potential. Then, the BdG Hamiltonian reads

\begin{equation}
\mathcal{H}_{\bm k} =
\Biggl[
\begin{array}{cc}
\frac{{\bm k}^2}{2m} - \mu  &  \Delta^\prime (k_x + ik_y) \\
\Delta^\prime (k_x - ik_y) & - (\frac{{\bm k}^2}{2m} - \mu)
\end{array}
\Biggr],
\label{BdG_ESP}
\end{equation}
or, in the basis of the Pauli matrices ${\bm \tau} =[\tau_1, \tau_2, \tau_3]$,

\begin{equation}
\mathcal{H}_{\bm k} = {\bm \tau} \cdot {\bm d}_{\bm k}, 
\quad 
{\bm  d}_{\bm k} = 
\left[
\Delta^\prime\, k_x,\, -\Delta^\prime\, k_y, \, \frac{{\bm k}^2}{2m} -\mu
\right].
\label{BdG_d}
\end{equation}
This equation resembles the spin Hamiltonian of a ferromagnet, with vector ${\bm d}_{\bm k}$ 
playing the role of the magnetization. In this case, ${\bm d}_{\bm k}$ defines the texture of the Nambu pseudospin ${\bm \tau}/2$ in the 2D momentum space 
(see Figs. \ref{Top_fig} and \ref{Ord_fig}).

The topology of the pseudospin texture can be characterized by the winding number 

\begin{equation}
C = \frac{1}{4\pi} \int {\bm  n}_{\bm k} \cdot \left( \frac{\partial {\bm  n}_{\bm k} }{\partial k_x} \times \frac{\partial {\bm  n}_{\bm k} }{\partial k_y} \right) dk_xdk_y, 
\quad 
{\bm n}_{\bm k} = \frac{ {\bm d}_{\bm k} }{ |{\bm d}_{\bm k}| },
\label{C}
\end{equation}
which, in differential geometry, is the first Chern invariant of a principal $U(1)$ bundle over a torus \cite{Thouless1982, Kohmoto85}.
Geometrically, $|C|$ is the number of times the unit vector ${\bm n}_{\bm k}$ sweeps a unit sphere as ${\bm k}$ covers the entire 
momentum space. The Chern invariant acquires nontrivial values $C=\pm 1$ under condition $\mu m > 0$. 
In this case, the configuration of the vector ${\bm d}_{\bm k}$ defines the skyrmion, a topological defect that compactifies the physical space on a sphere (see Fig. \ref{Top_fig}).
The superconducting state is, therefore, topologically nontrivial as opposed to the case $\mu m < 0$ (Fig.  \ref{Ord_fig}) in which no $d$ - vector winding takes place, hence $C=0$. 

\begin{figure}[b]
	\begin{center}
		\includegraphics[width=85mm]{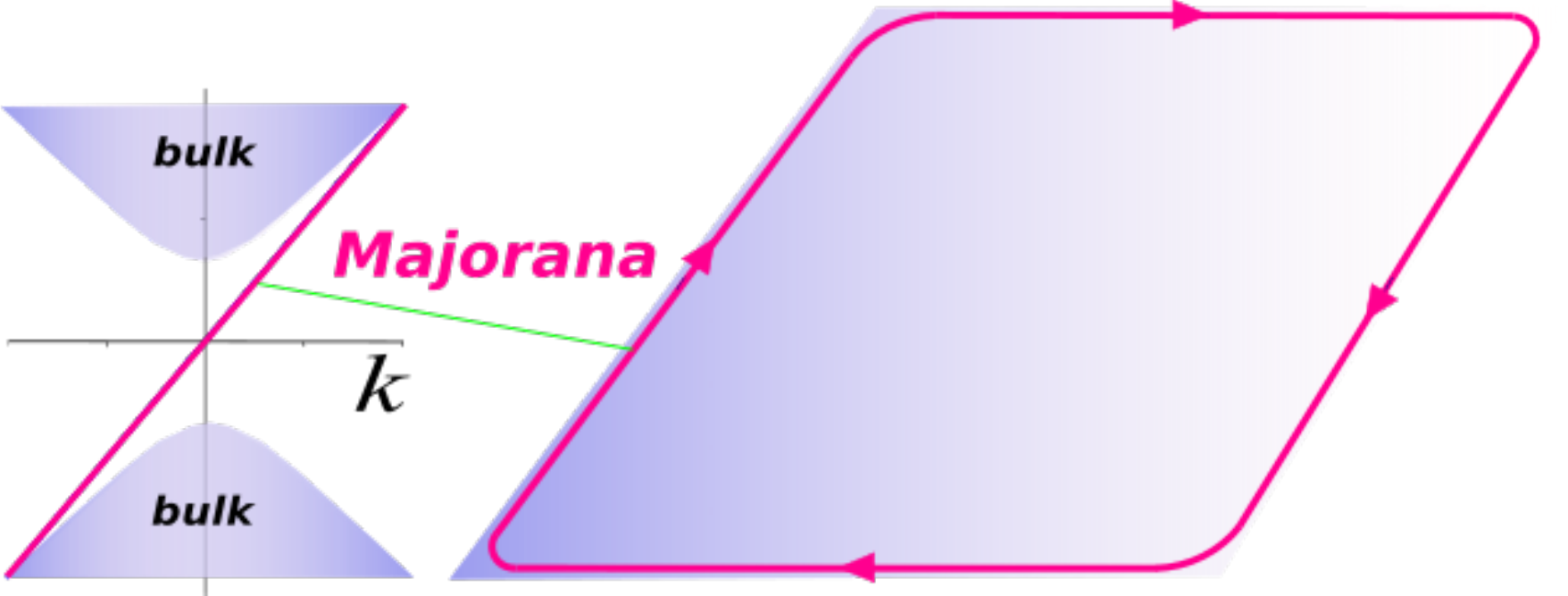}
	\end{center}
	\caption{
		Schematic of a chiral Majorana edge state in a TS [see also Eqs. (\ref{Conjugate}) -- (\ref{CMEM2})].
	}
	\label{Edge_fig}
\end{figure}

The above considerations have clear parallels with Chern insulators and the quantum anomalous Hall effect in zinc-blende materials (see recent review in Ref. \cite{Liu16}).
In a Chern insulator, $C$ corresponds to the TKNN invariant \cite{Thouless1982, Kohmoto85} and 
can be obtained from the Berry curvature of the electronic bands as

\begin{equation}
C = \frac{1}{2}[ {\rm sgn}(-\mu) - {\rm sgn}(m)].
\label{C_Chern}
\end{equation}
This analytical result confirms the picture of the winding of ${\bm d}_{\bm k}$ in momentum space shown in Figs. \ref{Top_fig} and \ref{Ord_fig}.
As in Chern insulators, the nontrivial number $C$ indicates a chiral gapless state at the edge of a TS (see Fig. \ref{Edge_fig}). 
Furthermore, due to the generic particle-hole symmetry

\begin{equation}
\mathbb{C} \mathcal{H}(x,y) \mathbb{C}^\dagger = - \mathcal{H}(x,y),
\label{PH}
\end{equation}
the edge modes of TSs mimic Majorana fermions of the relativistic quantum theory. Above, $\mathcal{H}(x,y)$ is the real-space BdG Hamiltonian, 
and $\mathbb{C}$ is the particle-hole conjugation operation. In non-topological superconductors, $\mathbb{C}$ converts a particle into a hole and vice versa. 
The Majorana edge state, $\Psi_M(x,y,t)$, transforms to itself under $\mathbb{C}$:

\begin{equation}
\Psi^\mathbb{C}_M(x,y,t) = \Psi_M(x,y,t),
\label{Conjugate}
\end{equation}
that is, emergent Majorana fermions in TSs are particles and holes at the same time. 
For the BdG Hamiltonian (\ref{BdG_d}), the conjugation operation is $\mathbb{C} = \tau_1 K$, 
so the Majorana edge solution is a real eigenstate of $\tau_1$. For a "hard-wall" boundary \cite{Liu16}, it is given by

\begin{eqnarray}
\Psi_M(x,y,t) = 
\Biggl[
\begin{array}{c}
1 \\
1
\end{array}
\Biggr]
\sum\limits_{k_x}
&
\mathcal{N}_{k_x}
\left[
e^{\varkappa_+(k_x)y} - e^{\varkappa_-(k_x)y}
\right]
&
\label{CMEM1}\\
&
\times
\cos\left[
k_x x - E(k_x)t/\hbar
\right].
&
\label{CMEM2}
\end{eqnarray}
It is assumed that the TS occupies the half-space $y \leq 0$, and the edge mode propagates along $x$. It has a linear dispersion 

\begin{equation}
E(k_x) = \Delta^\prime k_x, 
\label{E_kx}
\end{equation}
and is localized on the scale given by $\varkappa_\pm = m \Delta^\prime \pm \sqrt{ m^2\Delta^{\prime 2} + k^2_x - k^2_F}$. 
The coefficients $\mathcal{N}_{k_x}$ of the sum in Eq. (\ref{CMEM1}) are the normalizing factors. The normalizability requires that $|k_x| < k_F$.

\subsection{Search for intrinsic chiral superconductivity and related phenomena}

The model discussed above illustrates the principal possibility and essential attributes of the topological chiral $p$ - wave superconductivity. 
Perhaps the first material for which such a possibility has been considered is the layered perovskite ruthenate Sr$_2$RuO$_4$. 
It is a quasi - 2D material in which ruthenium oxides tend to become ferromagnetic, 
which favours a spin-triplet order parameter \cite{Mackenzie03, Rice95}. 
If the spin-triplet superconductivity has a definite chirality, the time-reversal symmetry should be broken 
below the critical temperature $T_c$.
The expectation that Sr$_2$RuO$_4$ harbors such chiral superconductivity was strengthened by 
the muon spin rotation experiments \cite{Luke98} which detected internal magnetic fields below $T_c$.
While the scenario of the chiral $p$ - wave superconductivity in Sr$_2$RuO$_4$ may be debatable, 
the tunneling experiment \cite{Kashiwaya11} has reported an enhanced zero-bias conductance consistent with the existence of the boundary modes.
The zero-bias peak in the tunneling spectrum was attributed to the surface Andreev bound states (ABSs), 
although this would not generally tell whether the order parameter is chiral or helical \cite{Scaffidi14}.
Initially, the theory of the tunneling spectroscopy of the surface ABSs was developed for $d$-wave superconductors \cite{Tanaka95,Kashiwaya00}.
Reference \cite{Yada14} has extended the theory to Sr$_2$RuO$_4$, using a three-band model and the recursive Green's function method.
The tunneling spectra with both zero-energy peaks and zero-energy dips were found, depending 
on the spatial dimensionality of the model and the presence or absence of SOC.
Another recent theoretical work \cite{Zhang16} has proposed to identify the chiral $p$-wave superconductivity 
by the electronic states on a domain wall between the order parameters with opposite chiralities $(k_x \pm i k_y)$.
If the superconducting order parameter breaks the time-reversal symmetry, domains with different chiralities and opposite edge currents
are expected to form akin to the ferromagnetic domains. An experimental attempt to detect such chiral domains has been reported in Ref. \cite{Wang17}.
A recent review on the point-contact spectroscopy as a means of detecting topological superconductivity is given in Ref. \cite{Wang18}.

Beside the chiral phase in Sr$_2$RuO$_4$, there has been a theoretical proposal for a topological crystalline superconductor phase in this material \cite{Ueno13}.
It is characterized by a pair of Majorana modes each protected by the mirror symmetry of the Sr$_2$RuO$_4$ crystal structure.
Reference \cite{Ueno13} discussed a magnetic-field-induced transition into the topological crystalline superconducting state accompanied by a rotation
of the Balian-Wertheimer ${\bm d}$ vector parametrizing the triplet order parameter.

Apart from Sr$_2$RuO$_4$, a number of other candidate materials to host topological superconductivity have been identified in the past decade, 
most notably Cu$_x$Bi$_2$Se$_3$ \cite{Hor10,Fu10,Sasaki11}, Sn$_{1-x}$In$_x$Te \cite{Sasaki12,Novak13} as well as 
some noncentrosymmetric superconductors in which the $p$ - wave gap is larger than the $s$ - wave one \cite{Sato09,Tanaka09_NCS,Schnyder15}. 
As opposed to the chiral superconductivity, these materials are expected to host time-reversal symmetric (helical) topological 
phases \cite{Yada11}~\footnote{The state of the art is well captured in the review article \cite{Sato17}.}.
If, however, the time-reversal symmetry is broken by an external magnetic field, noncentrosymmetric low-dimensional superconductors may turn into chiral TSs \cite{Loder15,Mazziotti18,Lei18}. 
Normally, this requires a magnetic field that is by far larger than the upper critical field $H_{c2}$. 
The way to overcome this problem is to apply the field parallel to the basal plane, reducing the Meissner currents in favour of the Zeeman splitting.
The theory  \cite{Loder15} has examined such a possibility for the superconducting interface between LaAlO$_3$ and SrTiO$_3$, 
assuming the Rashba SOC and a three-band model. More recently, 1D structures at LaAlO$_3$/SrTiO$_3$ oxide interfaces  have been found to support
Majorana modes \cite{Mazziotti18}. In fact, metallic superconducting films grown on a substrate and subject to an in-plane magnetic field 
may have all the ingredients required to achieve chiral superconductivity, i.e. Copper pairing, broken inversion symmetry, and broken time-reversal symmetry.
This expectation has been supported by the density functional theory calculations for ultrathin Pb and $\beta$-Sn \cite{Lei18}.

\subsection{TI materials as platform for topological superconductivity and Majorana fermions}

\begin{figure}[t]
	\begin{center}
		\includegraphics[width=70mm]{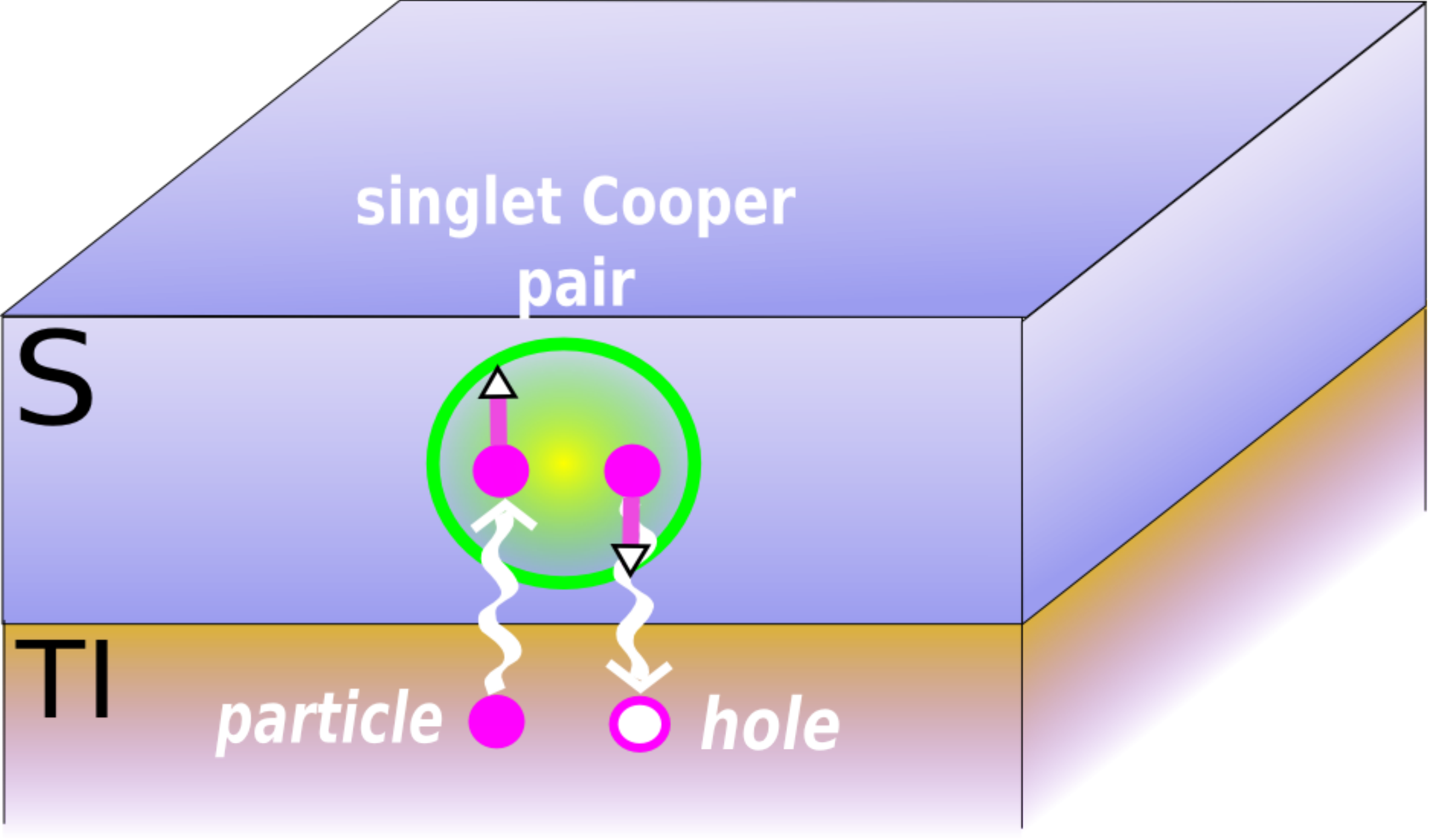}
	\end{center}
	\caption{
		Schematic of a hybrid structure created by placing a singlet superconductor (S) on top of a topological insulator (TI) material.
		Andreev reflection at the S/TI boundary gives rise to the superconducting proximity effect.
	}
	\label{Proxy_fig}
\end{figure}

The above discussion pertains to intrinsic superconductivity when the symmetry breaking order parameter 
occurs spontaneously below a certain $T_c$ ($\sim 1.5$ K for Sr$_2$RuO$_4$). 
An alternative to that is the induced superconductivity which occurs in a normal conductor brought into electric contact with
an intrinsic superconductor (see also Fig. \ref{Proxy_fig}). Although the normal conductor has no pairing interaction of its own, 
it acquires the superconducting correlations through the proximity effect.
Microscopically, this can be understood in terms of Andreev reflection \cite{Andreev64,Blonder82} 
whereby a particle in the normal system is converted into a hole (and vice versa), 
while a Cooper pair passes through the interface, as sketched in Fig. \ref{Proxy_fig}.
The particle-hole conversion is most efficient when the thickness of the normal region is smaller than 
the phase coherence length.

The superconducting proximity effect offers an attractive alternative to intrinsic superconductivity, 
as topological phases can be "engineered", using broken symmetries of the normal system. 
A prominent example is the theoretical proposal \cite{Fu08} for a chiral TS and Majorana fermions at the surface of a 3DTI proximitized by 
a conventional (singlet $s$-wave) superconductor. In the past decade, impressive progress has been achieved in fabricating and characterizing 
hybrid structures of superconductors and TI materials 
\cite{Zhang11,Koren11,Sacepe11,Veldhorst12,Qu12,Wang12,Williams12,Yang12,Maier12,Cho13,Oostinga13,Koren13,Sochnikov13,
	Hart14,Snelder14,Molenaar14,Galletti14,Kurter14,Finck14,Sochnikov15,Xu15,Stehno16,Wiedenmann16,Tikhonov16,Pang16,Dayton16,
	Hart16,Bocquillon16,Deacon17,He17,Charpentier17,Jauregui18,Ghatak18,Klett18,Snyder18,Lyu18,Shen18,Kurter19,Kayyalha19,Kayyalha19_Non}.
Most of these experiments have used the tetradymite compounds Bi$_2$Se$_3$ and Bi$_2$Te$_3$, ternary tetradymites (e.g., Bi$_2$Te$_2$Se) 
or later generations of Bi-based compounds such as Bi$_{2-x}$Sb$_x$Te$_{3-y}$Se$_y$ \cite{Taskin11}.
Other types of the TI materials include thick strained HgTe layers \cite{Maier12,Oostinga13,Sochnikov15,Wiedenmann16}, 
HgTe quantum wells \cite{Hart14,Hart16,Bocquillon16,Deacon17}, topological crystalline insulator SnTe \cite{Klett18,Snyder18},
and Cr-doped (Bi,Se)$_2$Te$_3$ thin films \cite{He17,Shen18}. 
The latter are magnetic topological insulators that, in the absence of the superconducting pairing, exhibit the quantum anomalous Hall effect \cite{Chang13}.
That is, Cr-doped (Bi,Se)$_2$Te$_3$ thin films with Nb contacts, such as in Refs. \cite{He17,Shen18}, are prototypes of the quantum anomalous Hall insulator (QAHI) - superconductor devices.  

A related paper \cite{Menard17} has reported an observation of 2D topological superconductivity in a Pb/Co/Si(111) structure. It was modeled as a Rashba system with a mixed singlet-triplet pairing and an exchange interaction.  

In fact, TI materials with induced superconducting and magnetic orders 
have long been a fertile ground for theoretical modeling of the chiral TS and related 
phenomena (see, e.g., \cite{Fu08,Fu09,Akhmerov09,Fu09_JJ,Law09,Tanaka09_TI,Linder10,Qi10,Ioselevich11,Chung11,Ii11,Badiane11,Golub11,
	Ii12,Weithofer13,He14,Wang15,Akzyanov15,Shapiro16,Lian16,Shapiro17,Virtanen18}).
This includes the issues of the observability of neutral Majorana fermions in quantum interferometry \cite{Fu09,Akhmerov09}, 
fractional Josephson effect \cite{Fu09_JJ}, resonant Andreev reflection \cite{Law09},
magnetic proximity effect \cite{Tanaka09_TI}, backscattering processes \cite{Chung11}, current noise \cite{Badiane11}, tunneling spectroscopy \cite{Ii12}, crossed Andreev reflection \cite{He14}, half-integer longitudinal conductance \cite{Wang15}, edge-state-induced Andreev oscillations \cite{Lian16},
just to name a few.

In particular, in the series of papers \cite{Qi10,Chung11} and \cite{Wang15}, 
a $k_x+ik_y$ phase with a chiral Majorana edge mode has been discovered theoretically in QAHI/superconductor structures.
Its expected transport signatures are Majorana backscattering and the half-integer longitudinal conductance.
In the following, these ideas are discussed in some more detail.

\subsection{Chiral TS with a single Majorana edge mode in QAHI/supercondictor structures}

To set the scene, we define the normal-state Hamiltonian for a magnetic thin film, 
$\hat{H} = \sum_{\bm k} c^\dagger_{\bm k} H_{\bm k} c_{\bm k}$, 
where $H_{\bm k}$ is an effective 4-band Hamiltonian \cite{Wang15}:

\begin{equation}
H_{\bm k} = A (\sigma_x k_y - \sigma_y k_x) \nu_3 + M_{\bm k} \nu_1 + \lambda \sigma_z - \mu, 
\label{H_QAHI}
\end{equation}
and $c_{\bm k} =[c^t_{\bm k \uparrow}, c^t_{\bm k \downarrow}, c^b_{\bm k \uparrow}, c^b_{\bm k \downarrow}]$.
Here, the operator $c^{t,b}_{\bm k \sigma}$ annihilates an electron with momentum ${\bm k}$ and spin $\sigma = \uparrow, \downarrow$, 
and the superscripts $t$ and $b$ refer to the top and bottom surface layers of the film, respectively.
$\sigma_i$  (with $i = x, y, z$) and $\nu_j$  (with $j = 1, 2, 3$) are Pauli matrices in spin and layer subspaces, respectively. 
The first term in Eq. (\ref{H_QAHI}) is the Hamiltonian of the two surface layers, where the constant $A$ (assumed positive throughout) 
determines the surface velocity. The second term $M_{\bm k} \nu_1$ introduces the coupling between the layers, 
opening a hybridization gap at the $\Gamma$ point (${\bm k}=0$).
The hybridization energy is $M_{\bm k} = M_0 + M_1(k^2_x + k^2_y)$, 
where $M_0$ yields the half of the gap between the conduction and valence bands, while $M_1$ accounts for the band curvature.
The third term  $\lambda \sigma_z$ is the mean-field exchange Hamiltonian due to the ferromagnetic ordering,
with $\lambda$ being the exchange energy. 

The Hamiltonian (\ref{H_QAHI}) decouples into two Chern subsystems with the Dirac masses $\lambda \pm M_{\bm k}$,
where the signs $\pm$ are dictated by the time-reversal symmetry. The corresponding Chern number is the sum 
of the Chern numbers for the two subsystems [cf. Eq. (\ref{C_Chern})],

\begin{eqnarray}
C &=& \frac{1}{2} \sum_{\nu =\pm} [{\rm sgn}(\lambda +  \nu M_0) - {\rm sgn}(\nu M_1)]
\label{C_QAHI_1}\\
&=& \frac{1}{2} [{\rm sgn}(\lambda + M_0 )  +  {\rm sgn}(\lambda - M_0)].
\label{C_QAHI_2}
\end{eqnarray}
A similar topological number is encountered in Haldane's model \cite{Haldane_ParAnom_PRL88}.
Taking for simplicity $M_0 > 0$ and $\lambda > 0$, we see that the system undergoes a topological phase transition from an ordinary 
insulator with $C=0$ for $\lambda < M_0$ to a QAHI with $C=1$ for $\lambda > M_0$:

\begin{equation}
C =
\left\{
\begin{array}{cc}
0 & \lambda < M_0, \\
1, &  \lambda > M_0.
\end{array}
\right.
\label{C_01}
\end{equation}
In the latter case, there is a chiral edge mode realizing a gapless Dirac fermion.

In contact with a conventional superconductor, placed on top of the structure, the magnetic film can be described at low energies by 
the BdG Hamiltonian (\ref{BdG}) with the singlet pairing 

\begin{equation}
\Delta_{\bm k} =
\Biggl[
\begin{array}{cc}
\Delta_t i\sigma_y & 0 \\
0 & \Delta_b i\sigma_y
\end{array}
\Biggr],
\label{Delta_QAHI}
\end{equation}
see also Sec. \ref{US} for a microscopic theory of the superconducting proximity effect.
Here, $\Delta_{\bm k}$ is a matrix in the layer subspace where $\Delta_t$ and $\Delta_b$ denote the pair potentials in the top and bottom layers.
As found in Ref. \cite{Wang15}, the essential condition for realizing a chiral TS is to have unequal pairing amplitudes $\Delta_t \not= \Delta_b$.
This point is best illustrated in the special case where

\begin{equation}
\Delta_t = \Delta, \qquad \Delta_b = - \Delta, \qquad \mu = 0,
\label{Delta_t,b}
\end{equation}
and $\Delta$ is real. In this case, the BdG Hamiltonian takes a compact form

\begin{equation}
\mathcal{H}_{\bm k} = A (\Sigma_x k_y - \Sigma_y k_x)  + (\lambda + M_{\bm k} \mathcal{V}_1 + \Delta \mathcal{T}_1 )\Sigma_z,
\label{BdG_QAHI}
\end{equation}
where $\Sigma_x, \Sigma_y, \Sigma_z$, $\mathcal{V}_1$, and $\mathcal{T}_1$ are the $8 \times 8$ matrices

\begin{eqnarray}
&&
\Sigma_x = \tau_3 \nu_3 \sigma_x, \quad \Sigma_y = \tau_3 \nu_3 \sigma_y, \quad \Sigma_z = \tau_0 \nu_0 \sigma_z,
\label{Sigmas}\\
&&
\mathcal{V}_1 = \tau_0 \nu_1 \sigma_z, \quad \mathcal{T}_1= \tau_1 \nu_0 \sigma_z,
\label{S_and_T}
\end{eqnarray}
and $\nu_0$ and $\tau_0$ are the unit matrices in layer and Nambu subspaces.
Now, the sum $\lambda + M_{\bm k} \mathcal{V}_1 + \Delta \mathcal{T}_1$ in Eq. (\ref{BdG_QAHI}) is the Dirac mass matrix.
Furthermore, $\mathcal{V}_1$ and $\mathcal{T}_1$ commute with each other and with any of $\Sigma_i$, 
so in the basis of the common eigenstates of $\mathcal{V}_1$ and $\mathcal{T}_1$ the mass matrix has a diagonal structure with the entries

\begin{equation}
\lambda + \nu M_{\bm k} + \tau \Delta, \qquad \nu, \tau = \pm 1.
\label{Mass_matrix}
\end{equation}
$\nu$ and $\tau$ are the eigenvalues of $\mathcal{V}_1$ and $\mathcal{T}_1$, respectively. 
Therefore, the BdG model for the QAHI decouples into four Chern subsystems or, equivalently, four species of the 2D $k_x +ik_y$ - superconductor. 
Accordingly, the total Chern number is

\begin{eqnarray}
N &=& \frac{1}{2} \sum_{\nu,\tau = \pm} [{\rm sgn}(\lambda + \nu M_0 + \tau \Delta) - {\rm sgn}(\nu M_1)]
\label{N_1}\\
&=& \frac{1}{2} [{\rm sgn}(\lambda + M_0 + \Delta)  +  {\rm sgn}(\lambda - M_0 + \Delta)
\nonumber\\
&&
+ {\rm sgn}(\lambda + M_0 - \Delta) +  {\rm sgn}(\lambda - M_0 - \Delta)].
\label{N_2}
\end{eqnarray}
To distinguish the superconducting case, we use here the notation $N$ instead of $C$ (cf. Ref. \cite{Wang15}).

\begin{figure}[t]
	\begin{center}
		\includegraphics[width=80mm]{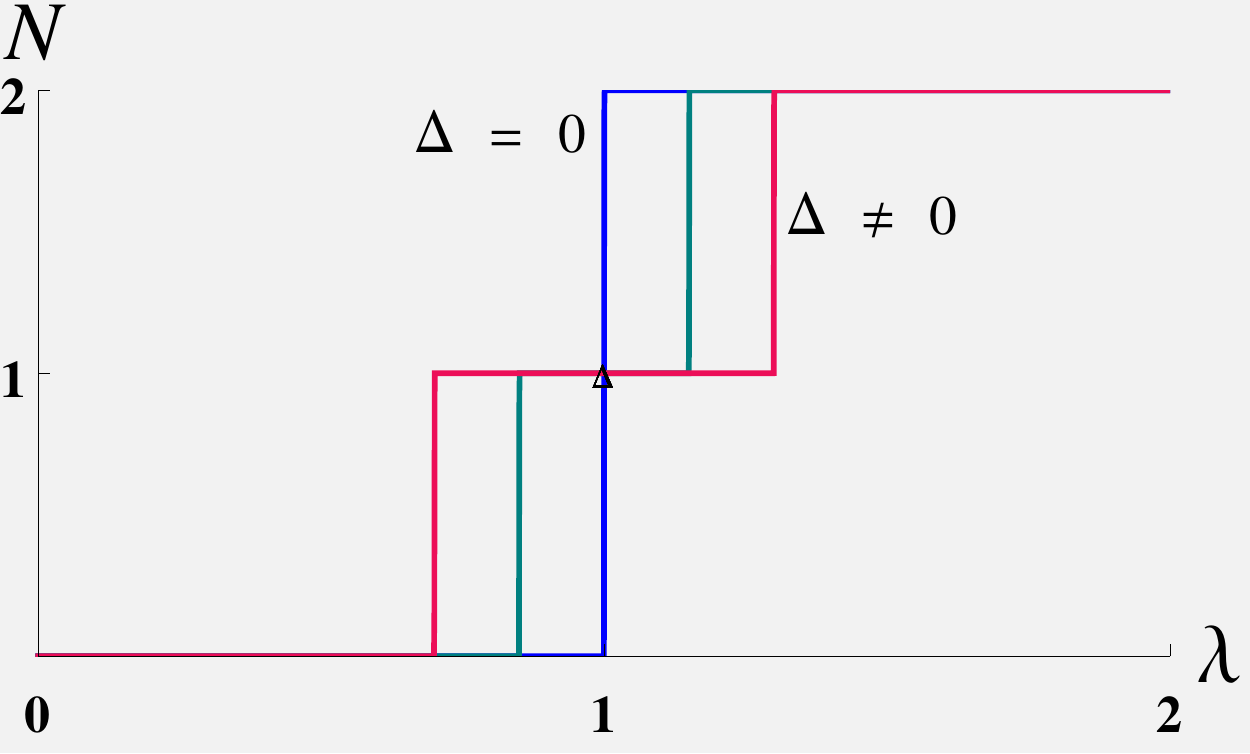}
	\end{center}
	\caption{
		Chern number of a QAHI/superconductor hybrid $N$ (\ref{N_012}) as function of exchange energy $\lambda$ in units of $M_0$. 
		For a finite pairing energy $\Delta$, a plateau develops at $N=1$, 
		corresponding to a chiral $k_x +ik_y$ state with a single Majorana edge mode (\ref{CMEM2}).
	}
	\label{N_fig}
\end{figure}

The superconducting pairing $\Delta$ allows for specific phase transitions that are absent in the normal case (see Fig. \ref{N_fig}).
For low-$T_c$ superconductor structures, we can safely assume $\Delta < M_0$. 
Then, for positive parameters, the possible values of the Chern number (\ref{N_2}) are 

\begin{equation}
N =
\left\{
\begin{array}{ccc}
0 & \lambda < M_0 - \Delta, \\
1 & M_0 - \Delta < \lambda <  M_0 + \Delta, \\
2, &  \lambda > M_0 + \Delta.
\end{array}
\right.
\label{N_012}
\end{equation}
In other words, an increasing exchange field $\lambda$ induces a series of phase transitions from an ordinary superconductor with $N=0$ to 
the topological phases with $N=1$ and $N=2$. The latter has two chiral Majorana edge modes which correspond to a single Dirac mode, 
so the $N=2$ phase matches the QAHI with $C=1$. The truly new phase is that with the odd Chern number $N=1$ (\ref{N_012}), 
which is nothing else as the $k_x +ik_y$ TS with a single chiral Majorana edge mode such as discussed above [cf. Eqs. (\ref{BdG_d}) and (\ref{CMEM1})].
By reversing the magnetization $\lambda \to -\lambda$, one can also access the opposite-chirality states $N=-1$ and $N=-2$.

\begin{figure}[t]
	\begin{center}
		\includegraphics[width=85mm]{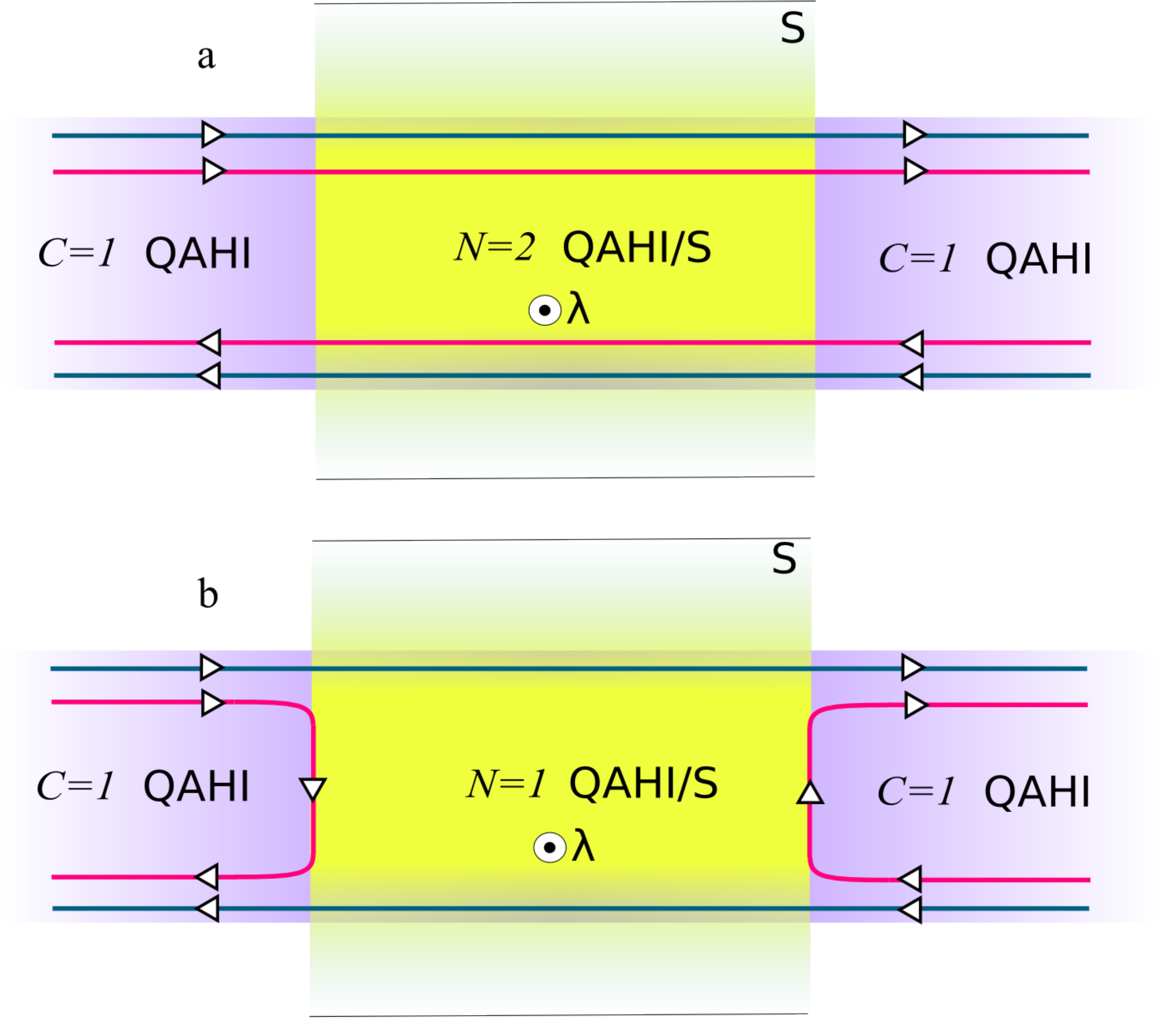}
	\end{center}
	\caption{Schematic of the chiral edge modes and Majorana backscattering in QAHI/superconductor (S) structure (after Refs. \cite{Wang15,He17}).
		(a) For a large enough exchange field $\lambda$, 
		a pair of incident Majorana edge modes (making up a single Dirac edge mode) match those in the superconducting region and pass almost perfectly through the device.
		(b) As the exchange field $\lambda$ decreases,
		the state of the superconducting region switches to the chiral ($N=1$) TS, 
		such that one of the paired Majorana edge modes vanishes.
		Consequently, only one of the incident Majorana edge modes is transmitted to the superconducting region, 
		whereas the other Majorana mode is almost perfectly reflected, 
		resulting in the half-integer $0.5\, e^2/h$ longitudinal conductance \cite{Wang15,He17}.
	}
	\label{QAHI_S_fig}
\end{figure}

As argued in Ref. \cite{Wang15}, the $N=1$ phase can be identified by a half-integer plateau $0.5\, e^2/h$ 
in the longitudinal conductance as a function of the exchange field. 
The proposal relies on the backscattering of Majorana edge modes \cite{Chung11}
which is expected in a QAHI/superconductor device at the topological transition to the $N=1$ phase. 
The basic setup consists of a magnetic thin film and a superconducting bar place across it, as depicted in Fig. \ref{QAHI_S_fig}. 
For a large enough exchange field ($\lambda > M_0 + \Delta$, see Fig. \ref{QAHI_S_fig}a), 
the normal regions are the $C=1$ QAHI with a Dirac edge state propagating along the sample boundary,
while the superconducting region supports two Majorana edge modes, forming the phase with $N=2$. 
Since a Dirac fermion is composed of two Majorana ones,  
we can think of two incident Majorana edge modes which match those in the superconducting region, getting transmitted almost perfectly through the device.
Consequently, the longitudinal conductance, $\sigma_{12}$, between contacts 1 and 2 reaches the quantum  $e^2/h$.

\begin{figure}[t]
	\begin{center}
		\includegraphics[width=85mm]{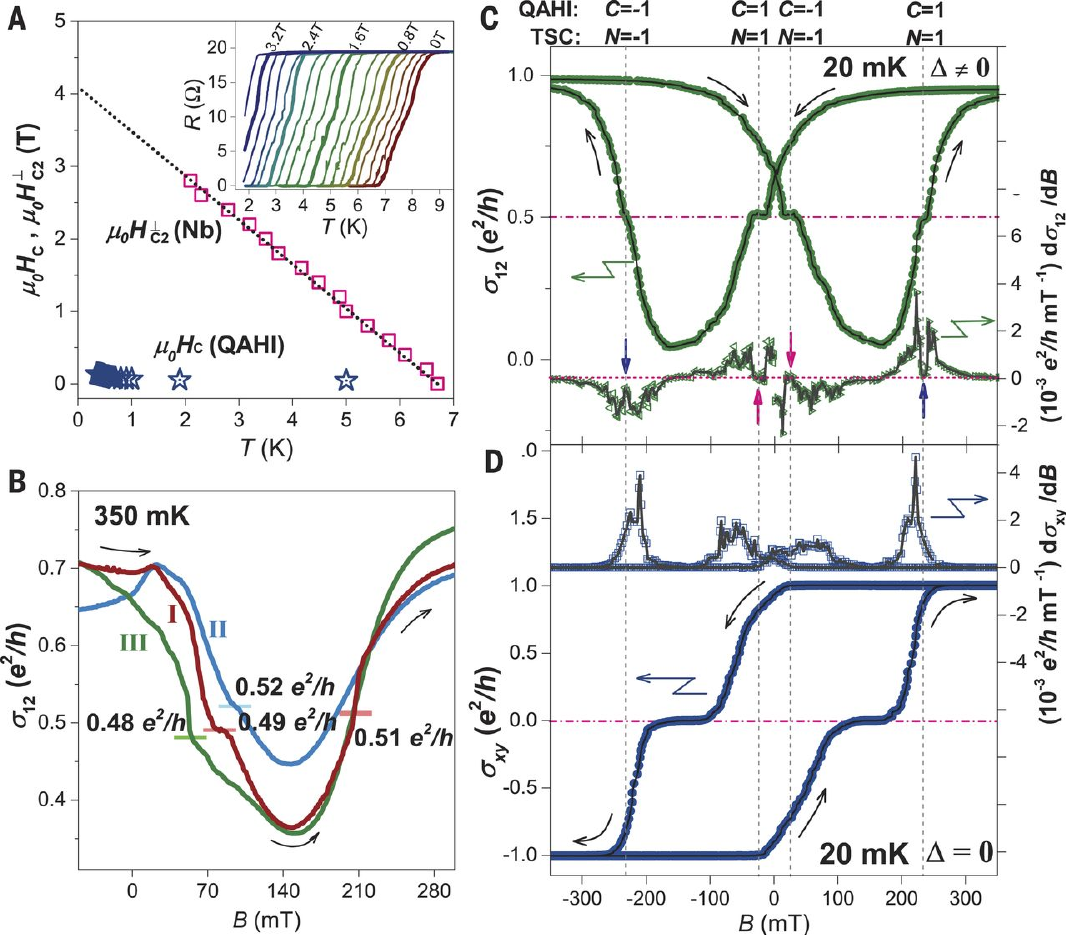}
	\end{center}
	\caption{
		Half-integer longitudinal conductance as a signature of single chiral Majorana edge modes 
		(From \cite{He17}. Reprinted with permission from the American Association for the Advancement of Science.)
		In particular, {\bf C} shows the longitudinal conductance $\sigma_{12}$ as a function of external perpendicular magnetic field measured at 20 mK. 
		When superconductivity is induced on the top surface of the QHAI, $\sigma_{12}$ shows additional half-integer plateaus ($~0.5 e^2/h$) 
		between the transitions of the $C=\pm 1$ QAHI and the normal insulator. 
		(Lower plot in {\bf C}) Derivative of $\sigma_{12}$ with respect to the magnetic field. 
		Topological transitions are indicated by dashed lines and arrows. 
		For full details of the presented data, see Ref. \cite{He17}.
	}
	\label{He17_fig}
\end{figure}

Upon lowering the exchange field to $M_0  < \lambda <  M_0 + \Delta$, 
the state of the superconducting region switches to the chiral TS with $N=1$, 
such that one of the paired Majorana edge modes vanishes, while the normal regions are still the $C=1$ QAHI (see Fig. \ref{QAHI_S_fig}b).
In this regime, only one of the incident Majorana edge modes can be transmitted to the superconducting region, 
whereas the other Majorana mode is almost perfectly reflected. 
This is what Ref. \cite{Wang15} called the separation of the two Majorana modes at the superconducting boundary, 
resulting in a half-integer plateau $0.5\, e^2/h$ in the longitudinal conductance.
Following this theoretical prediction, the experimental paper \cite{He17} reported the observation of the half-integer longitudinal conductance plateaus 
close to the coercive fields during magnetization reversals in an external out-of-plane magnetic field.
Some of the experimental data on the sample characterization and the occurrence of the half-integer plateaus
are shown in Fig. \ref{He17_fig}.

Following the experiment \cite{He17}, alternative theoretical and experimental interpretations of the half-integer longitudinal conductance have appeared.
Reference \cite{Ji18} suggested a mechanism for the $0.5\, e^2/h$ conductance plateau without 1D chiral Majorana fermions.
It was argued that such plateaus could be a feature of a good electric contact between quantum Hall and superconducting films, 
and could therefore indicate neither the existence nor absence of 1D chiral Majorana fermions.
The experiment \cite{Kayyalha19_Non} reported high-probability Andreev reflection in QAHI/superconductor structures, 
attributing their findings to high contact transparency and interpreting in this context the origin of the $0.5\, e^2/h$ conductance plateau.
The theory  \cite{Huang18} argued that a nearly flat conductance plateau, similar to that in Ref. \cite{He17}, could also arise from the percolation of 
quantum Hall edges well before the onset of the topological superconductivity or at temperatures much above the TS gap.

Another line of the theoretical research has dealt with the issues of control and manipulation of chiral Majorana fermions for possible practical applications of 
realistic QAHI/superconductors devices. Reference \cite{Chen18} has shown that quasi-1D QAHI structures could exhibit a broad topological regime supporting localized Majorana zero energy modes and proposed to implement networks of such quasi-1D QAHI systems for scalable topological quantum computation.
Since the Majorana fermion is a charge-neutral particle, the direct effect of an electric field on them should fail.
The recent study \cite{Zhou18} has proposed a magnetic flux control of the transport of chiral Majorana fermions in topological superconducting devices 
with Josephson junctions. Reference \cite{Lian18} has found that the propagation of chiral Majorana fermions could lead to the same unitary transformation 
as that in the braiding of Majorana zero modes, suggesting a platform to perform quantum computation with chiral Majorana fermions.
The theoretical work \cite{Li19} has suggested an interferometer for chiral Majorana modes where the interference effect was caused and controlled 
by a Josephson junction of proximity-induced TSs.
Another recent paper \cite{Beenakker19} elaborates on the deterministic creation and braiding of chiral edge vortices in hybrid structures.

\section{Topological weak superconductivity and the fractional Josephson effect.}
\label{TWS}

Striking manifestations of topological superconductivity are expected to occur in weak links between low-dimensional systems supporting Majorana modes.
Some illustrative examples of such systems are 2D $d$-wave superconductors \cite{Tanaka96},
1D $p$-wave superconductors \cite{Kitaev01,Kwon04a,Kwon04b}, superconductor/semiconductor wires 
\cite{Sau10,Alicea10,Lutchyn10,Oreg10}, Shiba chains \cite{Nadj-Perge13,Pientka13}, RKKY wires \cite{Klinovaja13}, just to name a few.
The boundaries of 1D TSs host a pair of Majorana zero modes (MZMs). 
These 0D cousins of the chiral Majorana edge mode appear at the midgap energy, i.e. at exactly zero energy relative to the Fermi level.

An interesting implication of the MZMs is the degeneracy of the ground state. 
For a pair of MZMs, there are two possible ground states corresponding to the eigenvalues $\pm 1$ of the hermitian operator $i\gamma_1\gamma_2$, 
where $\gamma_i$ are self-adjoint fermionic operators $\gamma_i = \gamma^\dagger_i$ that square to 1 (Majorana operators).
If we combine them into a usual fermion $c = (\gamma_1 + i\gamma_2)/2$, the two ground states have 
different occupation numbers 

\begin{equation}
c^\dagger c = \frac{1}{2} (1+i\gamma_1\gamma_2) = 
\left\{
\begin{array}{c}
1 \\ 0
\end{array}
\right.
, 
\label{Occupation}
\end{equation}
hence different fermion parities.
In Josephson junctions (JJs) of two TSs brought into electric contact,
a  change of the Josephson phase difference by $2\pi$ effectively causes swapping the MZMs and a transition between the ground states \cite{Kitaev01}.
This leads to the $4\pi$ - periodicity of the MZMs, as another phase advance of $2\pi$ is needed 
to recover the same ground state. 

Systems supporting the MZMs may prove useful in braiding-based topological quantum computation in which computing operations are performed by unitary transformations 
within a degenerate set of ground states \cite{Nayak08,Flensberg11,Bonderson11, Hyart13,DasSarma15}. 
However, the MZMs are not readily available in solids, and much effort has been put into engineering and detecting them in accessible materials and structures
(see review articles \cite{Alicea12_Rev,Beenakker13_Rev,Elliot15,DasSarma15,Sato17,Aguado17,Haim18}).
A growing number of experiments has been testing the existence of the MZMs in various superconducting structures, 
using tunneling spectroscopy (see Refs. \cite{Mourik12,Das12,Nadj-Perge14,Xu15,Deng16,Albrecht16,Jeon17,ZhangH18,Ren19}) 
and Josephson effects (JEs) (see Refs. \cite{Rokhinson12,Wiedenmann16,Bocquillon16,Deacon17,Li18_ABS,Wang18_Shapiro,LeCalvez19,Laroche19,Kayyalha19}).

The JE diagnostics of topological weak links relies on the so-called fractional JE associated with the ground state degeneracy. 
First proposed for model $p$-wave superconductors \cite{Kitaev01,Kwon04a,Kwon04b}
the fractional JE is also achievable in hybrid structures of conventional superconductors and 
normal SOC materials, which has caused the recent surge of interest in this and related phenomena.
Here, we review this topic from the theoretical perspective, aiming to give some background on the fractional JE and summarize the findings of 
different models.

\subsection{Majorana zero modes in a 1D chiral TS} 

As an exemplary model, we can choose a 1D version of the BdG Hamiltonian of the $k_x +ik_y$ TS  [see Eq. (\ref{BdG})].
Alternatively, we can think of the chiral TS in the QAHI/superconductor structures at the transition to the $N=1$ state 
[see Eqs. (\ref{BdG_QAHI}) and (\ref{N_012})]. The reported observability of such transitions \cite{He17,Shen18} 
gives us extra reason for this choice. With the $x$-axis parallel to the system, we can write

\begin{eqnarray}
\mathcal{H} &=&
\Biggl[
\begin{array}{cc}
\frac{k^2_x}{2m} - \mu  &  \Delta^\prime e^{i\varphi} k_x \\
\Delta^\prime e^{-i\varphi} k_x & - (\frac{k^2_x}{2m} - \mu)
\end{array}
\Biggr]
\label{BdG_kx}\\
&=&
\tau_3 \Bigl(\frac{k^2_x}{2m} - \mu \Bigr)  + \tau_1 \Delta^\prime e^{-i\tau_3 \varphi} k_x,
\label{BdG_kx_tau}
\end{eqnarray}
where $k_x = -i\partial_x$ and $\varphi$ is the phase of the order parameter.
It is convenient to make a unitary transformation of the BdG wave function,

\begin{equation}
\Psi(x) \to e^{i\tau_3 \varphi/2} \Psi(x),
\label{Axial}
\end{equation}
bringing the Hamiltonian to the form

\begin{eqnarray}
\mathcal{H} \to e^{-i\tau_3 \varphi/2} \mathcal{H} e^{i\tau_3 \varphi/2} 
=  \tau_3 \Bigl(\frac{k^2_x}{2m} - \mu \Bigr)  + \tau_1 \Delta^\prime k_x.
\label{BdG_kx_trans}
\end{eqnarray}
It maps to a 1D Dirac fermion model with a mass term $\frac{k^2_x}{2m} - \mu$.
Furthermore,  akin to Jackiw - Rebbi model \cite{Jackiw76} the chiral symmetry $\tau_2 \mathcal{H}(x) \tau_2 = - \mathcal{H}(x)$ 
ensures the existence of the MZMs as eigenstates of the chirality matrix $\tau_2$:

\begin{equation}
\tau_2 \Psi(x) = \tau \Psi(x), \qquad \tau_2 \Psi^\mathbb{C}(x) = \tau \Psi^\mathbb{C}(x),
\label{Eigenstates}
\end{equation}
where $\tau=\pm 1$ are the eigenvalues. The eigenstates of $\tau_2$ are self-adjoint [cf. Eq. (\ref{Conjugate})], 
satisfying the equation $\mathcal{H}(x) \Psi(x)= 0$ or, equivalently, 

\begin{equation}
(\partial^2_x +2\mu m - 2\tau_2 \Delta^\prime m \partial_x) \Psi(x) =0.
\label{Eq_MZMs}
\end{equation}
In view of Eq. (\ref{Eigenstates}), this is an ordinary differential equation
with simple solutions in the half space (say, $x \geq 0$) for the boundary condition $\Psi(0) =0$.
The substitution $\Psi(x) \propto e^{-\varkappa x}$ yields two solutions for the decay constant

\begin{equation}
\varkappa_{\pm} = -\tau \Delta^\prime m \pm \sqrt{(\Delta^\prime m)^2 - 2\mu m }.
\label{kappa_pm}
\end{equation}
For $\Delta^\prime m > 0$, the normalizability condition $\varkappa_{\pm} > 0$ is met for $\tau=-1$ in the parameter range 

\begin{equation}
0 < 2\mu m <  (\Delta^\prime m)^2.
\label{Top_regime}
\end{equation}
This defines the topological regime with a single MZM at $x=0$:

\begin{equation}
\Psi_M(x) = \mathcal{N} \Biggl[
\begin{array}{c}
1 \\ -i
\end{array}
\Biggr]
\left(
e^{-\varkappa_+ x} - e^{-\varkappa_- x}
\right),
\label{MZM}
\end{equation}
where $ \mathcal{N}$ is the normalizing factor. 
By the same token, the other eigenstate of $\tau_2$ would correspond to an MZM localized at the opposite end of the TS.

\subsection{Fractional Josephson effect. Phenomenology} 

The fractional JE is caused by the coupling of the MZMs across a weak link between two TSs.
The phenomenology is rather independent of the details of the TSs, and can be illustrated by an effective junction Hamiltonian

\begin{equation}
\mathcal{H}_J
=
\Biggl[
\begin{array}{cc}
\mathcal{H}_R & V^\dagger \\ 
V & \mathcal{H}_L
\end{array}
\Biggr], 
\qquad 
V =
\Biggl[
\begin{array}{cc}
V_0 & 0 \\ 
0 & -V^*_0
\end{array}
\Biggr],
\label{BdG_J}
\end{equation}
where $\mathcal{H}_R$ and $\mathcal{H}_L$ are the Hamiltonians of the right and left TSs, while operator $V$ models the coupling between them. 
$V$ is a diagonal Nambu matrix where $V_0$ is the normal-state coupling (generally complex, which would break time-reversal symmetry).
The unitary transformation (\ref{Axial}) in each TS yields 

\begin{equation}
\mathcal{H}_J
\to
\Biggl[
\begin{array}{cc}
\mathcal{H}_R & V^\dagger e^{-i\tau_3 \phi/2}\\ 
Ve^{i\tau_3 \phi/2} & \mathcal{H}_L
\end{array}
\Biggr], 
\,\,\, 
\phi = \varphi_R - \varphi_L,
\label{BdG_J_trans}
\end{equation}
where the coupling acquires the dependence on the phase difference $\phi$ between the TSs, whereas
the transformed $\mathcal{H}_R$ and $\mathcal{H}_L$ are both phase-independent akin to the Hamiltonian in Eq. (\ref{BdG_kx_trans}).
The energy spectrum of the JJ is obtained from the BdG equations

\begin{eqnarray}
E \Psi_R &=& \mathcal{H}_R \Psi_R + V^\dagger e^{-i\tau_3 \phi/2} \Psi_L,
\label{Eq_R}\\
E \Psi_L &=& V e^{i\tau_3 \phi/2} \Psi_R +  \mathcal{H}_L \Psi_L,
\label{Eq_L}
\end{eqnarray}
assuming the normalization condition $\langle \Psi_{L,R} |  \Psi_{L,R} \rangle =1$.
In the lowest order in $V$, the energy levels can be expressed through the MZMs by putting $\mathcal{H}_R \Psi_R =0$ and $\mathcal{H}_L \Psi_L =0$ 
and projecting the BdG equations on the bra states $\langle \Psi_{R,L} |$. Denoting the solution by $E_+$, we have  

\begin{equation}
E_+ = \langle \Psi_L | V e^{i\tau_3 \phi/2} | \Psi_R \rangle = \langle \Psi_R | V^\dagger e^{-i\tau_3 \phi/2} | \Psi_L \rangle.
\label{E+}
\end{equation}
Since $\langle \Psi_R | V^\dagger e^{-i\tau_3 \phi/2} | \Psi_L \rangle = \langle \Psi_L | Ve^{i\tau_3 \phi/2} | \Psi_R \rangle^*$, 
the second equality in Eq. (\ref{E+}) just means that the solution is real.

The particle-hole symmetry ensures the existence of another solution, $E_-$, irrespective of other symmetries of the system.
Replacing $\Psi_L \to \Psi^\mathbb{C}_L$ and $\Psi_R \to \Psi^\mathbb{C}_R$, we have

\begin{equation}
E_- = \langle \Psi^\mathbb{C}_L | V e^{i\tau_3 \phi/2} | \Psi^\mathbb{C}_R \rangle,
\label{E-C}
\end{equation}
where the matrix element can be evaluated as follows

\begin{eqnarray}
\langle \Psi^\mathbb{C}_L | V e^{i\tau_3 \phi/2} | \Psi^\mathbb{C}_R \rangle 
&=& 
\langle \Psi_L | \mathbb{C}^\dagger V e^{i\tau_3 \phi/2}\mathbb{C} | \Psi_R \rangle 
\label{PH_1}\\
&=& 
\langle \Psi_L | \mathbb{C}^\dagger V \mathbb{C} e^{i\tau_3 \phi/2} | \Psi_R \rangle
\label{PH_2}\\
&=& 
-\langle \Psi_L | V e^{i\tau_3 \phi/2} | \Psi_R \rangle.
\label{PH_3}
\end{eqnarray}
We use the particle-hole symmetry $\mathbb{C}^\dagger V \mathbb{C} = - V$, 
due to which the second level comes with the opposite sign

\begin{equation}
E_- = -\langle \Psi_L | V e^{i\tau_3 \phi/2} | \Psi_R \rangle.
\label{E-}
\end{equation}
Physically, $E_\pm$ are the levels of the Andreev bound states (ABSs) formed by the two hybridized MZMs. 
The corresponding wave functions $\Psi_\pm$ are the linear combinations of the right and left MZMs.
Remarkably, the topological ABSs are $4\pi$ - periodic in the Josephson phase difference $\phi$. 
This is a qualitative distinction from usual JJs where the ABSs are $2\pi$ - periodic \cite{Golubov04}.
For a time-reversal-invariant coupling $V= V_0 \tau_3$, the terms $\propto \sin(\phi/2)$ vanish because of
the orthogonality of the MZMs, so the phase dependence of the energy levels is reduced to 

\begin{figure}[t]
	\begin{center}
		\includegraphics[width=85mm]{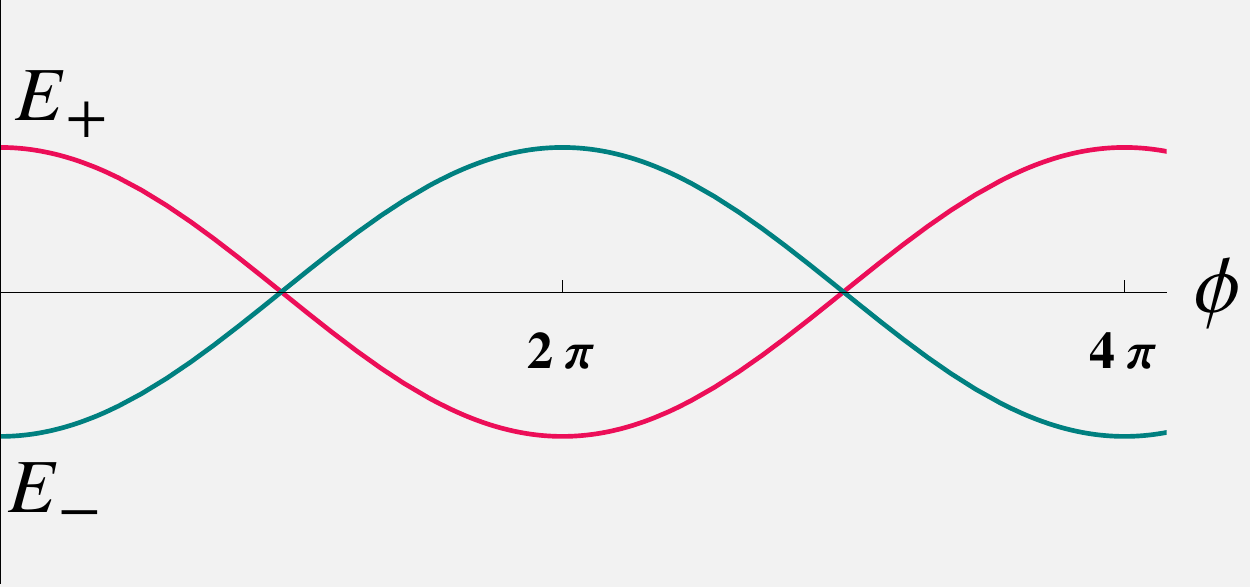}
	\end{center}
	\caption{
		$4\pi$ - periodic ABSs from hybridization of two MZMs [see Eq. (\ref{4pi_ABS})].
	}
	\label{ABS_fig}
\end{figure}

\begin{equation}
E_\pm (\phi) = \pm  \langle \Psi_L | V | \Psi_R \rangle \cos(\phi/2),
\label{4pi_ABS}
\end{equation}
see also Fig. \ref{ABS_fig}.
It is worth noting that microscopic calculations assuming a potential barrier in the JJ \cite{Tanaka96,Kwon04a,Kwon04b,Fu09_JJ} 
give a similar result for the ABS spectrum

\begin{equation}
E_\pm (\phi) = \pm  (\Delta^\prime k_F) \sqrt{D} \cos(\phi/2),
\label{4pi_ABS_micro}
\end{equation}
where $D$ is the barrier transparency, and $k_F$ is the Fermi wave number. 

Two aspects of the topological ABSs merit special attention. First, their $4\pi$ - periodicity harbors 
the topological degeneracy due to the underlying MZMs. 
A phase translation $\phi \to \phi + 2\pi$ brings the JJ into a state with the same energy

\begin{equation}
E_+(\phi + 2\pi) = E_-(\phi),
\label{Degeneracy}
\end{equation}
but with the opposite fermion parity, since the relative sign of $\Psi_R$ and $\Psi_L$ has changed [cf. Eq. (\ref{Axial})], 
which corresponds to switching the occupation number in Eq. (\ref{Occupation}).
Second,  each ABS carries a $4\pi$ - periodic supercurrent.
Since the phase and the particle number are conjugate dynamical variables \cite{Anderson66}, 
a phase dependent coupling energy gives rise to a current flow between the systems.  
In sufficiently short JJs, a major contribution to the current-phase relation (CPR) comes from the ABSs \cite{Golubov04}.
At equilibrium, the CPR $J(\phi)$ is given by the thermodynamic formula

\begin{eqnarray}
J(\phi) = J_+(\phi) + J_-(\phi),
\label{J_sum}\\
J_\pm(\phi)=\frac{e}{\hbar} \frac{\partial E_\pm (\phi)}{\partial \phi} n[E_\pm (\phi)],
\label{J_pm}
\end{eqnarray}
where $J_\pm(\phi)$ are the contributions of the two ABS levels occupied according to the Fermi distribution $n(E)$. 
Using Eq. (\ref{4pi_ABS}) and specially chosen occupations $n[E_+(\phi)] =0$ and $n[E_-(\phi)] =1$, one has 

\begin{eqnarray}
J(\phi)=J_-(\phi)= \frac{e\langle \Psi_L | V | \Psi_R \rangle}{2\hbar} \sin(\phi/2).
\label{4pi_J}
\end{eqnarray}
This example illustrates the fractional JE which is characterized by a subharmonic CPR with the frequency 1/2. 
References \cite{Kwon04a,Kwon04b} proposed that CPR (\ref{4pi_J}) could be observed in voltage-biased JJs where 
the phase difference evolves with time as
$
\phi(t) = 2eUt/\hbar,
$
producing the current 

\begin{eqnarray}
J(t)=\frac{e\langle \Psi_L | V | \Psi_R \rangle}{2\hbar} \sin(eUt/\hbar),
\label{4pi_Jac}
\end{eqnarray}
oscillating at half the usual AC Josephson frequency $2eU/\hbar$ at bias voltage $U$.
That is, in topological JJs the Josephson current is carried by single electrons, rather than by Cooper pairs, 
or, in other words, an MZM is, loosely speaking, half the fermion. 

\subsection{Recent theories of the fractional JE. Non-equilibrium dynamics}

More recent theories have revisited the fractional JE in the context of MZMs in hybrid structures combining conventional superconductors 
with topological insulators or semiconductor nanowires  
\cite{Fu09_JJ,Badiane11,Jiang11,Pikulin12_JETP,Dominguez12,San-Jose12,Pikulin12_PRB,Sau12,GT13_Rev,Beenakker13,Houzet13,Badiane13,Virtanen13,Potter13,Zhang14,Lee14,Shapiro16,Dmytruk16,Peng16,Hui16,Shapiro17,Vinkler-Aviv17,Dominguez17,Pico-Cortes17,Cayao17_TJJ,Trif18,Sun18,Li18_Shapiro,Sticlet18,Feng18,Mohanta18,Chiu19,Mota19,GT19_Soliton,GT19_Chiral}.
Theoretically, the $4\pi$ - periodic CPR (\ref{4pi_J}) has been proposed for ferromagnetic weak links 
in quantum spin-Hall insulator/superconductor structures under assumption of the local fermion-parity conservation \cite{Fu09_JJ}. 
An explicit calculation of the $4\pi$ - periodic CPR from a parity-constraint free energy has been carried out in Ref. \cite{Beenakker13}.
A related fractional JE occurs when the MZMs are spatially separated by a superconducting barrier \cite{Jiang11}.

A number of studies have opted for non-equilibrium dynamics as a more accessible alternative to fixing the fermion parity. 
Reference \cite{Badiane11} has identified signatures of the fractional JE in the finite-frequency current noise 
in a quantum spin-Hall insulator/superconductor structure with a ferromagnetic barrier. 
Particular attention has been paid to JJs between finite-length topological wires \cite{Pikulin12_JETP,Dominguez12,San-Jose12,Pikulin12_PRB,Cayao17_TJJ,Sun18,Feng18}
where the hybridization of the end MZMs opens a gap in the ABS spectrum, rendering it $2\pi$ - periodic \cite{Pikulin12_JETP}. 
The fractional JE is recovered by biasing the JJ and thereby inducing the Landau-Zener transitions \cite{Dominguez12,San-Jose12,Pikulin12_PRB,Sun18,Feng18}.
Several theoretical works have looked at the Shapiro steps in the current-voltage characteristics of dynamically driven JJs 
\cite{Dominguez12,Sau12,Houzet13,Virtanen13,Dominguez17,Pico-Cortes17,Li18_Shapiro}.
When a conventional JJ is exposed to an AC field with frequency $\omega$ 
a DC voltage develops, showing a series of steps at $U_{DC} = n \hbar \omega/2e$, where $n$ is an integer \cite{Shapiro63}. 
For topological JJs, where the current is carried by single electrons, the size of the Shapiro steps is expected to be twice larger  

\begin{equation}
U_{DC} = n \hbar \omega/e = (2n)\hbar \omega/2e,
\label{Shapiro}
\end{equation}
which corresponds to the even steps of conventional JJs.  
That expectation has been tested in the calculations using the resistively and capacitively shunted junction (RCSJ) model \cite{Dominguez12,Dominguez17,Pico-Cortes17} 
supplemented with an appropriate CPR. 
The even steps were reproduced along with some additional features, such as odd and fractional steps, depending 
on the details of the input CPR and the parameter choice. 
Instead of the RCSJ model and adiabatic analysis, Ref. \cite{Li18_Shapiro} has used 
the non-equilibrium Green's functions technique revealing a crossover from conventional Shapiro steps at high frequencies to 
a pattern with the missing odd steps at low frequencies. 

An applied bias leads to a finite lifetime and dynamics of the occupation of the ABS due to its non-adiabatic coupling to the continuum spectrum.
As argued in Refs. \cite{Houzet13,Badiane13}, the $4\pi$ periodicity manifests itself by an even-odd effect in Shapiro steps 
only if the ABS lifetime is longer than the phase adjustment time determined by the environment. 
However, another indicator of the $4\pi$ periodicity, a peak in the current noise spectrum at half the Josephson frequency, was found to be more robust
against the environment. Qualitatively, the predicted noise spectrum is   

\begin{equation}
S(\omega) \propto \frac{seU/\pi\hbar}{ (\omega \mp eU/\hbar)^2 + (seU/\pi\hbar)^2}, 
\label{Noise}
\end{equation}
for $|\omega \mp eU/\hbar| \ll eU/\hbar$. It has peaks at half the usual Josephson frequency $\omega = \pm eU/\hbar$ which manifest the fractional JE in the regime 
when the ABS occupation switches faster than the phase adjustment time (corresponding to a small parameter $s \ll 1$ in the equation above).

\subsection{Equilibrium tests of the fractional JE}

Driving JJs out of equilibrium brings about also unwanted effects that may hinder access to the topological physics.
One of them is Joule overheating. According to Ref. \cite{LeCalvez19}, it may be responsible for higher order odd Shapiro steps 
seen in the experiments (see, e.g., \cite{Rokhinson12,Wiedenmann16,Li18_ABS}), although all odd steps
should be missing in the fractional AC JE. 
Besides, Landau-Zener tunneling between the $2\pi$ - periodic branches of non-topological ABSs
can emulate the fractional JE (see, e.g., \cite{Sau12}). 
In order to rule out such a possibility and avoid heating, it would also be desirable to be able to test the fractional JE at equilibrium, 
ideally when the topological ABSs are decoupled from the continuum.

The difficulty is that the equilibrium CPR (\ref{J_sum}) is $2\pi$ periodic, as the two contributions there 
simply swap upon a $2\pi$ phase advance. 
Although the fractional JE cannot be easily inferred from such equilibrium CPRs, 
they, nevertheless, diagnose unconventional superconductivity in topological materials 
which has become a subject of intense effort on its own 
\cite{Linder10, Yokoyama12, Olund12, GT13_Rev, Snelder13, GT13_ABS, ZhangKH14, Bai14, Khezerlou15_IJMP, Khezerlou15_PhysC, GT15, Dolcini15, Lu15, Snel15, GT15_Book, Zyuzin15, Song16, Zyuzin16, Bobkova16, Choudhari17, GT17_EE, Yu17, Alidoust17, Charpentier17, Virtanen18, Blasi19, Ruoco19, Khezerlou19}.
In order to trace the fractional JE at equilibrium, 
one may look at the effect of an external magnetic field. 
Reference \cite{Potter13} has found an anomalous Fraunhofer pattern due to hybridized Majorana channels \cite{Fu08} at the top and bottom surfaces of a TI film.
Another specific interference effect has been proposed in Ref. \cite{Lee14} 
for two finite-length 1D TSs forming a loop thread by a magnetic flux.
The $4\pi$ periodicity translates into the magnetic-flux dependence with the period $2\Phi_0$, 
where $\Phi_0 = h/2e$ is the magnetic flux quantum. 
However, parity-switching events were found to spoil the $2\Phi_0$ -periodicity of the critical current,
causing instead a behaviour similar to that in $\pi$ - junctions.

\begin{figure}[t]
	\begin{center}
		\includegraphics[width=85mm]{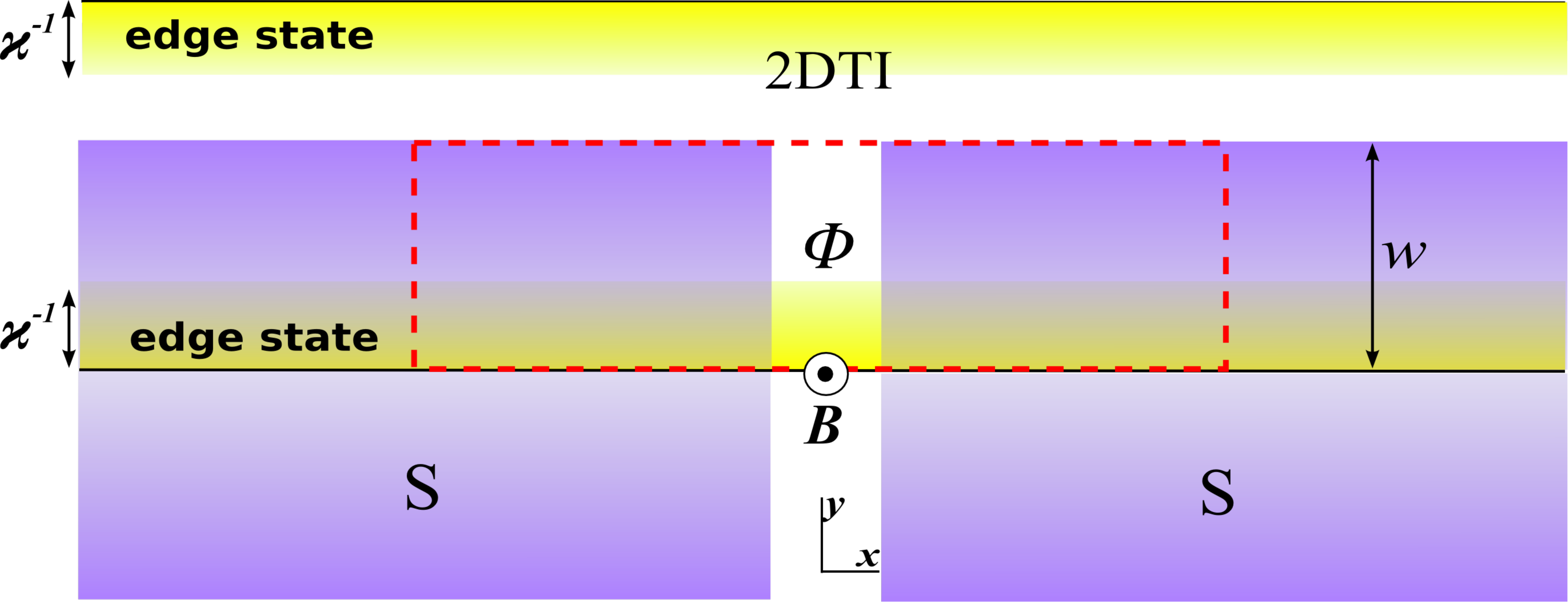}
	\end{center}
	\caption{
		Schematic of a topological JJ created by placing two superconducting films across the edge of a 2DTI.
		The spreading of the edge state into the 2DTI bulk (on length-scale $\varkappa^{-1}$) results in 
		the dependence of the Josephson transport on the magnetic flux $\Phi$ enclosed in the effective junction area 
		(indicated by the dashed contour), which shows the $2\Phi_0$ periodicity; 
		$w$ is the width of the superconducting contact to the 2DTI.
	}
	\label{S2DTI_fig}
\end{figure}

Very often, wire-like TSs are treated as strictly 1D systems with zero width.   
This approximation misses an orbital magnetic-field effect on the wire, thereby overlooking a possible mechanism 
for the $2\Phi_0$ periodicity in topological JJs \cite{GT19_Soliton}. 
Let us, for example, consider a weak link between two superconducting channels at the edge of a 2DTI/superconductor hybrid 
(see Fig. \ref{S2DTI_fig}). It is essential that in real space the edge states are quasi-2D, 
spreading exponentially into the 2DTI bulk. For typical band-structure parameters of the inverted HgTe quantum wells, 
the edge-state spreading can be estimated as $\varkappa^{-1} \sim 10$ nm. 
This finite length-scale makes a topological ABS nonlocal in the sense that it picks up a magnetic flux, $\Phi$, 
enclosed in the effective area of the JJ. Qualitatively, the critical current is given by 

\begin{figure}[t]
	\begin{center}
		\includegraphics[width=85mm]{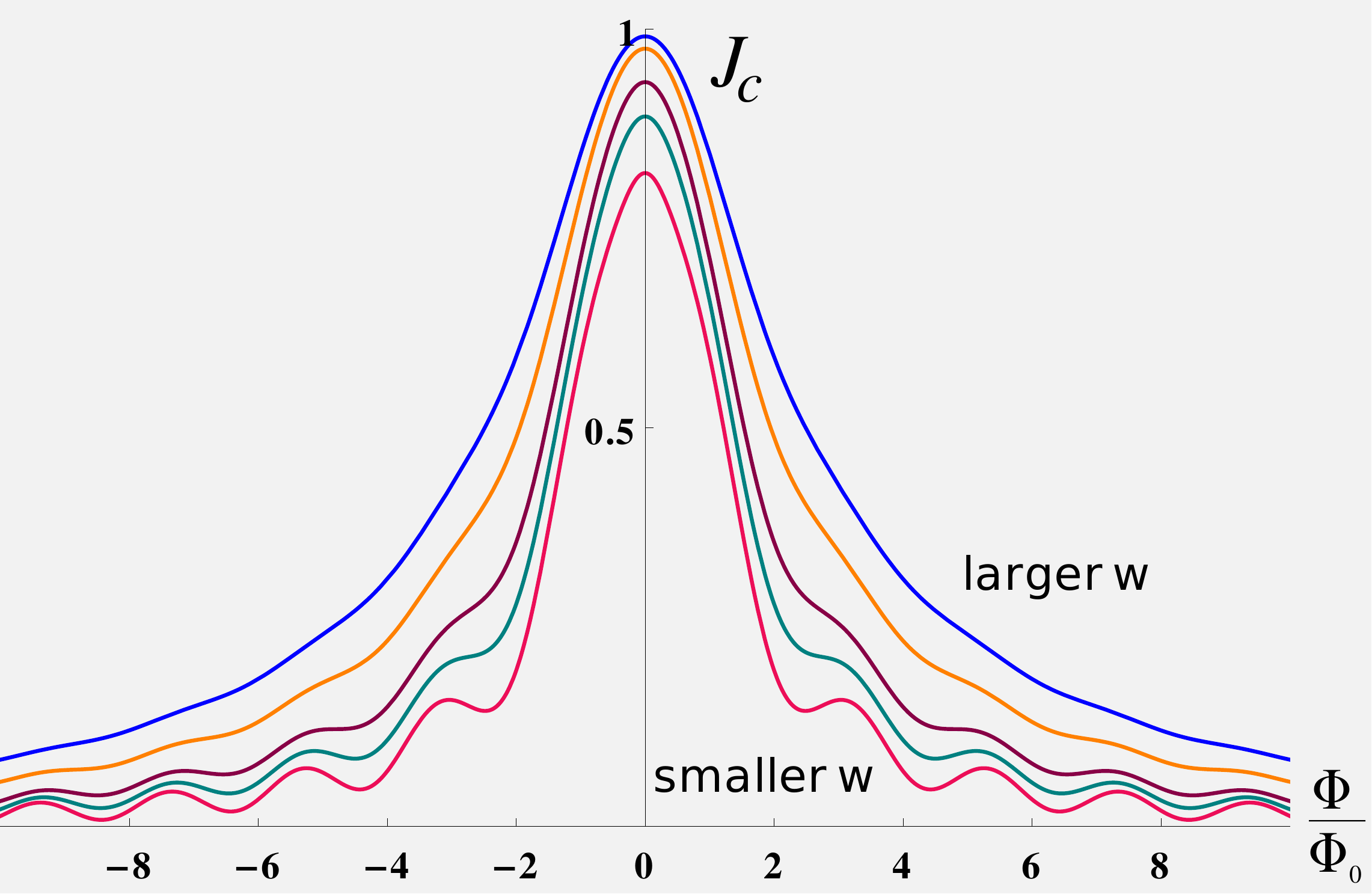}
	\end{center}
	\caption{
		Critical current of topological edge JJs for different values of the contact width $w$ (see also Fig. \ref{S2DTI_fig} and \cite{GT19_Soliton}).
		The topological $2\Phi_0$ - spaced oscillations are clearly visible if $w$ is not too large compared to the edge-state width $\varkappa^{-1}$. 
	}
	\label{Jc_fig}
\end{figure}

\begin{eqnarray}
J_c(\Phi) \approx \frac{e}{2\hbar} 
\left[
\Delta + \Delta_\varkappa \cos\left(\pi\frac{ \Phi}{\Phi_0}  \right)
\right],
\,
\Delta_\varkappa \sim \Delta e^{-\varkappa w},
\label{J_c}
\end{eqnarray}
where $\Delta$ is the proximity-induced $s$ - wave gap, while $\Delta_\varkappa$ accounts for the exponential spreading of the edge state 
underneath the superconducting contact. A more detailed analysis shows \cite{GT19_Soliton} that the $2\Phi_0$ - spaced oscillations of $J_c(\Phi)$ 
occur on top of a monotonic decrease, as depicted in Fig. \ref{Jc_fig}.  
The topological ABS levels show similar oscillations due to the gauge invariance of the $4\pi$ - periodic JE.

A different type of the magnetic-field dependence $J_c(B)$ has been predicted for semiconductor topological JJs \cite{Cayao17_TJJ}.
In that case, the Zeeman effect of the applied field leads to magnetic oscillations of the critical current 
indicating the splitting of the MZMs in finite-length wires.

\begin{figure}[t]
	\begin{center}
		\includegraphics[width=85mm]{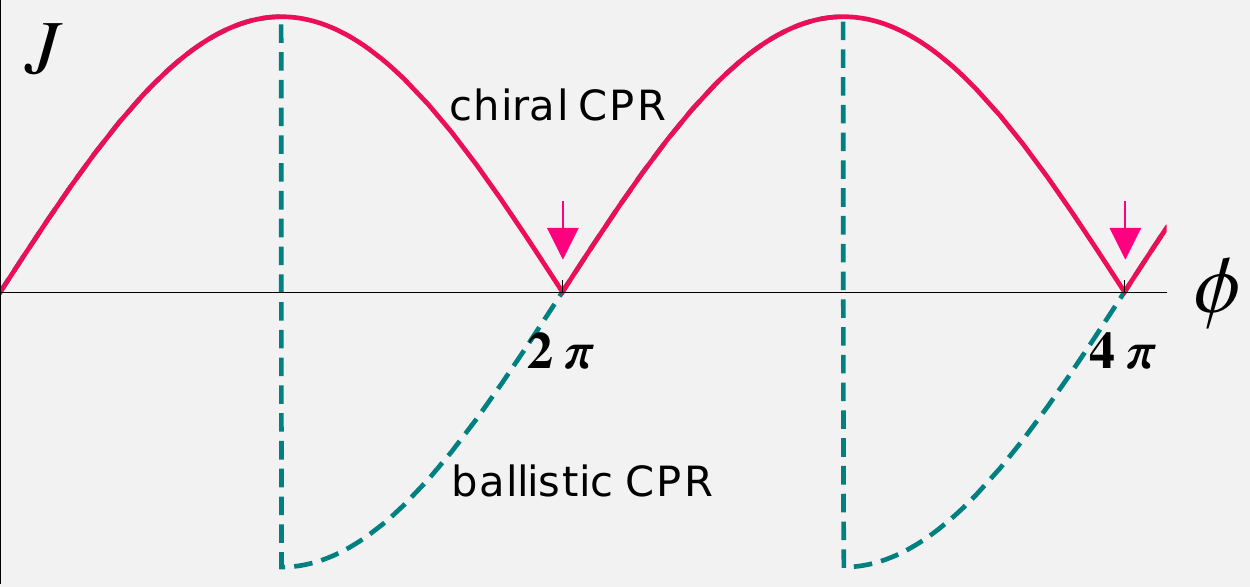}
	\end{center}
	\caption{
		Comparison between the chiral and ballistic CPRs, Eqs. (\ref{Chiral_J}) and (\ref{Ballistic_J}), respectively.
		The arrows indicate the discontinuities of the derivative $J^\prime(\phi)$ caused by 
		the fermion parity switching. 
	}
	\label{CPR_fig}
\end{figure}

An external magnetic field can also modify the shape of the equilibrium CPR $J(\phi)$,
exposing the hidden fractional JE despite the conventional $2\pi$ - periodicity in $\phi$. 
A recent example is the chiral CPR proposed in Ref. \cite{GT19_Chiral} for 2DTI-based JJs.
This is a CPR of the form 

\begin{equation}
J(\phi) =  \frac{e\Delta}{2\hbar} C  \left|\sin \frac{\phi}{2} \right|, \qquad C= \pm 1,
\label{Chiral_J}
\end{equation}
describing a unidirectional supercurrent with the chirality $C$ at $T=0$. 
Precisely speaking, $C$ coincides with the Chern number of the occupied spin band of the 2DTI.  
Noteworthy is a non-analytic phase dependence of Eq. (\ref{Chiral_J}) 
which clearly harbors the $4\pi$ - periodic CPR $J(\phi) \propto \sin(\phi/2)$.
This non-analyticity reflects a discontinuous topological transition 
associated with the change of the ground-state fermion parity and   
is inherent to the fractional JE. In Fig. \ref{CPR_fig}, 
we compare the chiral CPR (\ref{Chiral_J}) with the CPR of a 1D ballistic JJ at $T=0$: 

\begin{equation}
J(\phi) =  \frac{e\Delta}{2\hbar} \sin\frac{\phi}{2} {\rm sgn}\left( \cos\frac{\phi}{2} \right).
\label{Ballistic_J}
\end{equation}
The shape of this CPR is largely independent of the type of the superconductors provided that the JJ is fully transparent 
(cf. Refs. \cite{Tanaka97} and \cite{Kwon04b}). 
The V-shaped minima of the chiral CPR indicate the fermion parity switching at $2\pi, 4\pi, ...$, 
whereas the ballistic CPR is continuous at these points (see Fig. \ref{CPR_fig}).

The above discussion of topological weak superconductivity misses a number of factors that can be operational in realistic topological JJs.  
The recent work \cite{Chiu19} has scrutinized the role of various realistic physical effects, 
such as a finite wire length, gap suppression, non-topological Andreev bound states, or chemical potential variations, 
in Majorana nanowire systems. As argued in \cite{Chiu19}, the system may exhibit $2\pi$ or $4\pi$ JEs or a combination of both, 
without a clear indication of the topological physics or emphasizing only some aspects of it. Only in a rather idealized situation 
(a very long wire with no chemical potential fluctuations or gap suppression) one could establish the $4\pi$ (resp. $2\pi$) oscillations
in the Josephson effect as being reliable evidence for topological (resp. ordinary) superconductivity. 
These issues need to be understood better for the JEs as diagnostics of topological or trivial superconducting states.

\subsection{Beyond the $4\pi$ periodicity}

So war, we have discussed the fractional JE associated with a double ground state degeneracy
leading to a $4\pi$ periodicity due to the underlying MZMs. 
To conclude this section, let us mention an interesting generalization of the $4\pi$ periodic JE 
which comes into play when, in addition to Cooper pairing, other electronic interaction are present. 
Such electronic interactions can cause further fermion fractionalization 
due to induced many-body level splitting in topological JJs. 
The theory \cite{Zhang14} has proposed that electron-electron interactions lead to a fourfold ground state degeneracy 
and, consequently, to a $8\pi$ - periodic JE associated with the weak tunneling of charge $e/2$ quasiparticles.  
A series of theoretical papers \cite{Peng16,Hui16,Vinkler-Aviv17,Sticlet18} has addressed further aspects of electron interactions and "fractional" MZMs 
in topological JJs.

\section{Unconventional superconductivity}
\label{US}

\subsection{Mixed-parity superconducting order parameter. Phenomenology}

Intrinsic noncentrosymmetric superconductors (NCSs) as well as many proximity structures of conventional superconductors and topological materials  
lack a center of inversion symmetry. Such superconducting systems do not fit into 
traditional classification of superconducting states which invokes definite (even or odd) spatial parity of 
the Cooper-pair wave function. Two examples of odd-parity states, the $k_x + ik_y$ - and $k_x$ - TSs, were discussed in the preceding sections.
Both intrinsic NCSs and the mentioned proximity structures (dubbed, for convenience, 'proximity NCSs' here) 
exhibit an antisymmetric SOC which mixes the even-parity (spin-singlet) and the odd-parity (spin-triplet) Cooper pairs, 
producing an unconventional, mixed-parity superconducting order parameter.  

Unambiguous verification of the mixed-parity superconducting order remains one of the outstanding challenges in  
the NCS research \cite{Bauer12,Yip14,Samokhin15,Schnyder15,Smidman17}.
Among intriguing physical consequences of the parity mixing are magnetoelectric effects \cite{Levitov85} manifested 
in the conversion of a charge current into spin magnetization 
and vice versa (see, e.g., \cite{Edelstein95,Yip02,Konschelle15,Tokatly17}), 
the nonuniform (helical) superconducting order \cite{Mineev94,Mineev08} as well as
topological bulk and surface properties (see recent reviews in Refs. \cite{Schnyder15} and \cite{Smidman17}).

\begin{figure}[t]
	\begin{center}
		\includegraphics[width=85mm]{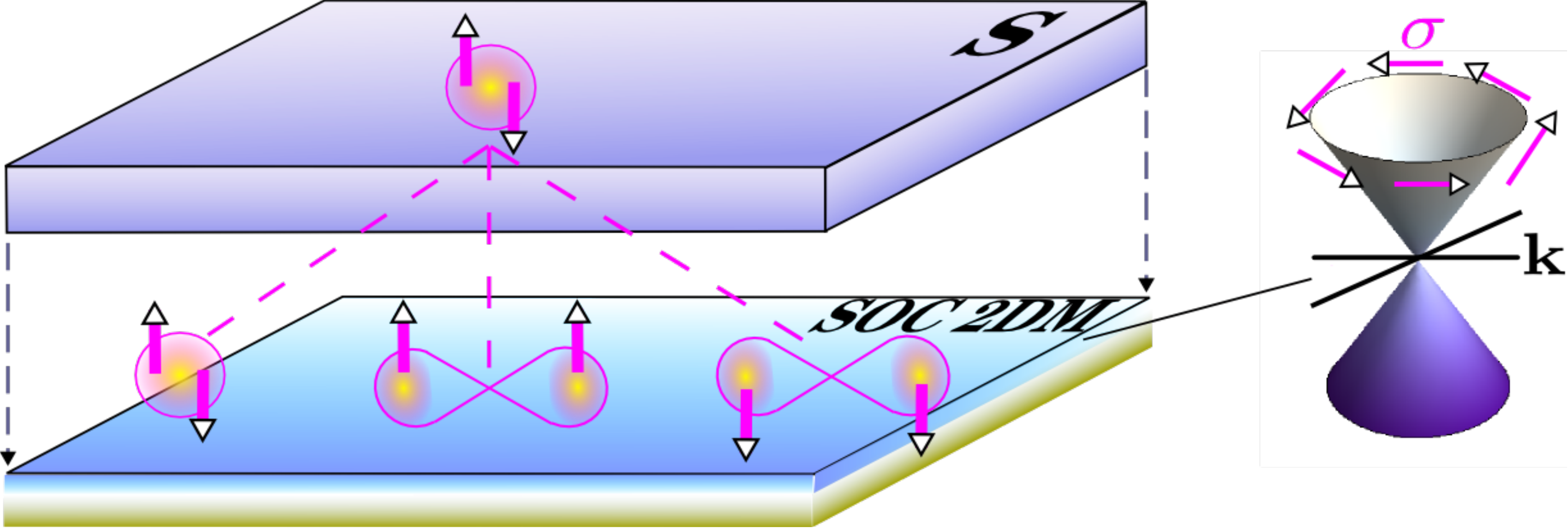}
	\end{center}
	\caption{
		A pictorial representation of a mixed-parity proximity effect in a SOC 2D material (2DM) contacted by a conventional ($s$-wave singlet) superconductor (S). 
		The in-plane spin-momentum-locking facilitates conversion of singlet Cooper pairs into a mixture of singlet and triplet states in the SOC 2DM.
	}
	\label{ProxyNCS_fig}
\end{figure}

For intrinsic NCSs, candidate order parameters can be classified according to their behaviour under the symmetry elements (space group) of the crystal. 
This is discussed extensively in literature (see, e.g., \cite{Sigrist09,Sergienko04,Bauer12,Yip14,Smidman17}). 
We will take a different route and derive the mixed-parity order parameter from a microscopic model for a proximity NCS.
It is assumed that superconductivity is induced in a 2D SOC system by an overlying $s$ - wave singlet superconductor, as depicted in Fig. \ref{ProxyNCS_fig}.
We can, for example, think of the surface of a 3DTI which should exhibit a pronounced parity mixing owing to extraordinary large SOC in these materials. 
In such proximity structures, the SOC forces the electron spins in a tunneling singlet Cooper pair to follow the electron momentum in the plane of the normal system,  
which causes a spin flip, hence an admixture of spin-triplet odd-parity Cooper pairs. Since each spin in a singlet pair can be flipped, 
both up - and down - spin pairs are induced with no net spin magnetization. This phenomenology is behind many microscopic studies of the proximity effect 
in systems with broken spin rotation symmetry, such as Rashba systems and TI materials\cite{Edelstein03,Stanescu10,Potter11,Labadidi11,Ito11,Khaymovich11,Yokoyama12,Virtanen12,Annunziata12,BlackSchaffer12,BlackSchaffer13,GT13_Proxy,Zyuzin14,Snel15,
	Burset15,Crepin15,Reeg15,GT15_Book,Yu16_SOC,Yu17_SOC,Bobkova17,GT17_MAE,Vasenko17a,Vasenko17b,Alidoust17,Hugdal17,Cayao17_Odd,Charpentier17,Lu18,Khezerlou19}.

\subsection{Theory of the mixed-parity proximity effect}
More insight can be gained from the weak-coupling model of the superconducting proximity effect used 
earlier for various low-dimensional systems without SOC (see, e.g., \cite{Volkov95,GT04,GT05,Fagas05,GT07,Rohlfing09,Kopnin11,Kopnin13}) 
and later for TI surface states (see, e.g., \cite{Stanescu10,Yokoyama12,BlackSchaffer13,GT13_Proxy}).
In this model, the proximity of the superconductor is accounted for by a tunneling self-energy $\Sigma_{_\mathcal{T}}$ 
in the equation of motion for the Green's function of the normal system $\mathcal{G}_{\bm k}$:

\begin{equation} 
[E - \mathcal{H}^{(0)}_{\bm k} - \Sigma_{_\mathcal{T}}]\mathcal{G}_{\bm k}= \mathcal{I}, \quad 
\mathcal{H}^{(0)}_{\bf k}=
\Biggl[
\begin{array}{cc}
H_{\bf k} &  0  \\
0 &  -H^*_{-\bf k}
\end{array}
\Biggr],
\label{Eq_G}
\end{equation}
where $\mathcal{H}^{(0)}_{\bf k}$ is the bare Hamiltonian of the normal system in the Nambu representation, 
and $\mathcal{I}$ is the corresponding unit matrix.
The self-energy is a matrix in the Nambu space with the following structure 

\begin{equation}
\Sigma_{_\mathcal{T}} =
\left[
\begin{array}{cc}
-i\Gamma(E)   &  \Delta(E) i\sigma_y   \\
-\Delta^*(E) i\sigma_y    &  -i\Gamma(E)  
\end{array}
\right].
\label{Sigma}
\end{equation}
Its off-diagonal entries yield the induced singlet pair potential, 
while the diagonal elements account for the shift of the spectrum due to the tunneling: 

\begin{eqnarray}
&&
\Gamma(E) = \Gamma_0 g_{_S}(E) = \Gamma_0 \frac{ E }{ \sqrt{ E^2 - \Delta^2_{_S} } }, \quad \Gamma_0 =\pi \mathcal{T}^2 \rho_{_S},
\label{Gamma_N}\\
&&
\Delta(E) = i \Gamma_0 f_{_S}(E) =  i \Gamma_0 \frac{ \Delta_{_S} }{ \sqrt{ E^2 - \Delta^2_{_S} } }. 
\label{Delta_N}
\end{eqnarray}
Here, $g_{_S}(E)$ and $f_{_S}(E)$ are the momentum-integrated quasiparticle and condensate Green functions of the overlying superconductor
which has the gap energy $\Delta_{_S}$. 
The energy $\Gamma_0$ is determined by the single-particle tunneling rate depending on the normal-state density of the states in the 
superconducting metal, $\rho_{_S}$, and the tunneling matrix element $\mathcal{T}$. 
At low energies $E \ll \Delta_{_S}$, in the main approximation $\Gamma = 0$ and 
$\Delta =\Gamma_0$, hence the effective BdG Hamiltonian

\begin{equation}
\mathcal{H}_{\bm k} = \mathcal{H}^{(0)}_{\bf k} + \Sigma_{_\mathcal{T}} =
\Biggl[
\begin{array}{cc}
H_{\bm k} &  \Delta i\sigma_y  \\
-\Delta i\sigma_y  & - H^*_{-{\bm k}}
\end{array}
\Biggr].
\label{BdG_N}
\end{equation}
The normal system is described by a 2D Hamiltonian

\begin{equation} 
H_{\bm k} = {\bm \sigma} \cdot {\bm \gamma}_{\bm k} - \mu, 
\label{H_N}
\end{equation}
with an antisymmetric SOC field ${\bm \gamma}_{\bm k}$.  

\begin{figure}[t]
	\begin{center}
		\includegraphics[width=40mm]{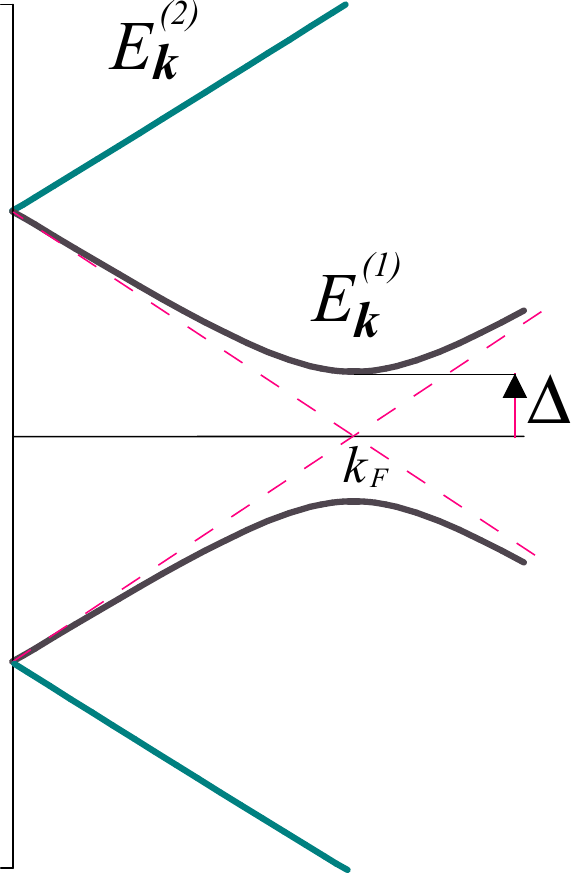}
	\end{center}
	\caption{
		Energy spectrum of a proximity NCS from Eqs. (\ref{E^1}) and Eqs. (\ref{E^2})
	}
	\label{Spectrum_fig}
\end{figure}

The pair potential in the self-energy (\ref{Sigma}) should not be confused with the induced order parameter.
The latter can be characterized by the matrix pair amplitude (the anomalous average), which is a $2\times 2$ spin matrix with the elements

\begin{eqnarray}
\Big\langle
\begin{array}{cc}
c_{\uparrow {\bm k} }(t) c_{\uparrow -{\bm k} }(t) &   
c_{\uparrow {\bm k} }(t) c_{\downarrow -{\bm k} }(t)   \\
c_{\downarrow {\bm k} }(t) c_{\uparrow -{\bm k} }(t) &   
c_{\downarrow {\bm k} }(t) c_{\downarrow -{\bm k} }(t)
\end{array}
\Big\rangle
&=& 
f_0(t,{\bm k}) i\sigma_y
\label{singlet}\\
&+& 
{\bm f}(t,{\bm k}) \cdot {\bm \sigma}i\sigma_y.
\label{triplet}
\end{eqnarray}
Here, the brackets $\langle...\rangle$ denote the ground-state expectation value. 
Also, we use the singlet-triplet basis, with the singlet pair amplitude $f_0(t,{\bm k})$ and the triplet vector

\begin{equation}
{\bm f}(t,{\bm k}) =
\left[ \frac{ f_{\downarrow\downarrow} - f_{\uparrow\uparrow} }{2}, \,\, 
\frac{ f_{\uparrow\uparrow} + f_{\downarrow\downarrow} }{2i}, \,\,
f_{\uparrow\downarrow +\downarrow\uparrow} \right],
\label{f}
\end{equation}
combining the amplitudes $f_{\uparrow\uparrow}(t,{\bm k})$, $f_{\downarrow\downarrow} (t,{\bm k})$, and $f_{\uparrow\downarrow +\downarrow\uparrow}(t,{\bm k})$
of the triplet pair states with the total spin projections $S_z=1,-1$, and $0$. 
In proximity NCSs, the role of the ${\bm f}$ vector is similar to that of the Balian-Wertheimer ${\bm d}$ vector in intrinsic NCSs.

\begin{figure}[t]
	\begin{center}
		\includegraphics[width=85mm]{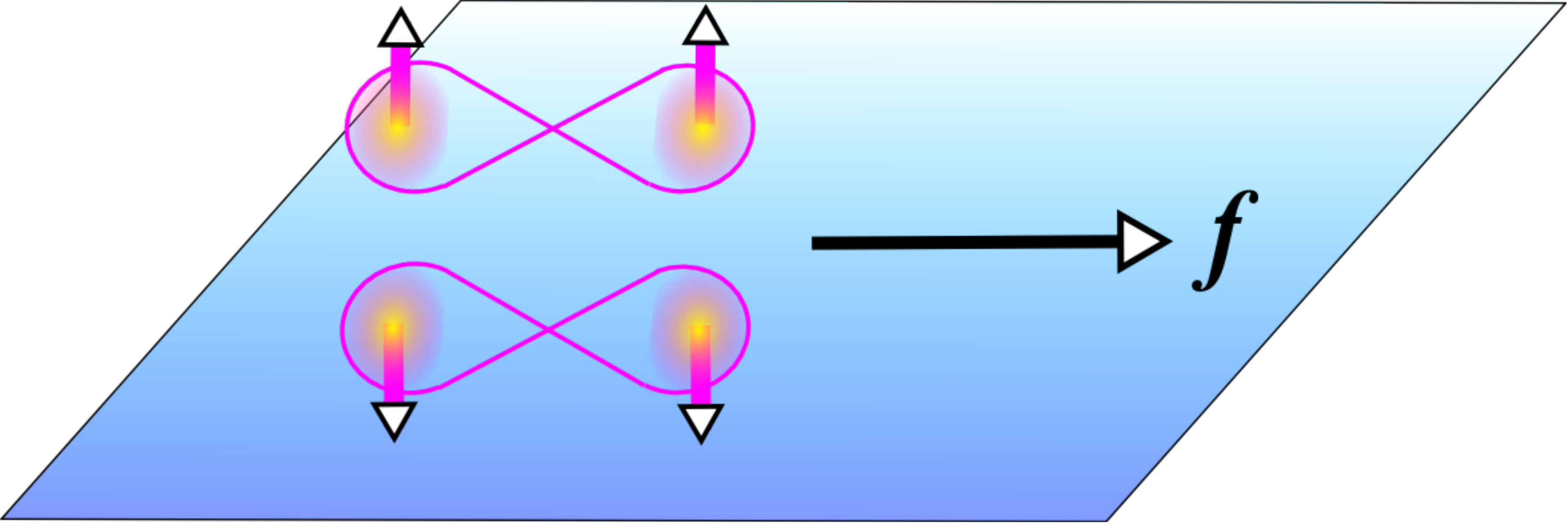}
	\end{center}
	\caption{
		Schematic of the triplet pairing in a 2D proximity NCS.  
		There are two species of opposite-spin pairs in orbital $(k_x \pm ik_y)$ states.
	}
	\label{2ESP_fig}
\end{figure}

All pairing amplitudes can be obtained from the Green function of Eq. (\ref{Eq_G}), which is the Nambu matrix 

\begin{equation}
\mathcal{G}(E,{\bm k})=
\left[
\begin{array}{cc}
G(E,{\bm k})  &  F(E,{\bm k}) \\
F^\dagger(E,{\bm k}) &  \overline{G}(E,{\bm k})
\end{array}
\right].
\label{G_ph}
\end{equation}
Here, each entry is a $2\times 2$ matrix in spin space: 
$G(E,{\bm k})$ and $\overline{G}(E,{\bm k})$ are the quasiparticle Green functions, while 
$F(E,{\bm k})$ is the anomalous (condensate) Green function given by

\begin{equation} 
F(E,{\bm k}) = [f_0(E,{\bm k}) + {\bm f}(E,{\bm k}) \cdot {\bm \sigma}] i\sigma_y,
\label{F}
\end{equation}
where $f_0(E,{\bm k})$  and ${\bm f}(E,{\bm k})$ are the orbital amplitudes for the singlet and triplet pairing at given energy.
From Eqs. (\ref{Eq_G}) and (\ref{BdG_N}) one readily finds (see, e.g., \cite{GT17_MAE})

\begin{eqnarray}
f_0(E, {\bm k}) &=& \frac{ \Delta }{\Pi(E, {\bm k})}
(E^2 - \mu^2 - \Delta^2 - {\bm \gamma}^2_{\bm k} ),
\label{f_s}\\
{\bm f}(E, {\bm k}) &=& -\frac{ 2 \mu \Delta  }{\Pi(E, {\bm k})} \, {\bm \gamma}_{\bm k},
\label{f_t}\\
\Pi(E, {\bm k}) &=& 
\left[
E^2 - ( \mu - |{\bm \gamma}_{\bm k}|^2 - \Delta^2
\right]
\nonumber\\
&\times&
\left[
E^2 - (\mu + |{\bm \gamma}_{\bm k}|)^2 - \Delta^2
\right].
\label{Pi}
\end{eqnarray}
Noteworthy is the information about the energy spectrum and the order parameter of a proximity NCS.
The energy spectrum is given by the roots of $\Pi(E, {\bm k})$ (\ref{Pi}) and consists of two spin-split BCS-like branches 

\begin{eqnarray}
E^{(1)}_{\bm k} &=& \pm \sqrt{ (\mu - |{\bm \gamma}_{\bm k}|)^2 + \Delta^2 },
\label{E^1}\\
E^{(2)}_{\bm k} &=& \pm \sqrt{ (\mu + |{\bm \gamma}_{\bm k}|)^2 + \Delta^2 }.
\label{E^2}
\end{eqnarray}
These are plotted in Fig. \ref{Spectrum_fig}. 
As for the order parameter, it has a mixed parity, with the even singlet $f_0(E, {\bm k})$ (\ref{f_s}) and odd triplet ${\bm f}(E, {\bm k})$ (\ref{f_t}) 
components. The triplet admixture is proportional to the induced singlet pair potential $\Delta$ and the SOC vector  ${\bm \gamma}_{\bm k}$.
That is, the SOC converts some of the tunneling singlet pairs into triplets, as depicted in Fig. \ref{ProxyNCS_fig}. 
For example, for the Rashba SOC with

\begin{equation}
{\bm \gamma}_{\bm k} = \alpha_{\rm so} ({\bm k} \times {\bm z}) = \alpha_{\rm so} [k_y, -k_x, 0]
\label{RSOC}
\end{equation}
the ${\bm f}$ vector lies in the basal plane, describing the equal-spin triplets with the orbital $(k_x \pm ik_y)$ symmetries  
(see also Fig. \ref{2ESP_fig}): 

\begin{equation}
f_{\uparrow\uparrow, \downarrow\downarrow} \propto \alpha_{\rm so} (k_x \mp i k_y), \qquad 
f_{\uparrow\downarrow +\downarrow\uparrow} =0,
\label{Helical}
\end{equation}
where $\alpha_{\rm so}$ is the SOC constant. The $S_z = 0$ triplet is absent by time-reversal symmetry.
A similar triplet admixture occurs in intrinsic NCSs of the tetragonal group \cite{Sigrist09, Tanaka09_NCS,Yip14}. 
We note that Eq. (\ref{f_t}) is not limited to the linear - in - ${\bm k}$ SOC. For example, the theory \cite{Vasenko17b} has discussed 
the role of the hexagonal warping of the Fermi surface, which is relevant for the surface states of tetradymite compounds. 

\begin{figure}[t]
	\begin{center}
		\includegraphics[width=85mm]{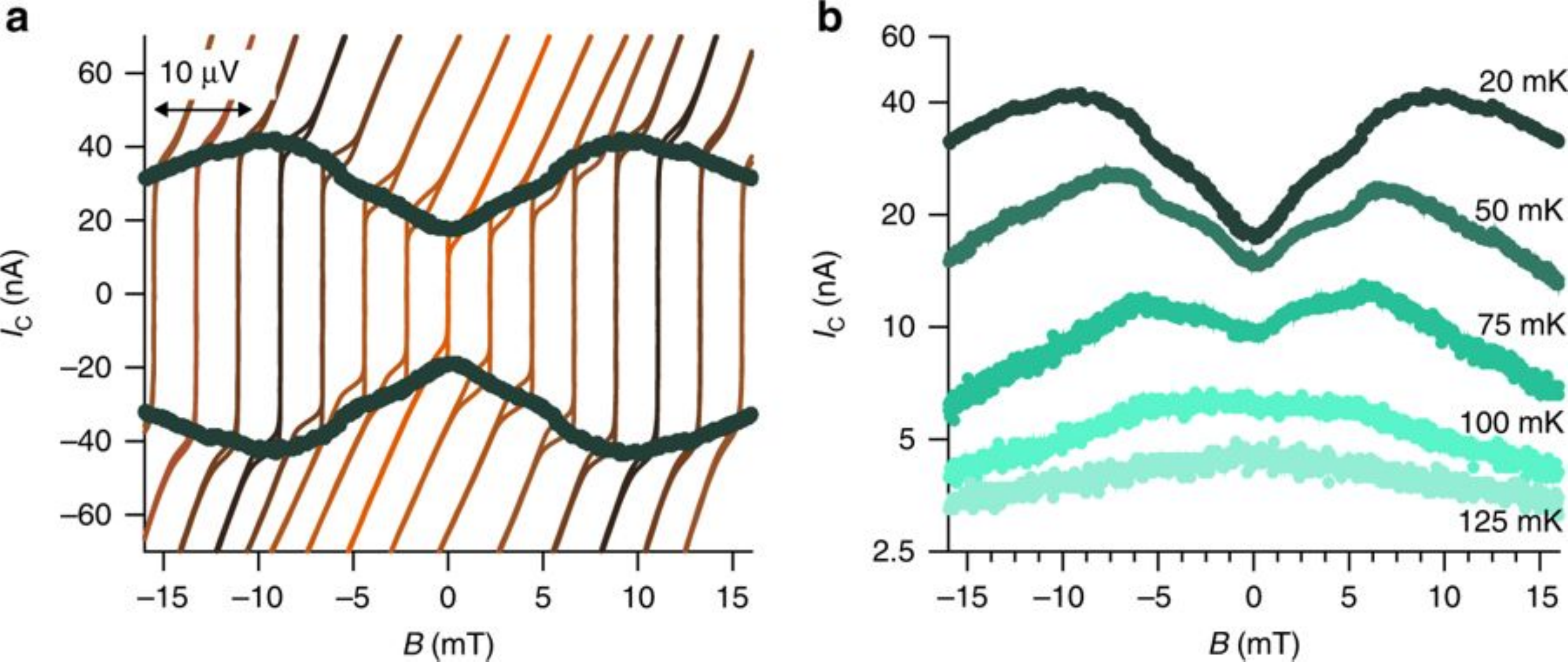}
	\end{center}
	\caption{
		(a) Current-voltage characteristics as a function of magnetic field and temperature for Al/Bi$_2$Te$_3$/Al JJs (from Ref. \cite{Charpentier17}). 
		The curves are shifted in voltage by a value proportional to the magnetic field. The dark points yield the magnetic pattern of the JJ with a pronounced dip at $B=0$.
		(b) Evolution of the magnetic pattern shown in (a) as a function of the temperature. The dip flattens out at a temperature close to 100 mK.
	}
	\label{Charpentier17_fig1}
\end{figure}

The role of disorder deserves separate comment. In dirty TIs, the $p$ - wave component (\ref{f_t}) was found to be suppressed relative to the $s$ - wave pairing  
when the elastic mean-free path was much smaller than the superconducting coherence length \cite{GT13_Proxy}. 
The suppression is due to the generic nonlocality of the odd-parity Cooper pairs, which makes them sensitive to the electron mean-free path in a disordered system. 
In cleaner TIs, however, the $p$ -wave component can be comparable to the $s$ - wave one and should therefore be observable despite the presence of a modest
amount of disorder (e.g., random impurity potential). Reference \cite{Hugdal17} has developed the quasiclassical theory for the proximity effect in impure Dirac materials.
Non-equilibrium Eilenberger and Usadel equations were derived to first order in quantities small compared to the Fermi energy for Dirac edge and surface states 
with spin-momentum locking.

\begin{figure}[t]
	\begin{center}
		\includegraphics[width=85mm]{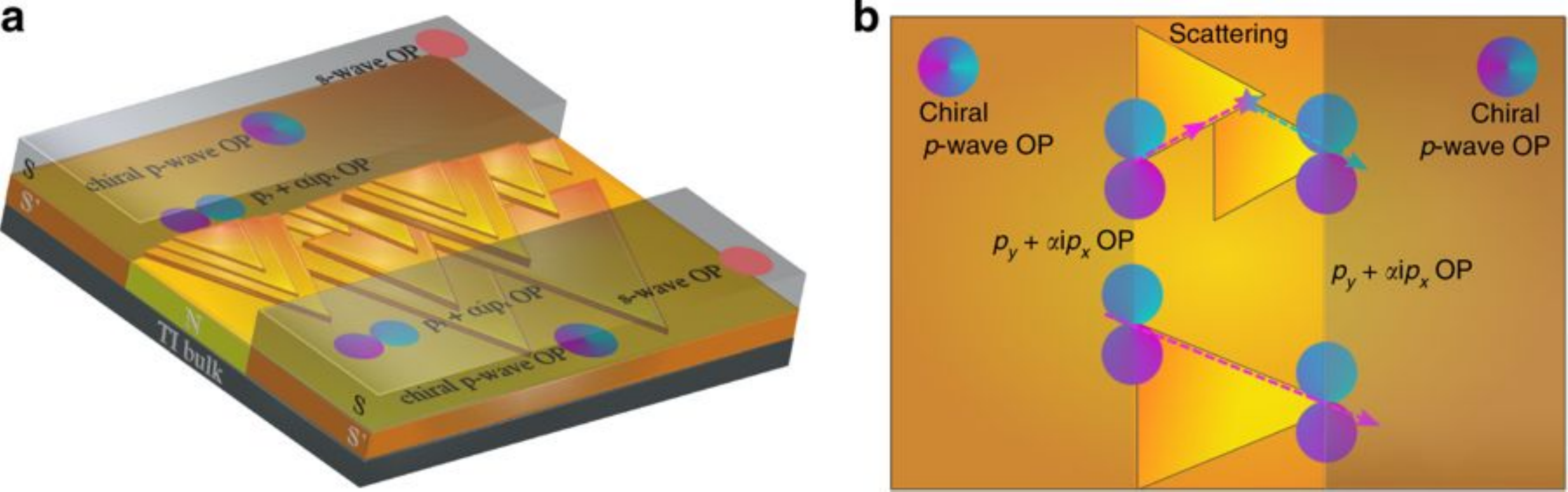}
	\end{center}
	\caption{
		Sketch of a JJ at the surface of Bi$_2$Te$_3$ to probe induced mixed-parity superconductivity (from Ref. \cite{Charpentier17}).   
		(a) The S electrodes (Al) induce a mixed $s + p$ - wave superconductivity at the surface of Bi$_2$Te$_3$ (only the $(p_x + ip_y)$ - component is shown). 
		In close proximity to low transparency interfaces, the $p_x + ip_y$ symmetry changes to a $p_y$ one. 
		(b) Top view of the device illustrating $\pi$ coupling. In the presence of scattering, a quasi-particle trajectory emerging from the negative $p_y$ - lobe 
		on one side of the JJ (blue arrow) couples to a trajectory associated with the positive $p_y$ - lobe on the other side of the JJ (red arrow). 
		In the case of a scattering-free transport, the quasi-particle trajectories probe the same phase in both electrodes.
		For more details, see Ref. \cite{Charpentier17}.
	}
	\label{Charpentier17_fig2}
\end{figure}

The experiment \cite{Charpentier17} has reported an observation of the induced unconventional superconductivity at the surface of Bi$_2$Te$_3$ 
in phase-sensitive measurements on nanoscale JJs. The magnetic field pattern of the junctions was found to have a dip at zero applied magnetic field 
(see Fig. \ref{Charpentier17_fig1}), presumably, due to the simultaneous existence of the $0$ and $\pi$ couplings across the junction 
provided by a mixed $s + p$ - wave order parameter. The $\pi$ coupling was attributed to the combined effect of a sign-changing $p$ - component of the order parameter 
and scattering in the JJ (see also Fig. \ref{Charpentier17_fig2}). 

\subsection{Odd-frequency triplet superconductivity}

Breaking the time-reversal symmetry enriches the unconventional proximity effect in topological materials.
An informative generalization of the above model is achieved by adding an exchange (or Zeeman) spin field ${\bm h}$: 

\begin{equation}
{\bm \gamma}^\prime_{\bm k} = {\bm \gamma}_{\bm k} + {\bm h}.
\label{SOC+Z}
\end{equation}
The total spin field ${\bm \gamma}^\prime_{\bm k}$ is no longer antisymmetric in momentum:

\begin{equation}
{\bm \gamma}^\prime_{-{\bm k}} \not= {\bm \gamma}^\prime_{\bm k},
\nonumber
\end{equation}
which leads to the following generalization of the triplet ${\bm f}$ vector in Eq. (\ref{f_t}):

\begin{eqnarray}
{\bm f}(E,{\bm k}) = \frac{\Delta}{\Pi(E,{\bm k})} 
[
-\mu ( {\bm \gamma}^\prime_{ {\bm k} } - {\bm \gamma}^\prime_{-{\bm k}} ) 
&+& E ({\bm \gamma}^\prime_{ {\bm k} } + {\bm \gamma}^\prime_{-{\bm k} })
\nonumber\\
& + & i {\bm \gamma}^\prime_{ {\bm k} } \times {\bm \gamma}^\prime_{-{\bm k}}
],
\label{f_h_1}
\end{eqnarray}
or, explicitly,

\begin{eqnarray}
{\bm f}(E,{\bm k}) =  \frac{2\Delta}{\Pi(E,{\bm k})} 
[ -\mu {\bm \gamma}_{\bm k} + E {\bm h} + i{\bm \gamma}_{\bm k} \times {\bm h} ].
\label{f_h_2}
\end{eqnarray}
We have now two new contributions. The term linear in energy $E$ corresponds 
to the odd-frequency triplet pairing found a while ago in proximity structures of 
ferromagnets and $s$-wave superconductors \cite{Bergeret01,Bergeret05,Bergeret13}. 
Such odd-frequency Cooper pairs have the $s$-wave spatial symmetry and are robust to disorder and other spatial inhomogeneities,
thus offering both new interesting physics and application potential \cite{Eschrig15,Linder15}. 

The triplet component (\ref{f_h_2}) was calculated in Ref. \cite{Yokoyama12} for a TI surface state with the linear Dirac spectrum.
A growing body of work has been dealing with different aspects of the odd-frequency pairing in hybrid structures involving 3DTIs 
\cite{BlackSchaffer12,BlackSchaffer13,Burset15,Vasenko17b,Lu18}, 2DTI \cite{Crepin15,Cayao17_Odd}
and related Rashba materials \cite{Reeg15,Bobkova17}. 
In particular, Ref. \cite{Cayao17_Odd} has looked into the emergence of the odd-frequency $s$-wave pairing at the edge of a 2DTI without any magnetism.

The odd-frequency proximity effects have been studied for superconductors of various symmetry classes 
\cite{Tanaka07,Yokoyama07}. Also, the connection to the Majorana modes has been pointed out (see, e.g., Refs. \cite{Asano13,Tamura19}).
Various other findings have been summarized in the review articles \cite{Bergeret05,Tanaka12_Rev,Linder17_Odd}.
While the literature on the odd-frequency superconductivity is abundant, 
the emergence and physical consequences of the imaginary term in Eqs. (\ref{f_h_1}) and (\ref{f_h_2}) have gone largely unnoticed.
This type of pairing is an analogue of the paradigmatic nonunitary pairing in triplet superfluids \cite{Leggett75} and superconductors \cite{Sigrist91} with a complex triplet order parameter. We elaborate on this point  below.

\subsection{Nonunitary triplet pairing and charge-spin conversion}

As mentioned above, the standard classification of the Cooper pairing which invokes the Balian-Wertheimer ${\bm d}$ vector 
is not applicable to proximity NCSs where no pairing interaction takes place. For that purpose, 
we employ the matrix condensate Green function  $F(E,{\bm k})$ (\ref{F}) which has proved useful in
diverse proximity structures \cite{Bergeret05,Buzdin05,Annunziata12,Fritsch14,Fritsch15} and driven superconductors \cite{Triola16}. 

Following Ref. \cite{Leggett75}, we call the pairing nonunitary if the product $\hat{F}\hat{F}^\dagger$ is not proportional to a unit spin matrix. 
Using Eq. (\ref{F}), we find
\begin{equation}
\hat{F}\hat{F}^\dagger = (|f_0|^2 + {\bm f} \cdot {\bm f}^* ) \hat{1} + (f^*_0 {\bm f} + f_0 {\bm f}^*) \cdot {\bm \sigma} + (i{\bm f} \times {\bm f}^*) \cdot {\bm \sigma},
\label{FF}
\end{equation}
where $\hat{1}$ stands for the unit spin matrix (the arguments $E$ and ${\bm k}$ are suppressed for brevity). 
The second and third terms above indicate the nonunitary pairing due to the lack of inversion and time-reversal, respectively. 
We are interested in the latter case where, by analogy with triplet superfluids, Cooper pairs have a net spin polarization:

\begin{equation}
\langle {\bm S} \rangle \propto i{\bm f} \times {\bm f}^*.
\label{S_CSP}
\end{equation}
For example, for a 2D NCS in a perpendicular spin field ${\bm h}$, Eq. (\ref{f_h_2}) yields the following result for the axial vector $i{\bm f}\times{\bm f}^*$
at the Fermi level:

\begin{eqnarray}
i{\bm f}(0,{\bm k}) \times  {\bm f}^*(0,{\bm k}) 
=  \frac{ 8 \mu \Delta^2 {\bm \gamma}^2_{\bm k} }{\Pi^2(0,{\bm k})} \, {\bm h}.
\label{CSP_h}
\end{eqnarray}
As expected, the pair spin polarization is parallel to the spin field ${\bm h}$, indicating an imbalance between the equal-spin triplets 
$\uparrow\uparrow$ and $\downarrow\downarrow$ (see also Fig. \ref{2ESP_fig}).

\begin{figure}[t]
	\includegraphics[width=85mm]{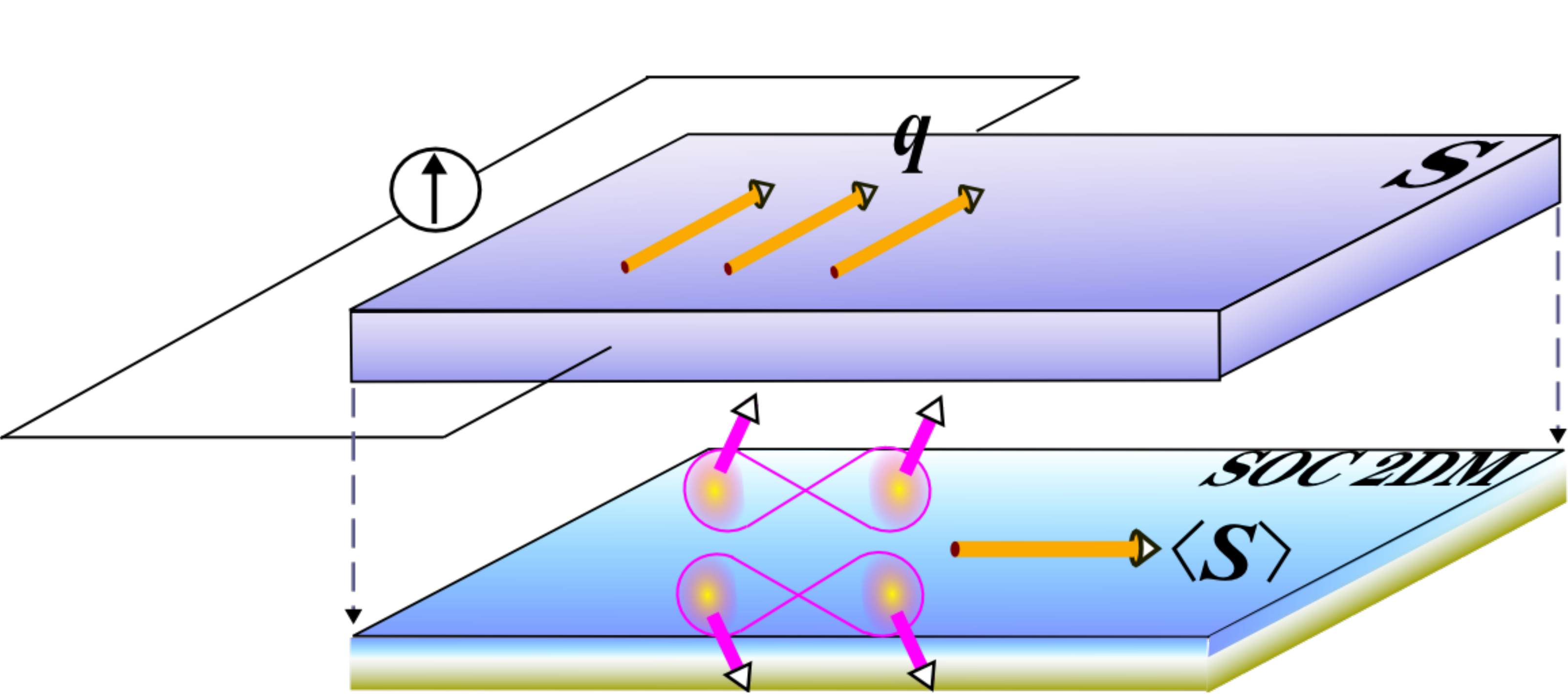}
	\caption{
		Sketch of charge-spin conversion in a SOC 2DM proximitized by a current-biased conventional superconductor.
		An applied electric current generates an in-plane spin polarization $\langle {\bm S} \rangle \propto i{\bm f} \times {\bm f}^*$ 
		reflecting the triplet pairing with the total spin projection $S_z=0$ on the SOC plane.
		This can be interpreted as tilting the pair spins  $\uparrow$ and $\downarrow$ such that they acquire a common in-plane projection 
		[see also Eqs. (\ref{f_q}) and (\ref{CSP_q})]. 
	}
	\label{Conversion_fig}
\end{figure}

In the above example, the ${\bm f}$ vector formalism allows us to extend the notion of the nonunitary pairing 
beyond its original context \cite{Leggett75}, viz. to treat proximiy-induced superconductivity. 
Furthermore, the nonunitary pairing does not generally require the spin field ${\bm h}$. 
The pair spin polarization $i{\bm f} \times {\bm f}^*$ can be induced just 
by an electric current via charge-spin conversion \cite{GT17_MAE,GT18_INUP}. 
To illustrate this point let us consider the BdG Hamiltonian

\begin{eqnarray}
\mathcal{H}_{\bm k}=
\left[
\begin{array}{cc}
{\bm \sigma} \cdot {\bm \gamma}_{ {\bm k} + {\bm q} }  - \mu & \Delta i\sigma_y \\
-\Delta i\sigma_y & - ({\bm \sigma} \cdot {\bm \gamma}_{-{\bm k} + {\bm q}} - \mu)^*
\end{array}
\right],
\nonumber
\end{eqnarray}
where the wave-vector shift ${\bm q}$ accounts for the presence of a superconducting phase gradient 
associated with a dissipationless electric current. The current is applied to the overlying superconductor 
and is weak enough to disregard the depairing effects in $\Delta$. 
The corresponding triplet ${\bm f}$ vector is given by \cite{GT17_MAE}

\begin{eqnarray}
{\bm f}(E,{\bm k}) = \frac{ 2\Delta }{ \Pi(E,{\bm k}) } 
[
- \mu {\bm \gamma}_{\bm k}  + E {\bm \gamma}_ {\bm q}  + i {\bm \gamma}_{\bm k} \times {\bm \gamma}_{\bm q}
],
\label{f_q}
\end{eqnarray}
where we used ${\bm \gamma}_{ \pm {\bm k} + {\bm q} } = \pm {\bm \gamma}_{\bm k} + {\bm \gamma}_{\bm q}$ for any linear SOC.
The combined effect of the SOC and the supercurrent produces a Zeeman-like field ${\bm \gamma}_ {\bm q}$, 
thereby generating both odd-frequency and nonunitary pairing akin to ${\bm h}$. 
Generally, the direction of ${\bm \gamma}_ {\bm q}$ depends on the type of the structural or lattice asymmetry behind the SOC,
so does the pair spin polarization

\begin{equation}
i {\bm f}(0,{\bm k}) \times {\bm f}^*(0,{\bm k}) = -
\frac{8 \mu \Delta^2 }{\Pi^2(0,{\bm k})}
{\bm \gamma}_{\bm k} \times ({\bm \gamma}_{\bm k} \times {\bm \gamma}_{\bm q}).
\label{CSP_q}
\end{equation}
For the Rashba SOC, the spin polarization is carried by the triplet state with the total spin projection $S_z = 0$ on the SOC plane. 
Loosely speaking, the supercurrent tilts the pair spins $\uparrow$ and $\downarrow$ such that they acquire a common in-plane projection, 
as sketched in Fig. \ref{Conversion_fig}. Importantly, the spin polarization does not vanish upon averaging 
over the directions of the wave vector ${\bm k}$. Using Eq. (\ref{RSOC}), one finds

\begin{figure}[t]
	\begin{center}
		\includegraphics[width=60mm]{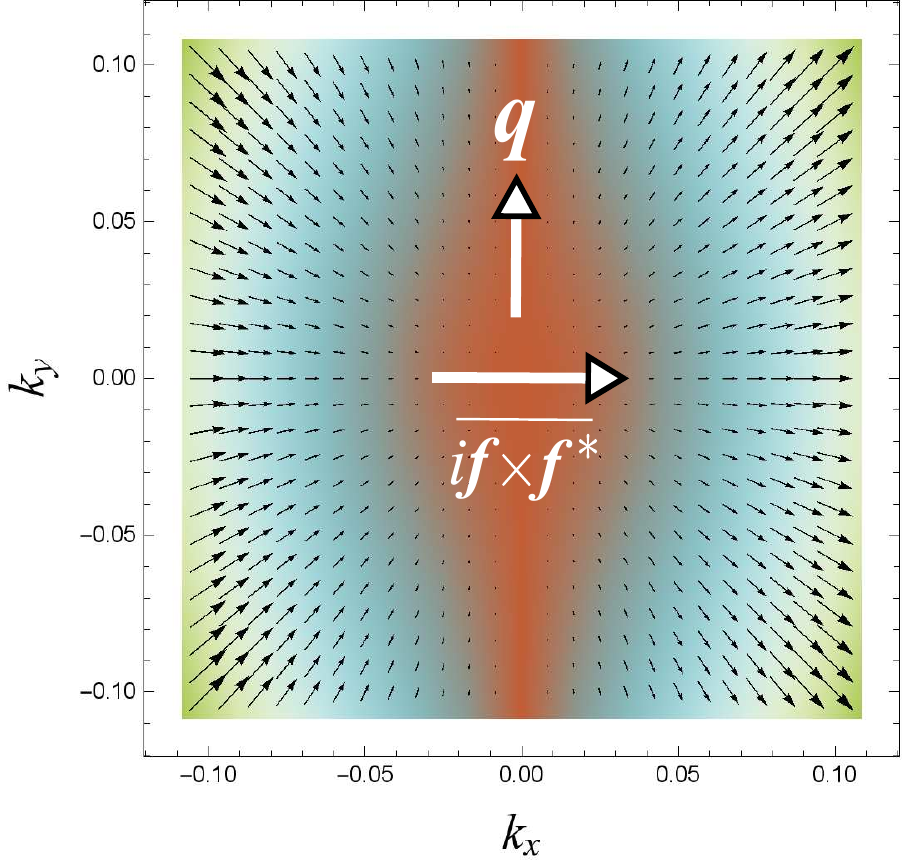}
	\end{center}
	\caption{
		Vector plot of the pair spin polarization (\ref{CSP_q}) for the Rashba SOC.  
		Vector $\overline{i {\bm f}\times {\bm f}^*}$ shows the average polarization (\ref{CSP_q_av}).
	}
	\label{CSP_fig}
\end{figure}

\begin{eqnarray}
\overline{i {\bm f}(0,{\bm k}) \times {\bm f}^*(0,{\bm k})} =
\frac{4 \mu \Delta^2 \alpha^3_{\rm so} {\bm k}^2}{\Pi^2(0,{\bm k})}\, ({\bm q} \times {\bm z}),
\label{CSP_q_av}
\end{eqnarray}
where the bar means the average value. 
This result just means that an unpolarized charge current is converted into spin magnetization of the superconducting condensate,
a form of the magnetoelectric effect pioneered in normal metals \cite{DP71} and studied later in NCSs \cite{Edelstein95}. 
Here, the magnetoelectric effect refers to the spin magnetization induced by a phase gradient of the order parameter, 
while in the inverse magnetoelectric effect the magnetic polarization causes charge and spin flows in a variety of situations
\cite{Yip02,Krive04,Buzdin08,Malshukov10,BlackSchaffer11,Yokoyama13,Yokoyama14,Alidoust15,Mironov15,Dolcini15,Malshukov17,Amundsen17}.  
In both effects, the magnetoelectric coupling is characterized by the SOC constant $\alpha_{\rm so}$.

Equation (\ref{CSP_q_av}) resembles the current-induced thermodynamic magnetization of a Rashba NCS \cite{Edelstein95,Yip02,Konschelle15,Tokatly17}. 
The averaged spin polarization retains the dependence on the structural or lattice asymmetry.
For the Rashba SOC, the average polarization direction is perpendicular to an applied supercurrent in the SOC plane (see also Fig. \ref{CSP_fig}). 
For NCSs of the cubic crystal group we expect a different result. In this case, the SOC vector is simply parallel to the momentum,
$
{\bm \gamma}_{\bm k} = \alpha_{\rm so} {\bm k} = \alpha_{\rm so} [k_x, k_y, k_z]
$,
and Eq. (\ref{CSP_q}) yields the following result

\begin{eqnarray}
\overline{i {\bm f}(0,{\bm k}) \times {\bm f}^*(0,{\bm k})} =
\frac{16 \mu \Delta^2 \alpha^3_{\rm so} {\bm k}^2}{3\Pi^2(0,{\bm k})}\, {\bm q}.
\label{CSP_q_av_cub}
\end{eqnarray}
As we see, in cubic NCSs the pair spin polarization is locked parallel to the applied current.

The charge-spin conversion is an indicator of the unconventional, mixed-parity order parameter. 
Still, the direct magnetoelectric effect has not been verified experimentally yet despite a diverse range of other observed properties 
\footnote{A review article \cite{Smidman17} gives a detailed account of the ongoing theoretical and experimental studies of intrinsic NCSs.}.
On the other hand, a growing body of theoretical predictions may help in planning a decisive experiment. 
Some specific predictions include spin Hall effects and nonequilibrium spin accumulation in superconducting structures 
\cite{Malshukov08,Malshukov11,Bergeret16,Bobkova16}, 
electrically controllable spin filtering in TI surfaces states \cite{Bobkova17}, 
equal-spin Andreev reflection due to the induced nonunitary pairing \cite{GT17_MAE,GT18_INUP},
magnetoelectric $0 - \pi$ transitions in quantum spin Hall insulators \cite{GT17_EE}, 
the generation of a transverse spin supercurrent by a charge supercurrent \cite{Linder17}, 
and the long-range effect of a Zeeman field on the electric current in an Andreev interferometer \cite{Malshukov18}.

\section{Outlook}

We give a brief overview of future directions. The contrast between quantum anomalous Hall research, which focuses on dissipationless transport at very low temperatures, and spin-orbit torque devices, which are aiming for room-temperature operation, is noteworthy. As a result of the latter several gaps remain in our understanding of spin-orbit torques at very low temperatures, for example the role of quantum interference effects such as weak localisation and anti-localisation. The exact origin of the anti-damping torque continues to be hotly debated, with possibilities including the spin Hall effect, the Berry curvature anomalous Hall term, which may however be overwhelmed by disorder, and spin-orbit scattering mechanisms that have not been fully explained. A related question, to date unsettled, is whether it is possible for an anti-damping torque to exist without the spin-Hall effect. in simple analytical models, the final result to this question is very sensitive to the starting Hamiltonian. 

Furthermore, whereas the bulk of attention has focused on ferromagnets, antiferromagnets are also generating excitement \cite{Ghosh_PRB2017, TAFM_NP2018} and will no doubt witness considerable growth. Traditional problems from magnetism, such as current-driven domain wall motion, have yet to take off in topological materials. Spin-momentum locking offers new functionalities for magneto-resistive devices such as spin valves, since electrons travelling in a specific direction have a fixed spin orientation determined by their momentum. Experimentally, there is a lot of space to investigate van der Waals heterostructures where experiments are just beginning \cite{Novoselov_vdW_Sci2016}.

A fundamental gap in the theoretical approach to spin-orbit torques as well as non-linear response is the method used to handle the spin current. It is well known that in spin-orbit coupled systems the proper definition of the spin current is not $(1/2) \{ \hat{s}_i, \hat{v}_j \}$, where $\hat{\bm s}$ and $\hat{\bm v}$ represent the spin and velocity operators respectively, as that current is not conserved, but $(1/2) (d/dt) \{ \hat{s}_i, \hat{r}_j \}$, with $\hat{\bm r}$ the position operator. This is motivated by spin non-conservation in the presence of spin-orbit coupling (an enlightening strategy for circumventing these ambiguities is described in \cite{Tatara_PRB2018}). Whereas the position operator is a difficult quantity to handle, in particular in bases of Bloch states, in which most such calculations are attempted, a complete understanding of spin-orbit torques in topological materials will remain elusive until the magnitude of the proper spin current is determined.

Important unanswered questions in anomalous Hall transport include the role of spin-charge correlations, which has only recently begun to receive attention \cite{Aydin_2019}. Moreover, according to a number of experiments, the sign of anomalous Hall conductivity can be the same \cite{Zhang1582, Zhang2014, Chang_adv_mat} as or differ \cite{Checkelsky2012,Zhang1582,ChangPRL2014,LeePRB2014,Liu2018} from that of the intrinsic contribution~\cite{QAHE_exp1,QAHE_exp2,QAHE_exp3,experimental_AHE}, depending on the magnetic doping concentration. This puzzling observation is thus far unexplained. 
Likewise, the role of disorder in the anomalous Hall contribution due to Fermi arcs in WSMs has not been elucidated. The anomalous Hall effect is being explored in Dirac systems with spin, pseudospin and valley degrees of freedom \cite{Offidani_PRL2018}, whose behaviour is qualitatively different from TIs. With the possibility of achieving strong spin-orbit coupling in graphene, one can envisage further research on this topic. In skyrmions studies it is assumed the skyrmion texture remains unaffected by the electrons at the interface, an assumption that remains to be verified by further research. Finally, anomalous Hall transport affects Coulomb drag \cite{Hong_AHE_Drag_PRB2017}, which must be explored further. 

The understanding of interband coherence effects on the non-linear optical response is in its infancy. This applies both to interband transitions induced by Berry curvature terms as well as the role of scattering, whether by disorder, phonons or magnons, and the examination of effects known to be important in transport such as skew scattering, side-jump, localisation and Kondo physics. To add to this, the vast majority of work on optical systems has focused on the case of an undoped conduction band, while potential Fermi surface effects in doped systems have not been explored. 

Another exciting avenue for future research is the exploration of chiral superconductivity, non-Abelian excitations and unconventional Cooper pairing in topological materials. These topics remain a subject of intense experimental and theoretical effort.
The experimental quest for chiral superconductivity in both intrinsic and proximity-induced superconductors continues.
Speaking of hybrid proximity structures, we have seen that the chiral superconductivity can be understood as the duality between a $k_x + ik_y$ - superconductor 
and a Chern insulator. Is there more new physics beyond this intricate duality? The answer to this question is not only of theoretical interest, but may also uncover 
hitherto unexplored routes towards topological quantum computation and superconducting spintronics.

\ack

DC would like to thank Branislav Nikoli\'c for a series of instructive discussions and to acknowledge the stimulating input of Di Xiao, Shulei Zhang, Giovanni Vignale, Allan MacDonald, and Oleg Tretiakov. YL thanks N. Liu, X. F. Niu and Z. C. Wang for technical assistance, and X. Dai, Z. Fang, K. He, H. Z. Lu, A. D. Mirlin, S. Q. Shen, J. R. Shi, H. M. Weng, P. Xiong and X. C. Xie for valuable discussions. GT thanks Ulrich Eckern, Sebastian Bergeret, Alexander Golubov, Yukio Tanaka and Dieter Weiss for their valuable comments. This work was supported by the Australian Research Council Centre of Excellence in Future Low-Energy Electronics Technologies funded by the Australian Government, by the National Science Foundation of China (Project No. 61425015), National Basic Research Program of China (Project No. 2015CB921102), National Key Research and Development Program (Project No. 2016YFA0300600), and Strategic Priority Research Program of Chinese Academy of Sciences (Project No. XDB28000000) and 
by the German Research Foundation (DFG) through TRR 80.

\section*{References}

	
	 \bibliography{2D_Materials_IOP.bbl}

\end{document}